\newif\ifshowtodoos 
\showtodoostrue 
\ifshowtodoos
\newcommand{\todoo}[1]{{\color{red}{\textbf{TODOO: #1}}}}
\else
\newcommand{\todoo}[1]{}
\fi

\pdfoutput=1

\documentclass[superscriptaddress,nofootinbib,floatfix,notitlepage,aps,prd]{revtex4-1}

\usepackage{amsmath,amssymb,epsfig}
\usepackage[unicode]{hyperref}

\usepackage{natbib,ifthen}

\usepackage{mathrsfs}
\usepackage{simplewick}
\usepackage[lofdepth,lotdepth,caption=false]{subfig}
\usepackage{color}

\begin{document}


\newcommand{\jcap}{JCAP}
\newcommand{\apjl}{APJL~}


\newcommand{\lbar}[1]{\underline{l}_{#1}}
\newcommand{\drm}{\mathrm{d}}
\renewcommand{\d}{\mathrm{d}}
\newcommand{\gaensli}[1]{\lq #1\rq$ $}
\newcommand{\bartilde}[1]{\bar{\tilde #1}}
\newcommand{\barti}[1]{\bartilde{#1}}
\newcommand{\ti}{\tilde}
\newcommand{\oforder}[1]{\mathcal{O}(#1)}
\newcommand{\D}{\mathrm{D}}
\renewcommand{\[}{\left[}
\renewcommand{\]}{\right]}
\def\<{\left\langle}
\def\>{\right\rangle}
\newcommand{\mycaption}[1]{\caption{\footnotesize{#1}}}
\newcommand{\hattilde}[1]{\hat{\tilde #1}}
\newcommand{\mycite}[1]{[#1]}
\newcommand{\mnras}{Mon.\ Not.\ R.\ Astron.\ Soc.}
\newcommand{\apjs}{Astrophys.\ J.\ Supp.}
\newcommand{\physrep}{Physics Reports}

\newcommand{\fixme}[1]{{\textbf{Fixme: #1}}}
\newcommand{\detD}{{\det\!\cld}}
\newcommand{\clh}{\mathcal{H}}
\newcommand{\ud}{{\rm d}}
\renewcommand{\eprint}[1]{\href{http://arxiv.org/abs/#1}{#1}}
\newcommand{\adsurl}[1]{\href{#1}{ADS}}
\newcommand{\ISBN}[1]{\href{http://cosmologist.info/ISBN/#1}{ISBN: #1}}
\newcommand{\vort}{\varpi}
\newcommand\ba{\begin{eqnarray}}
\newcommand\ea{\end{eqnarray}}
\newcommand\be{\begin{equation}}
\newcommand\ee{\end{equation}}
\newcommand\lagrange{{\cal L}}
\newcommand\cll{{\cal L}}
\newcommand\cln{{\cal N}}
\newcommand\clx{{\cal X}}
\newcommand\clz{{\cal Z}}
\newcommand\clv{{\cal V}}
\newcommand\cld{{\cal D}}
\newcommand\clt{{\cal T}}

\newcommand\clo{{\cal O}}
\newcommand{\cla}{{\cal A}}
\newcommand{\clp}{{\cal P}}
\newcommand{\clr}{{\cal R}}
\newcommand{\uD}{{\mathrm{D}}}
\newcommand{\calE}{{\cal E}}
\newcommand{\calB}{{\cal B}}
\newcommand{\curl}{\,\mbox{curl}\,}
\newcommand\del{\nabla}
\newcommand\Tr{{\rm Tr}}
\newcommand\half{{\frac{1}{2}}}
\newcommand\fourth{{1\over 8}}
\newcommand\bibi{\bibitem}
\newcommand{\kf}{\beta}
\newcommand{\kfprod}{\alpha}
\newcommand\calS{{\cal S}}
\renewcommand\H{{\cal H}}
\newcommand\K{{\rm K}}
\newcommand\mK{{\rm mK}}
\newcommand\synch{\text{syn}}
\newcommand\opacity{\tau_c^{-1}}

\newcommand{\Psil}{\Psi_l}
\newcommand{\bsigma}{{\bar{\sigma}}}
\newcommand{\bI}{\bar{I}}
\newcommand{\bq}{\bar{q}}
\newcommand{\bv}{\bar{v}}
\renewcommand\P{{\cal P}}
\newcommand{\numfrac}[2]{{\textstyle \frac{#1}{#2}}}

\newcommand{\la}{\langle}
\newcommand{\ra}{\rangle}
\newcommand{\lla}{\left\langle}
\newcommand{\rra}{\right\rangle}

\newcommand{\Omtot}{\Omega_{\mathrm{tot}}}
\newcommand\xx{\mbox{\boldmath $x$}}
\newcommand{\phpr} {\phi'}
\newcommand{\gam}{\gamma_{ij}}
\newcommand{\sqgam}{\sqrt{\gamma}}
\newcommand{\delk}{\Delta+3{\K}}
\newcommand{\dph}{\delta\phi}
\newcommand{\om} {\Omega}
\newcommand{\dom}{\delta^{(3)}\left(\Omega\right)}
\newcommand{\rar}{\rightarrow}
\newcommand{\Rar}{\Rightarrow}
\newcommand\gsim{ \lower .75ex \hbox{$\sim$} \llap{\raise .27ex \hbox{$>$}} }
\newcommand\lsim{ \lower .75ex \hbox{$\sim$} \llap{\raise .27ex \hbox{$<$}} }
\newcommand\bigdot[1] {\stackrel{\mbox{{\huge .}}}{#1}}
\newcommand\bigddot[1] {\stackrel{\mbox{{\huge ..}}}{#1}}
\newcommand{\Mpc}{\text{Mpc}}
\newcommand{\Al}{{A_l}}
\newcommand{\Bl}{{B_l}}
\newcommand{\eAl}{e^\Al}
\newcommand{\ix}{{(i)}}
\newcommand{\ixp}{{(i+1)}}
\renewcommand{\k}{\beta}
\newcommand{\HD}{\mathrm{D}}

\newcommand{\nonflat}[1]{#1}
\newcommand{\Cgl}{C_{\text{gl}}}
\newcommand{\Cgltwo}{C_{\text{gl},2}}
\newcommand{\He}{{\rm{He}}}
\newcommand{\Mhz}{{\rm MHz}}
\newcommand{\vx}{{\mathbf{x}}}
\newcommand{\ve}{{\mathbf{e}}}
\newcommand{\vv}{{\mathbf{v}}}
\newcommand{\vk}{{\mathbf{k}}}
\newcommand{\vL}{{\mathbf{L}}}
\newcommand{\vn}{{\mathbf{n}}}

\newcommand{\vnhat}{{\hat{\mathbf{n}}}}
\newcommand{\vkhat}{{\hat{\mathbf{k}}}}
\newcommand{\taueps}{{\tau_\epsilon}}

\newcommand{\vgrad}{{\mathbf{\nabla}}}
\newcommand{\fbarln}{\bar{f}_{,\ln\epsilon}(\epsilon)}

\newcommand{\secref}[1]{Section \ref{#1}}
\newcommand{\expt}{\mathrm{expt}}
\newcommand{\eq}[1]{(\ref{eq:#1})} 
\newcommand{\fig}[1]{Fig.~\ref{fig:#1}} 
\renewcommand{\to}{\rightarrow}
\renewcommand{\[}{\left[}
\renewcommand{\]}{\right]}
\renewcommand{\vec}[1]{\mathbf{#1}}
\newcommand{\vy}{\vec{y}}
\newcommand{\vz}{\vec{z}}
\newcommand{\vq}{\vec{q}}
\newcommand{\VPsi}{\vec{\Psi}}
\newcommand{\vecv}{\vec{v}}
\newcommand{\vnabla}{\vec{\nabla}}
\newcommand{\vl}{\vec{l}}
\newcommand{\VL}{\vec{L}}
\newcommand{\dl}{\d^2\vl}
\newcommand{\valpha}{\vec{\alpha}}
\renewcommand{\L}{\mathscr{L}}

\newcommand{\abs}[1]{\lvert #1\rvert}
\newcommand{\ul}{\underline{l}}


\thispagestyle{empty}

\title{On the joint analysis of CMB temperature and lensing-reconstruction power spectra}

\author{Marcel M.~Schmittfull}
\affiliation{DAMTP, Centre
for Mathematical Sciences, Wilberforce Road, Cambridge CB3 0WA, U.K.}

\author{Anthony Challinor}
\affiliation{Institute of Astronomy and Kavli Institute for Cosmology, Madingley Road, Cambridge CB3 0HA, U.K.}
 \affiliation{DAMTP, Centre
for Mathematical Sciences, Wilberforce Road, Cambridge CB3 0WA, U.K.}

\author{Duncan Hanson}
\affiliation{Department of Physics, McGill University, Montreal QC H3A 2T8, Canada}

\author{Antony Lewis}
\homepage{http://cosmologist.info}
\affiliation{Department of Physics \& Astronomy, University of Sussex, Brighton BN1 9QH, U.K.}

\date{\today}

\begin{abstract}
Gravitational lensing provides a significant source of cosmological information in modern CMB parameter analyses. It is measured in both the power spectrum and trispectrum of the temperature fluctuations. These observables are often treated as independent, although as they are both determined from the same map this is impossible. In this paper, we perform a rigorous analysis of the covariance between lensing power spectrum and trispectrum analyses.
We find two dominant contributions coming from: (i) correlations between the disconnected noise bias in the trispectrum measurement and sample variance in the temperature power spectrum; and (ii) sample variance of the lenses themselves.
The former is naturally removed when the dominant $N^{(0)}$ Gaussian bias in the reconstructed deflection spectrum is dealt with via a partially data-dependent correction, as advocated elsewhere for other reasons. The remaining lens-cosmic-variance contribution is easily modeled but can safely be ignored for a Planck-like experiment, justifying treating the two observable spectra as independent. We also test simple likelihood approximations for the deflection power spectrum, finding that a Gaussian with a parameter-independent covariance performs well. 
\end{abstract}

\maketitle

\section{Introduction}
\label{sec:introduction}

Weak gravitational lensing by large-scale structure leaves subtle imprints in the temperature anisotropies of the cosmic microwave background (CMB); see~\cite{phys-repts}
for a review. These imprints can be detected in surveys with resolution better than a few arcminutes and used to reconstruct the lensing deflection field~\cite{1999PhRvD..59l3507Z,2001ApJ...557L..79H}. Since the lensing deflections depend on the growth of structure and geometry at much lower redshifts ($z \sim 2$) than the CMB last-scattering surface, lens reconstructions can be used to constrain parameters that are largely degenerate in the primary anisotropies sourced at last-scattering. Examples include sub-eV neutrino masses, spatial curvature, dark energy and modifications to gravity (e.g.~\cite{2002PhRvD..65b3003H,2003PhRvL..91x1301K,2002PhRvD..65d3007V,2004PhRvD..70b3515A,calabrese0803LensingAmplitude,2012MNRAS.425.1170H}). 

Lensing is an emerging frontier of observational cosmology. 
The first direct measurements of the deflection power spectrum were reported recently by
the ACT~\cite{ACTLensingDetectionDas1103,2013arXiv1301.1037D}, SPT~\cite{SPTLensingDetectionVanEngelen1202} and Planck~\cite{2013arXiv1303.5077P} teams with significances of $4.6$, $6.3$ and
$25\,\sigma$,
respectively. These measurements provide the first evidence for dark energy from the
CMB \emph{alone}.  Since lens reconstructions are
quadratic in the temperature anisotropies, the power spectrum of the reconstruction
is probing the 4-point non-Gaussianity of the CMB induced by lensing~\cite{huLensedTTrispectrum0105}.
The statistical power of lens reconstructions is expected to improve rapidly with ongoing analyses of the full $2500\,\mathrm{deg}^2$ SPT survey and the full-mission data from Planck, which 
also allow for polarization-based lensing reconstruction.
Lensing also affects the power spectrum (or 2-point function) of the temperature
anisotropies, smoothing the acoustic peaks and transferring power from large to small scales (e.g.~\cite{1996ApJ...463....1S}). The smoothing effect has been detected at nearly $10\,\sigma$ in current temperature power spectrum measurements~\cite{2012arXiv1210.7231S,2013arXiv1303.5076P}.

A question that has received only limited attention to date is how one should
model the likelihood of the lensed CMB anisotropies when deriving constraints on
cosmological parameters. As the unlensed CMB and deflection field can be approximated as Gaussian on the scales relevant for CMB lensing, it is
straightforward to write down a formal expression for the likelihood of the lensed
temperature~\cite{2003PhRvD..67d3001H}. However, this is very difficult to work with directly. Indeed,
working with the exact likelihood even for Gaussian fields in mega-pixel maps is
computationally prohibitive. Instead, in the Gaussian case, at high multipoles
the data is usually compressed to an empirical power spectrum (or set of cross-spectra)
and an approximate likelihood is constructed based on this spectrum and its covariance.
Such an approach is both computationally feasible and allows for robust treatment of instrumental effects such as beam asymmetry. For non-Gaussian fields,
like the lensed CMB, working with only the empirical power spectrum is clearly lossy. Instead, we should include further empirical connected $n$-point functions in our
compressed data. In the context of CMB lensing, the 4-point function is the most
relevant moment and the information it carries is captured in the estimated power spectrum of the reconstructed deflection field.

Parameter analyses involving lens reconstruction to date have followed the route
described above. 
However, they have simply combined estimates of the temperature power spectrum and the lensing power spectrum as if they were independent~\cite{2011PhRvL.107b1302S,SPTLensingDetectionVanEngelen1202,2013arXiv1303.5077P}. Since both power spectra are derived from the same CMB temperature map, one might question the validity of this approach, raising the concern that lensing information is inadvertently being double counted. While early lensing forecasts \cite{2002PhRvD..65b3003H,2003PhRvL..91x1301K,2006PhRvD..73d5021L} addressed this by using \emph{unlensed} CMB power spectra, an optimal combination of the observed lensed CMB 2- and 4-point functions should model their cross-covariance. Intuitively, we might expect two effects to be relevant.
 First, the statistical noise in the reconstruction, due to chance alignments in the unlensed CMB which mimic the locally-anisotropic effects of lensing, is dependent on the CMB fluctuations themselves. If, due to cosmic
variance, the unlensed temperature anisotropies fluctuate high at some particular scale, the noise in the lens reconstruction will also fluctuate high. The mode-coupling nature of lens reconstruction, whereby modes near the resolution limit of the observation dominate the reconstruction on much larger scales, will lead to broad correlations
between the temperature and lensing power spectra. Since the reconstruction is quadratic, the correlation of the power spectra involves the CMB 6-point function, the disconnected part of which arises from the effect just decribed. The second
effect is due to the cosmic variance of the lenses. If the lensing field fluctuates high
at some scale, the reconstructed power will also fluctuate high at the same scale.
In parallel, there will be more smoothing of the acoustic peaks in the measured
CMB power spectrum giving an anti-correlation with the reconstucted lensing
power at the location of acoustic peaks and a positive correlation at troughs.
This effect is
second order in the deflection power and derives from the connected
part of the CMB 6-point function. 
As we shall show, the induced \emph{correlations} are rather small (a few percent) for a Planck-like experiment for broad-band measures of the lensing power such as a lensing power spectrum amplitude.
Essentially, this is because there are a limited number of modes of
the lensing power spectrum that influence the acoustic part of the temperature power spectrum, 
and the correlation due to cosmic variance of these modes is diluted by the significant 
noise due to cosmic variance of the CMB and instrumental noise 
(i.e.~the fact that lensing measurements from the CMB 2- and 4-point function are not limited by cosmic variance of the lenses).
Moreover, the first effect mentioned above produces rather small lensing amplitude correlations since CMB modes at different scales fluctuate independently, and most of the information on peak smearing in the CMB power spectrum comes from modes near the acoustic peaks and troughs, whereas the reconstruction of (large-scale) lenses is most effective at CMB scales where the CMB spectrum changes most rapidly, i.e.~between acoustic peaks and troughs.
Note also that even the unlensed CMB anisotropies are correlated with
the lensing field due to the late-time integrated Sachs-Wolfe effect~\cite{1999PhRvD..59j3002G}.
This mostly produces a diagonal covariance between the power spectra
of the lens reconstruction and the lensed CMB which falls rapidly at small scales ($L>100$).
One of our main aims in this paper
is to quantify these arguments with detailed calculations of the 6-point function,
verified with simulations,
and to assess whether the correlations are small enough to be safely ignored in the
likelihood.

A further issue concerns the form of the likelihood for the power spectrum estimates.
This has been well studied for Gaussian fields, e.g.~\cite{2000ApJ...533...19B,2003ApJS..148..195V,2006PhRvD..73b3517S,2008PhRvD..77j3013H}, and simple
approximate forms are known to give reliable parameter constraints when applied
on all but the largest scales. However, the lensing reconstruction is
quadratic in the nearly Gaussian CMB fluctuations and therefore highly non-Gaussian.
Here, we test a particularly simple form of the lensing power likelihood, a Gaussian
with model-independent covariance. 
Constraining the
amplitude and tilt of a fiducial deflection power spectrum, we demonstrate that
the fiducial-Gaussian approximation performs well, returning maximum-likelihood
points that scatter across simulations in a manner consistent with the width of the
likelihood.

The paper is organised as follows. We review CMB lensing
reconstruction in Sec.~\ref{se:fullSkyLensing} and we describe our
simulations of lensed CMB maps and the mechanics of our reconstructions in Sec.~\ref{se:sims}.  Section~\ref{se:powerAutoCorrels} surveys known results for the
auto-correlations of the lensed CMB temperature power spectrum and the
reconstruction power spectrum. In Sec.~\ref{se:temperatureLensingCrossCorrelation} we present results for
the cross-correlation of these power spectra and assess the
importance of correlations for estimating cosmological parameters. We test likelihood
approximations for the lensing reconstruction (in isolation) in Sec.~\ref{se:towardsLensingLikelihood} and conclude in Sec.~\ref{se:conclusions}.
In Appendix \ref{se:plausibilityOfSmallCorrel} we provide 
intuitive arguments for the magnitude of the
temperature-lensing power correlation. A series of further appendices
provide calculational details, and motivate some 
of the approaches taken in the main text. Table~\ref{tab:MathSymbolsLensingLikelihood} summarises the key quantities, and their definitions and symbols, used in this paper.

\begin{table*}[t]
\centering
\renewcommand{\arraystretch}{1.0}
{\footnotesize
\begin{tabular}{@{}lllll@{}}
\hline 
Symbol & \phantom{i}
 & Description &\phantom{i} & Definition \\
\hline 
$T$ && Unlensed CMB temperature && Eq.~\eqref{eq:remapping} \\
$\tilde T$ && Lensed CMB temperature && Eq.~\eqref{eq:remapping} \\
$\phi_\text{in}$ && Input lensing potential field in simulations && \\
$\hat\phi$ && Lensing reconstruction (quadratic in $\tilde T$) && Eq.~\eqref{eq:phi_esti} \\
$A_L$, $\tilde g$ && Normalisation and weights for $\hat\phi$ && Eqs.~\eqref{eq:A_L_def}, \eqref{eq:gtilde_def}
\vspace*{2mm} \\
$C^{XX}_l$ && Fiducial theoretical power spectrum of $X=T$, $\tilde T$ or $\phi$ (without noise/beam) && \\
$C^{\tilde T\tilde T}_{l,\expt}$ && Lensed temperature power spectrum including beam-deconvolved noise && Eq.~\eqref{eq:cl-expt} \\
$\hat C^{XX}_l$ && Empirical power spectrum of a realisation of $X=T$, $\tilde T$, $\phi_\text{in}$ or $\hat\phi$ && Eq.~\eqref{eq:powerspecDef} \\
$N^{(0)}_l$ && Gaussian, fully disconnected noise bias of $\hat C^{\hat\phi\hat\phi}_l$ && Eq.~\eqref{eq:N0} \\
$N^{(1)}_l$, $N^{(2)}_l$ && Biases of $\hat C^{\hat\phi\hat\phi}_l$ at $\mathcal{O}(C^{\phi\phi})$ and $\mathcal{O}[(C^{\phi\phi})^{2}]$ && Ref.~\cite{hanson1008}  \\
$\hat N^{(0)}_l$ && Data-dependent, empirical bias subtraction term && Eqs.~\eqref{eq:empirical_N0}, \eqref{eq:empN0_subtraction_with_derivatives} 
  \vspace*{2mm} \\
$\mathrm{cov}(\hat C^{\hat\phi\hat\phi}_L,\hat C^{\tilde T\tilde T}_{L',\expt})$ && Covariance of CMB temperature and lensing reconstruction power spectra && Eq.~\eqref{eq:cov_CTT_Cphiphi_rec}  \\
$\mathrm{cov}(\hat C^{\hat\phi\hat\phi}_L,\hat C^{\tilde T\tilde T}_{L',\expt})_\text{disconn.}$ && Noise contribution (from Gaussian, fully disconnected CMB 6-point function) && Eqs.~\eqref{eq:noisecontri_cov_CTT_Cphiphi_rec}, \eqref{eq:noise_correlation_heuristic} \\
$\mathrm{cov}(\hat C^{\hat\phi\hat\phi}_L,\hat C^{\tilde T\tilde T}_{L',\expt})^{\text{conn.4pt.}X}_\text{[(non-)primary]}$ && Trispectrum contributions of type $X=A,B$ [from (non-)primary coupling] && Appendix \ref{se:Conn4ptAppendix} \\
$\mathrm{cov}(\hat C^{\hat\phi\hat\phi}_L,\hat C^{\tilde T\tilde T}_{L',\expt})_{\la\tilde T_1\cdots\tilde T_6\ra_{c,\mathrm{dom.}}^{(4)}}$ && Matter cosmic variance contribution [from $\mathcal{O}(\phi^4)$ connected CMB 6-point function]  && Eqs.~\eqref{eq:cov_CTT_Cphiphi_IN}, \eqref{eq:matter_contri_theo} 
\vspace*{2mm}\\
$A$ && Overall lensing amplitude of a fiducial lensing power spectrum && 
$C^{\phi\phi}_l$$=$$AC^{\phi\phi}_l|_\text{fid}$ \\
$\hat A$ && Estimator for $A$ based on reconstruction power spectrum, i.e.~CMB trispectrum && Eq.~\eqref{eq:Aphiphi_estimator}  \\
$\hat A'$ && Estimator for $A$ based on CMB power spectrum && Eq.~\eqref{eq:ATT_estimator} \\
$n$ && Tilt of a fiducial lensing power spectrum $C^{\phi\phi}_l|_\text{fid}$ && Eq.~\eqref{eq:lensing_tilt} 
 \vspace*{0.5mm}\\
\hline 
\end{tabular}
} 
\caption{Summary of the main quantities used throughout this paper (roughly ordered after first appearance).}   
\label{tab:MathSymbolsLensingLikelihood}
\end{table*}

\section{CMB lensing reconstruction}

\label{se:fullSkyLensing}

The lensed CMB temperature $\tilde{T}(\vnhat)$ along direction $\vnhat$ is related to the unlensed temperature $T(\vnhat)$ by the deflection field $\valpha(\hat\vn)$\footnote{The notation here is rather symbolic on the spherical
sky: the point $\vnhat + \valpha(\vnhat)$ is understood to be obtained
from $\vnhat$ by displacing through a distance $|\valpha(\vnhat)|$ along the geodesic
that is tangent to $\valpha$ at $\vnhat$~\cite{2002PhRvD..66l7301C}.}
\begin{equation}
  \label{eq:remapping}
  \tilde T(\hat\vn) =  T[\hat\vn+\valpha(\hat\vn)].
\end{equation}
The deflection angle can be written in the Born approximation as the angular gradient of a lensing potential, $\valpha(\vnhat) = \vnabla \phi(\hat\vn)$, which is
given by an integral along the (unperturbed) line of sight of the gravitational potentials
(e.g.~\cite{phys-repts}).

On the full sky, the effect of lensing on the CMB temperature can be expressed
perturbatively by Taylor expanding Eq.~(\ref{eq:remapping}). Expanding in
spherical harmonics, the multipoles of the lensed CMB, $\tilde{T}_{lm}$, are related
to those of the unlensed CMB $T_{lm}$ and the lensing potential $\phi_{lm}$ via~\cite{huHarmonicCMBLensing0001303}
\begin{equation}
  \label{eq:deltaT_expansion}
  \tilde T_{lm}=T_{lm}+ \delta T_{lm}+\delta^{2}T_{lm}+\cdots,
\end{equation}
where changes due to lensing $\delta^n T_{lm}$ are of order $\mathcal{O}(\phi^n)$  and linear in the
unlensed temperature $T$:
\begin{eqnarray}
  \label{eq:deltaT_definitions}
  \delta T_{lm}&=&\sum_{\underline l_1, \underline l_2}\phi_{\underline
    l_1}T_{\underline l_2}I^{mm_1m_2}_{ll_1l_2},\\
  \delta^2 T_{lm}&=&\sum_{\underline l_1, \underline l_2, \underline l_3}\phi_{\underline
    l_1}\phi_{\underline l_2}T_{\underline l_3}J^{mm_1m_2m_3}_{ll_1l_2l_3},
\end{eqnarray}
where we have introduced the notation $\underline l\equiv (lm)$ and $I$
denotes an angular integral over a product of (derivatives of)
spherical harmonics given by \cite{huHarmonicCMBLensing0001303}
\begin{eqnarray}
  \label{eq:IJ_definitions}
  I^{mm_1m_2}_{ll_1l_2} &=& (-1)^m
\left( \begin{matrix} 
l & l_1 & l_2 \\ 
-m & m_1 & m_2
\end{matrix} \right)
F_{ll_1l_2}.
\end{eqnarray}
Expressions for $J^{mm_1m_2m_3}_{ll_1l_2l_3}$ can be found in
\cite{hanson1008}. The geometrical factor $F_{l_1Ll_2}$, which
is symmetric in the last two indices, is given by
\begin{equation}
  \label{eq:F_def}
  F_{l_1Ll_2} = \left[L(L+1)-l_1(l_1+1)+l_2(l_2+1)\right]
\sqrt{\frac{(2l_1+1) (2L+1) (2l_2+1)}{16\pi}} 
\left( \begin{matrix} 
l_1 & L & l_2 \\ 
0 & 0 & 0
\end{matrix} \right) ,
\end{equation}
and describes the rotationally-invariant part of the coupling between the three
multipoles.

It is possible to reconstruct the lensing potential $\phi(\hat\vn)$ from the observed CMB by exploiting the fact that fixed lenses introduce correlations between temperature modes~\cite{1999PhRvD..59l3507Z,2002ApJ...574..566H}.
Following the non-perturbative calculations in~\cite{lewis1101}, we have
\begin{equation}
\left\langle \frac{\partial}{\partial \phi_{LM} }(\tilde{T}_{l_1 m_1} \tilde{T}_{l_2 m_2}) \right\rangle
\approx (-1)^M \left( \begin{matrix} 
L & l_1 & l_2 \\ 
-M & m_1 & m_2
\end{matrix} \right)  \tilde{f}_{l_1 L l_2} ,
\label{eq:couple}
\end{equation}
where $\tilde f_{l_1Ll_2}$ is symmetric in $l_1$ and $l_2$ and
contains the lensed temperature power spectrum, $C^{\tilde T\tilde T}_l$:
\begin{equation}
  \label{eq:smallftilde_def}
  \tilde f_{l_1Ll_2}=C^{\tilde T\tilde T}_{l_2}F_{l_1Ll_2}+C^{\tilde
    T\tilde T}_{l_1}F_{l_2Ll_1}.
\end{equation}
The angle brackets in Eq.~(\ref{eq:couple}) denote the expectation value over
$\phi$ and $T$ (and noise) and we have neglected the $T$--$\phi$ correlation which is generally
a  good approximation for CMB lensing since we are usually interested in small-scale ($l>600$)
modes of the temperature where the correlation is small. More generally, we shall
neglect the $T$--$\phi$ correlation throughout this paper, except where explicitly
stated otherwise. Equation~(\ref{eq:couple}) motivates forming a quadratic estimator
for the lensing potential~\cite{okamotoHu0301},
\begin{equation}
  \label{eq:phi_esti}
  \hat\phi_{LM}=A_L\sum_{\underline l_1\underline l_2} (-1)^M
\left( \begin{matrix} 
l_1 & l_2 & L \\ 
m_1 & m_2 & -M
\end{matrix} \right)
\tilde g_{l_1l_2}(L)\tilde T_{\underline l_1}\tilde T_{\underline l_2} .
\end{equation}
For any choice of weights $\tilde{g}_{l_1l_2}(L)$, we determine the normalisation $A_L$
by demanding that $\langle \partial \hat{\phi}_{LM} / \partial \phi_{L'M'} \rangle =
\delta_{LL'} \delta_{MM'}$ which gives
\begin{equation}
  \label{eq:A_L_def}
  A_L=(2L+1)\left(\sum_{ l_1 l_2}
\tilde{f}_{l_1Ll_2}\tilde{g}_{l_1l_2}(L)
\right)^{-1} .
\end{equation}
One can determine optimal weights by minimising the variance of the estimator
to find\footnote{The normalisation \eqref{eq:A_L_def} and weight \eqref{eq:gtilde_def} differ a little from that in the original works of Hu
and coworkers~\cite{2001ApJ...557L..79H,okamotoHu0301} since $\tilde{f}_{l_1Ll_2}$ involves the power
spectrum of the lensed rather than unlensed CMB. The form here is particularly suited to power
spectrum estimation. }
\begin{equation}
  \label{eq:gtilde_def}
  \tilde g_{l_1l_2}(L)=\frac{\tilde{f}_{l_1 L l_2}}{2C_{l_1,\expt}^{\tilde T\tilde T}   C_{l_2,\expt}^{\tilde T\tilde T}},
\end{equation}
where the numerator contains the lensed temperature power
$C^{\tilde T\tilde T}_l$, while the denominator involves
the total power spectrum for the experiment $C^{\tilde T\tilde T}_{l,\expt}$
including beam-deconvolved noise; see Eq.~\eqref{eq:cl-expt} below. 

The expectation value of the simple power spectrum estimate
\begin{equation}
\hat{C}_L^{\hat{\phi}\hat{\phi}} \equiv \frac{1}{2L+1} \sum_{M=-L}^L \hat{\phi}_{LM}
\hat{\phi}_{LM}^\ast , 
\end{equation}
involves the 4-point function of the lensed CMB. The connected part of the
4-point function can be written in terms of the fully-reduced trispectrum
$\mathbb{T}^{l_1 l_2}_{l_3 l_4}(L)$ as~\citep{huLensedTTrispectrum0105}
\begin{equation}
\langle \tilde T_{l_1 m_1} \tilde T_{l_2 m_2}\tilde T_{l_3 m_3}\tilde T_{l_4 m_4} \rangle_c = \frac{1}{2} \sum_{LM} (-1)^M \left( \begin{array}{ccc} l_1 & l_2 & L \\
m_1 & m_2 & M\end{array}\right) 
\left( \begin{array}{ccc} l_3 & l_4 & L \\
m_3 & m_4 & -M\end{array}\right) \mathbb{T}^{l_1 l_2}_{l_3 l_4}(L)
+ \mathrm{perms}\, .
\label{eq:4pt}
\end{equation}
By evaluating the trispectrum correct to $\clo[(C_L^{\phi\phi})^2]$, Ref.~\cite{hanson1008} shows that
the dominant terms for lensing reconstruction can be approximated by
\begin{equation}
\mathbb{T}^{l_1 l_2}_{l_3 l_4}(L) \approx \frac{1}{4} C_L^{\phi\phi}
\tilde{f}_{l_1 L l_2} \tilde{f}_{l_3 L l_4} ,
\label{eq:lensing_trispectrum}
\end{equation}
which involves the lensed $C_l^{\tilde T\tilde T}$ (via $\tilde{f}_{l_1 L l_2}$ etc.). The normalisation
of Eq.~(\ref{eq:A_L_def}) therefore correctly normalises the power spectrum
$\hat{C}_l^{\hat{\phi}\hat{\phi}}$, avoiding a bias [$N^{(2)}$] of around $-15\%$ on large scales
that results from using the unlensed spectra, i.e.\ $f_{l_1 L l_2}$, in the normalisation
$A_L$.

Taking the expectation value $\langle \hat{C}_L^{\hat{\phi}\hat{\phi}} \rangle$,
and using Eq.~\eqref{eq:4pt}, gives~\cite{KCK0302N1,hanson1008}
\begin{equation}
  \label{eq:reconstruction_power_biases}
  \la\hat C^{\hat\phi\hat\phi}_L\ra = C_L^{\phi\phi}+N^{(0)}_L+N^{(1)}_L+\mathcal{O}((C^{\phi\phi})^3),
\end{equation}
where $N^{(n)}_L$ is of order $(C^{\phi\phi})^n$ if we do not count
appearances of $C^{\phi\phi}$ in the lensed temperature power
spectrum. The disconnected part of the lensed temperature $4$-point function
leads to the Gaussian bias 
\begin{equation}
  \label{eq:N0}
N^{(0)}_L=\frac{2A_L^2}{2L+1}\sum\limits_{l_1,l_2}\tilde g^2_{l_1l_2}(L) C^{\tilde
  T\tilde T}_{l_1,\mathrm{expt}} C^{\tilde T\tilde
  T}_{l_2,\mathrm{expt}}= A_L,
\end{equation}
where the last equality holds only if the weights $\tilde g_{l_1 l_2}(L)$ are given by
Eq.~\eqref{eq:gtilde_def}. Since this bias is present even in the absence of lensing, it corresponds
to the power spectrum generated by the statistical noise of the
lensing reconstruction (see Appendix~\ref{se:N0BiasAppendix} for
a more detailed discussion of how this bias is generated).
The $N^{(1)}$ bias is due to those permutations in Eq.~\eqref{eq:4pt} that
mix multipoles between the primary $(l_1 m_1, l_2 m_2)$ and
$(l_3 m_3,l_4 m_4)$ couplings, e.g.\ $1 \leftrightarrow 3$. It
has been computed in \cite{KCK0302N1,hanson1008}; see also \fig{fig1-recphi}.
An unbiased estimate for the lensing
potential power spectrum can be obtained by subtracting $N^{(0)}+N^{(1)}$
from $\hat C^{\hat\phi\hat\phi}$. Here, for $N^{(1)}$ we subtract the bias evaluated
in the fiducial cosmology used for our simulations. In practice, the variation of
$N^{(1)}$ with the cosmological model should be included in the likelihood
analysis. However, since direct evaluation of this term is slow, a faster alternative is
to include the uncertainty in $N^{(1)}$ as a correlated error in the covariance matrix
for $\hat{C}_L^{\hat{\phi}\hat{\phi}}$\footnote{Note also that the normalisation $A_L$ is dependent on the cosmological model
through $\tilde{f}_{l_1 L l_2}$. However, the lensed temperature power spectrum,
and hence $A_L$, generally varies much less across models consistent with the
measured CMB power spectrum than $C_L^{\phi\phi}$ (and so $N^{(1)}$).
In practical applications, if a fiducial model is assumed to normalise the
4-point function, the parameter-dependent $C_L^{\phi\phi}$ can easily be renormalised
in the likelihood by the ratio of the fiducial $A_L$ to that at the current location
in parameter space~\cite{2013arXiv1303.5077P}.}.

The subtraction of the Gaussian $N^{(0)}$ bias can be done in various ways. The
simplest is to subtract a fiducial model. For temperature reconstructions,
$N^{(0)}$ generally exceeds the signal power on all scales, and by around two orders
of magnitude at multipoles $l \sim 1000$, so that accurate $N^{(0)}$ subtraction
is critical. For the idealised isotropic surveys considered here, this is perhaps not
too problematic since only $C_{l,\expt}^{\tilde T\tilde T}$ is required to calculate
$N^{(0)}$ and this can be estimated from a smoothed version of the measured power
spectrum. However, in the presence of survey anisotropies, the Gaussian bias
must generally be subtracted via simulations and this requires an accurate procedure
for simulating maps including all relevant real-world effects such as noise
inhomogeneities and correlations, beam asymmetry and unresolved foreground
emission. A workaround to these issues is to use alternative data-dependent
forms of $N^{(0)}$ (see~\cite{2012arXiv1209.0091N} and references therein) whose expectation value
either equals $N^{(0)}$ or closely approximates it. Here, we use the
form (for an isotropic survey) $\hat C_L^{\hat\phi\hat\phi}-(2\hat N_L^{(0)}-N_L^{(0)})$,
advocated in~\cite{hanson1008,2012arXiv1209.0091N}, where
\begin{equation}
  \label{eq:empirical_N0}
\hat N^{(0)}_L=\frac{2A_L^2}{2L+1}\sum\limits_{l_1,l_2}\tilde
g^2_{l_1l_2}(L) \hat C^{\tilde
  T\tilde T}_{l_1,\mathrm{expt}} C^{\tilde T\tilde T}_{l_2,\mathrm{expt}},
\end{equation}
i.e.~we replace one occurence of $C^{\tilde T\tilde T}_{l,\mathrm{expt}}$ in $N^{(0)}$ with
the empirical temperature power spectrum of our given sky. Note that the
expectation value $\langle \hat N^{(0)}_L \rangle = N_L^{(0)}$. As well as reducing
the impact of modelling errors, this form of $N^{(0)}$ subtraction has the benefit of
greatly reducing correlations between the $\hat{C}_L^{\phi\phi}$ estimates that arise
from the disconnected part of the CMB 8-point function; see Sec.~\ref{subsec:recon_autocov} and Ref.~\cite{hanson1008}. As we shall see in Sec.~\ref{subsec:noise},
it also eliminates the Gaussian contribution to the covariance between $\hat{C}_L^{\hat{\phi}\hat{\phi}}$ and the measured temperature power spectrum.
Further motivation for the $2 \hat N^{(0)}_L - N^{(0)}_L$ construction comes
from considering optimal trispectrum estimation for weakly non-Gaussian
fields~\cite{2010PhRvD..82b3520R}; these arguments are discussed further in the context of CMB lensing in Appendix~\ref{app:edgeworth}.

Throughout this paper we assume that the instrumental beam has been
deconvolved from the $\tilde{T}_{lm}$. The total power spectrum of the experiment
is then of the form
\begin{equation}
  \label{eq:cl-expt}
  C_{l,\mathrm{expt}}^{\tilde T\tilde T} =
  C_l^{\tilde T\tilde T} + \sigma_N^2 
  \exp[l(l+1)\sigma^2_{\mathrm{FWHM}}/(8\ln 2)],
\end{equation}
where $\sigma_N^2$ is the white noise power spectrum and $\sigma_{\mathrm{FWHM}}$
is the full width at half-maximum (FWHM) of the optical beam.
Following~\cite{hanson1008}, we will use $\sigma_N=27\,
\mu\mathrm{K}\,\mathrm{arcmin}$ and $\sigma_\mathrm{FWHM}=7\,
\mathrm{arcmin}$, which is roughly appropriate for Planck.

\section{Simulations}

\label{se:sims}

We use simulations to verify our analytic arguments. These are based on
$1000$ realisations of a flat $\Lambda$CDM cosmology with
WMAP7+BAO+$H_0$ parameters \cite{wmap-lambda-website}
$h=0.704$, $\Omega_bh^2=0.0226$, $\Omega_ch^2=0.1123$, $\tau=0.087$,
$n_s=0.963$, $\Delta_\mathcal{R}^2=2.441\times 10^{-9}$ at $k_0=0.002
\mbox{ Mpc}^{-1}$ and three massless neutrino species. We start with
realisations of the unlensed temperature and lensing potential up
to $l_\mathrm{max}^\mathrm{in}=3000$ (including the $C^{T\phi}$ correlation
on large scales), which are then lensed using Lenspix \cite{lenspix}. 
The lensed temperature up to $l^T_\mathrm{max}=2750$
is used to reconstruct the lensing potential up to
$l^\phi_\mathrm{max}=2650$ with the full-sky simulation setup of
Hanson \emph{et al.}~\cite{hanson1008}.
The
 convolution in harmonic space in Eq.~\eqref{eq:phi_esti} is evaluated as a product in pixel
 space, where spin-1 spherical harmonic transforms are taken using HEALPix
 \cite{healpix}.
 As a slight modification to
\cite{hanson1008} we use the reconstruction normalisation $A_L$ and
weights $\tilde g$ given in Eqs.~\eqref{eq:A_L_def} and \eqref{eq:gtilde_def} to avoid the $N^{(2)}$ 
bias. 

\begin{figure}[tp]
\includegraphics[width=0.7\textwidth]{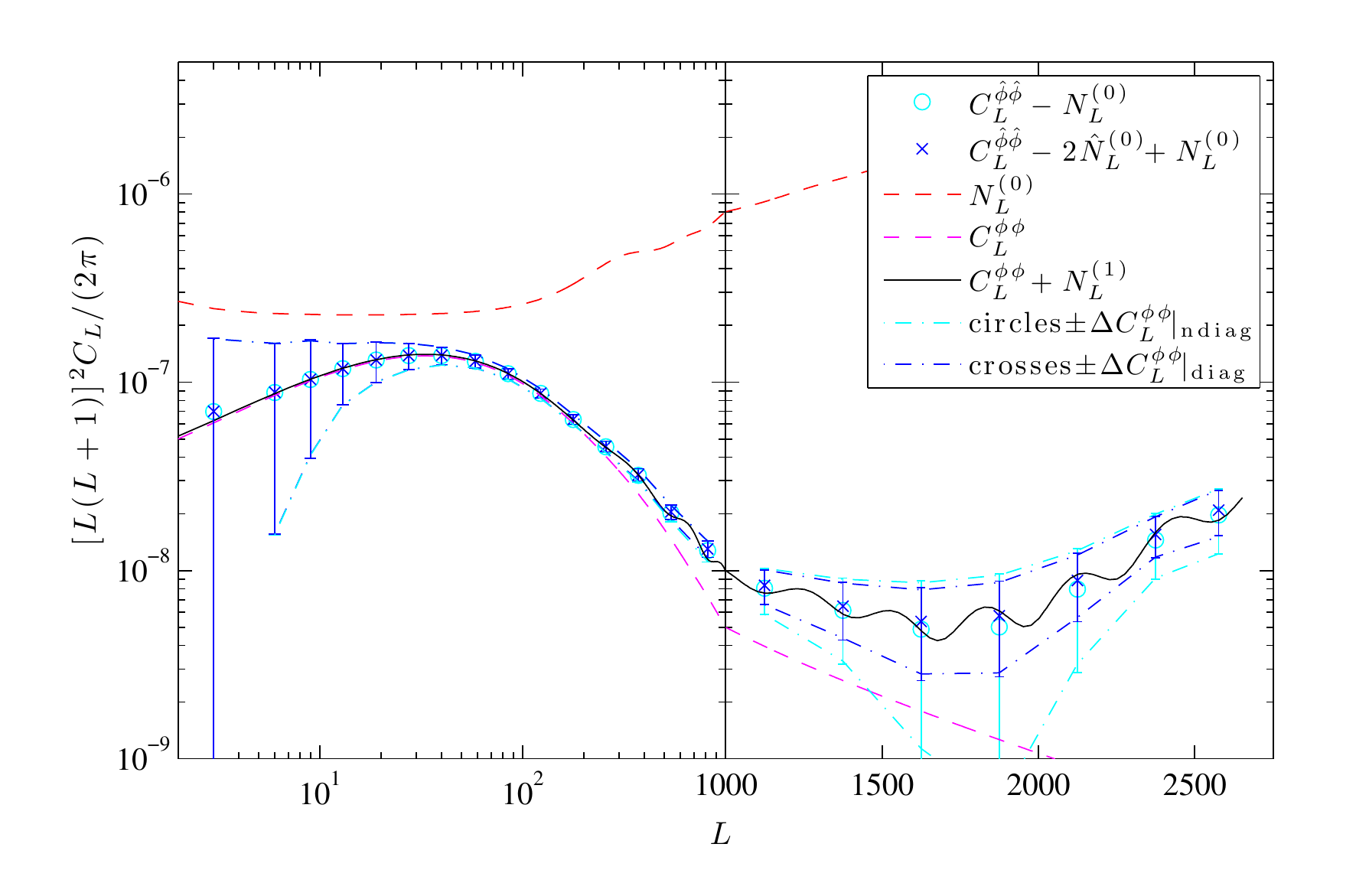}
\caption{Reconstructed lensing potential power spectrum
  $C_L^{\hat\phi\hat\phi}$ averaged over $1000$ simulations, after
  subtracting the analytically-calculated $N^{(0)}$ bias (cyan
  circles). The magenta line is the input lensing potential power.
  Blue symbols include the empirical reconstruction bias correction of Eq.~\eqref{eq:empirical_N0}.
  The error bars show the estimated standard deviation of the binned
  power spectrum  for a single realisation of the lensed CMB. Cyan error
  bars are only visible when they disagree from the blue error bars.  Cyan
  dash-dotted lines show theoretical error bars obtained by binning
  Eqs.~\eqref{eq:cov-phi-power} and \eqref{eq:var-phi-power} below.  Blue
  dash-dotted lines are theoretical error bars from
  Eq.~\eqref{eq:var-phi-power} only.  The biases $N^{(0)}$ (red) and
  $N^{(1)}$ (black, which also includes $C_L^{\phi\phi}$)
 are calculated analytically.  Note that the left-hand panel uses a log scale for multipole
$L$, whereas the right-hand panel uses a linear scale.}
\label{fig:fig1-recphi}
\end{figure}

\begin{figure}[tp]
\includegraphics[width=0.65\textwidth]{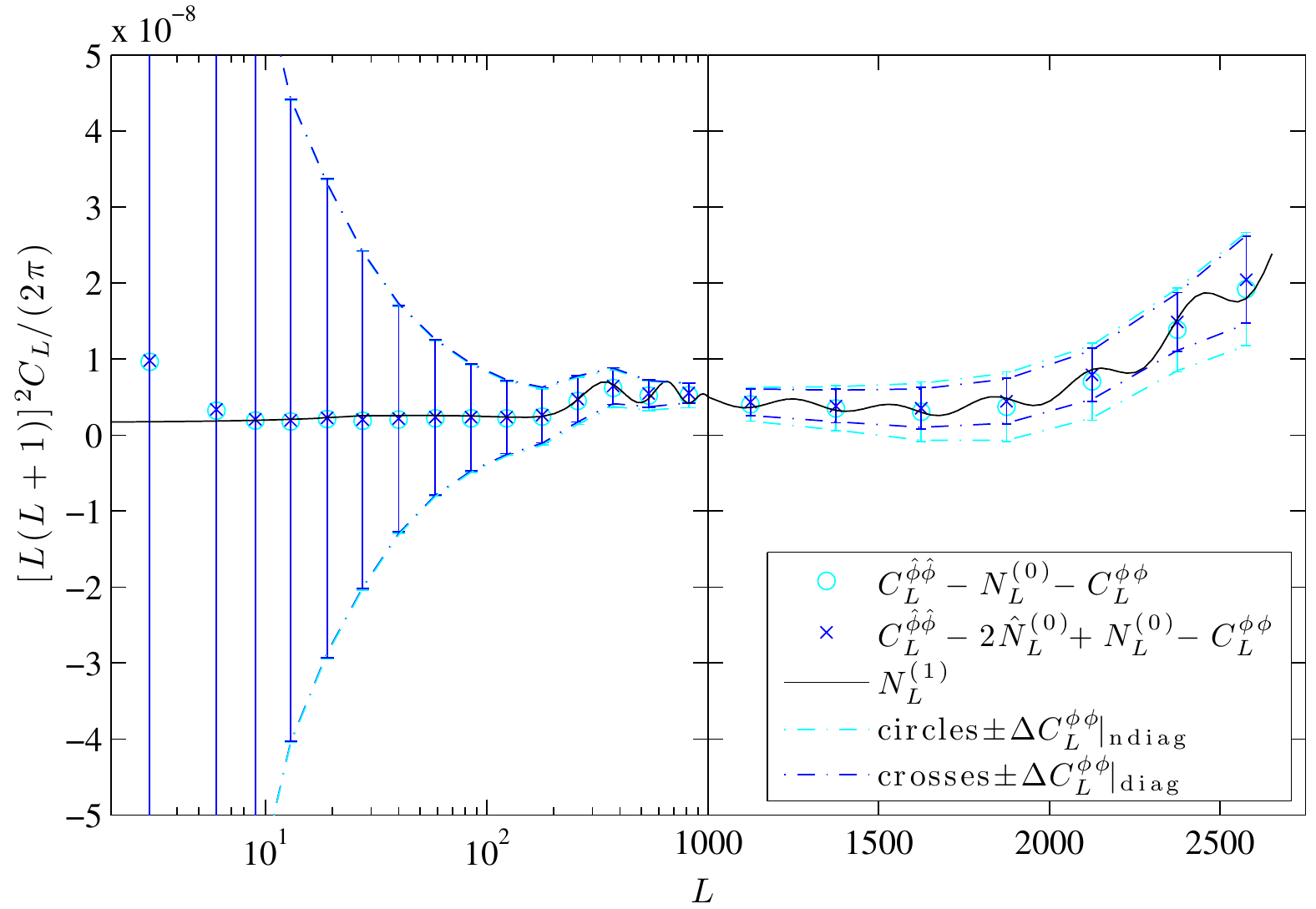}
\caption{Same as \fig{fig1-recphi} after subtracting the theoretical
  lensing power spectrum $C_L^{\phi\phi}$.}
\label{fig:fig2-recphi}
\end{figure}

Figures~\ref{fig:fig1-recphi} and~\ref{fig:fig2-recphi} confirm that the
power spectrum of the lensing reconstruction agrees with the input
lensing power spectrum if the $N^{(0)}$ and $N^{(1)}$ biases are taken
into account. Similar plots in \cite{hanson1008} contain the $N^{(2)}$
bias because their simulations used reconstruction weights and normalisation with
unlensed instead of lensed temperature power spectra.
The realisation-dependent $\hat N^{(0)}$ bias
correction  does not change the
expectation value of the reconstruction power spectrum but reduces its
covariance.

\section{Auto-correlations of power spectra}

\label{se:powerAutoCorrels}

We will argue later that parameter estimation with the lensing
reconstruction should be based on empirical power spectra defined by 
\begin{equation}
  \label{eq:powerspecDef}
  \hat C_l^{XX}\equiv \frac{1}{2l+1}\sum_{m=-l}^l (-1)^m \hat
  X_{lm}\hat X_{l,-m},
\end{equation}
where $\hat X_{lm}$ are the multipole coefficients of the realisation $\hat{X}
(\hat{\vec{n}})$  of a field on the sphere. Here, $X=\tilde{T}$ or $\phi$.
To construct  a likelihood for the empirical power
spectra we will model auto- and cross-correlations of the empirical
power spectra of the observed lensed temperature and the reconstructed
lensing potential.
If $X$ is a statistically-isotropic, Gaussian field,
then the power covariance is diagonal with
\begin{equation}
  \label{eq:gaussian_power_covariance}
  \mathrm{cov}(\hat C^{XX}_l,\hat
  C^{XX}_{l'})=\delta_{ll'}\frac{2}{2l+1} \la \hat C_l^{XX}\ra^2.
\end{equation}
In the following we will abbreviate the Gaussian variance with
\begin{equation}
  \label{eq:gaussian_power_variance_def}
  \mathrm{var}_G(C^{XX}_l)\equiv \frac{2}{2l+1}\left(C^{XX}_l\right)^2.
\end{equation}

We demonstrate in Appendix~\ref{se:CTPhiEeffect} that for most applications
we can neglect the effect of $C^{T\phi}$ on the covariance between the
power spectra of the lensed temperature and the lens reconstruction.
Because the 
Taylor expansion of Eq.~\eqref{eq:deltaT_expansion} is linear in the unlensed temperature
 all odd  $n$-point functions of the lensed
temperature vanish.

\subsection{Lensed temperature}
The auto-correlation of the lensed temperature power spectrum has been
computed at first order in $C^{\phi\phi}$ in
\cite{smith_hu_kaplinghat0607} under the flat-sky approximation and in
\cite{li_smith_cooray0607} on the full sky.  A contribution at second
order in $C^{\phi\phi}$ was recently identified in \cite{aurelien1205}.
   The power covariance is given by
\begin{equation}
  \mathrm{cov}(\hat C^{\tilde T\tilde T}_{l,\expt},\hat C^{\tilde T\tilde
    T}_{l',\expt})
  \label{eq:cov_TT_TT_defn}
=
\delta_{ll'}
\mathrm{var}_G(C^{\tilde T\tilde T}_{l,\expt})
+
\frac{1}{(2l+1) (2l'+1)}\sum_{m,m'} (-1)^{m+m'}
\la \tilde T_{lm}\tilde T_{l,-m}\tilde T_{l'm'}\tilde T_{l',-m'}\ra_c,
\end{equation}
where $\la\,\ra_c$ denotes the connected part of the $4$-point
function, which is at $\mathcal{O} (C^{\phi\phi})$
\cite{huLensedTTrispectrum0105} 
\begin{equation}
  \label{eq:connected_T4_calc3}
      \la\tilde T_{\lbar{1}}\tilde T_{\lbar{2}}\tilde T_{\lbar{3}}\tilde T_{\lbar{4}}\ra_c
= 
\frac{1}{2}C^{TT}_{l_2}C^{TT}_{l_4}\sum_{LM} (-1)^MC_L^{\phi\phi}
\left( \begin{matrix} 
l_1 & l_2 & L \\ 
m_1 & m_2 & -M
\end{matrix} \right)
\left( \begin{matrix} 
l_3 & l_4 & L \\ 
m_3 & m_4 & M
\end{matrix} \right)
F_{l_1Ll_2}F_{l_3 Ll_4} + \mbox{all perms.}
\end{equation}
Here and in the following `all perms' denotes permutations in all non-contracted multipole indices, i.e.~permutations of $1$, $2$, $3$, and $4$
in Eq.~\eqref{eq:connected_T4_calc3}.  If we also include the contribution
at second order in $C^{\phi\phi}$ from \cite{aurelien1205} we get
\cite{li_smith_cooray0607,aurelien1205}\footnote{The $\mathcal{O} (\phi^4)$ part of Eq.~\eqref{eq:CTTexpt-cov} was not derived in a rigorous perturbative analysis, which would imply additional corrections. For example, the
unlensed $f_{lLl'}$ in the second term on the right could be replaced by its lensed counterparts, as in Eq.~\eqref{eq:lensing_trispectrum}. We do not
  investigate such corrections to the temperature power auto-covariance further because corrections to the leading Gaussian term are
 negligible for all applications in this paper.}
\begin{eqnarray}
\nonumber
    \mathrm{cov}(\hat C^{\tilde T\tilde T}_{l,\expt},\hat C^{\tilde T\tilde T}_{l',\expt})&=&
\delta_{ll'} \mathrm{var}_G(C^{\tilde T\tilde T}_{l,\expt})
+
\frac{2}{(2l+1) (2l'+1)}\sum_L C^{\phi\phi}_L f^2_{lLl'}
\\
\label{eq:CTTexpt-cov}
&&\;\;+
\sum_{L}\frac{\partial C^{\tilde T\tilde T}_{l}}{\partial
  C^{\phi\phi}_{L}} 
\frac{2}{2L+1}\left(C^{\phi\phi}_{L}\right)^2
\frac{\partial C^{\tilde T\tilde T}_{l'}}{\partial
  C^{\phi\phi}_{L}}+\mathcal{O}[(C^{\phi\phi})^3],
\end{eqnarray}
where $f_{l_1Ll_2}$ is from the unlensed version of Eq.~\eqref{eq:smallftilde_def}: $f_{l_1Ll_2}=C^{TT}_{l_2}F_{l_1Ll_2}+C^{TT}_{l_1}F_{l_2Ll_1}$.
The third term on the right of Eq.~\eqref{eq:CTTexpt-cov} arises from cosmic variance of the lenses. Fluctuations at lens multipole $L$ produce fluctuations in the empirical lensed temperature power spectrum over a range of multipoles. The fluctuations in the lens power, $\Delta C_L^{\phi\phi}$, propagate to the empirical temperature power spectrum approximately as $\partial C^{\tilde{T}\tilde{T}}_l / \partial C_L^{\phi\phi}$. The power derivative here can be calculated perturbatively by
noting that at $\mathcal{O}(C^{\phi\phi})$ \cite{huHarmonicCMBLensing0001303}
\begin{equation}
  \label{eq:CTT_expansion}
  C^{\tilde T \tilde T}_{l}=C_{l}^{TT}[1- 
  l(l+1)R]+\sum\limits_{l_1,l_2}
C_{l_1}^{\phi\phi}C_{l_2}^{TT}\frac{F^2_{l l_1l_2}}{2l+1},
\end{equation}
where
\begin{equation}
  \label{eq:R_defn}
  R=\frac{1}{8\pi}\sum\limits_l (2l+1)l(l+1)C_l^{\phi\phi}
\end{equation}
is half the mean-squared deflection. 
Therefore 
\begin{equation}
  \label{eq:CTT_Cphiphi_deriv}
 \frac{\partial C^{\tilde{T}\tilde{T}}_{l}}{\partial C^{\phi\phi}_{L}}
 =  \sum_{l'} C_{l'}^{TT}\, \frac{F_{lLl'}^2}{2l+1} -
\frac{L(L+1)(2L+1)}{8\pi} \,l(l+1)C_{l}^{TT}.
  \end{equation}
  While Ref.~\cite{aurelien1205} included higher-order corrections to this
  expression by taking numerical derivatives of lensed power spectra computed non-perturbatively with the \textsc{CAMB} code~\cite{camb,2005PhRvD..71j3010C}, these corrections are not expected to be important
  for our purposes here.

\begin{figure}[t]
\subfloat[][$\mathcal{O}(C^{\phi\phi})$ contribution from \cite{li_smith_cooray0607}]{
\includegraphics[width=0.45\textwidth]{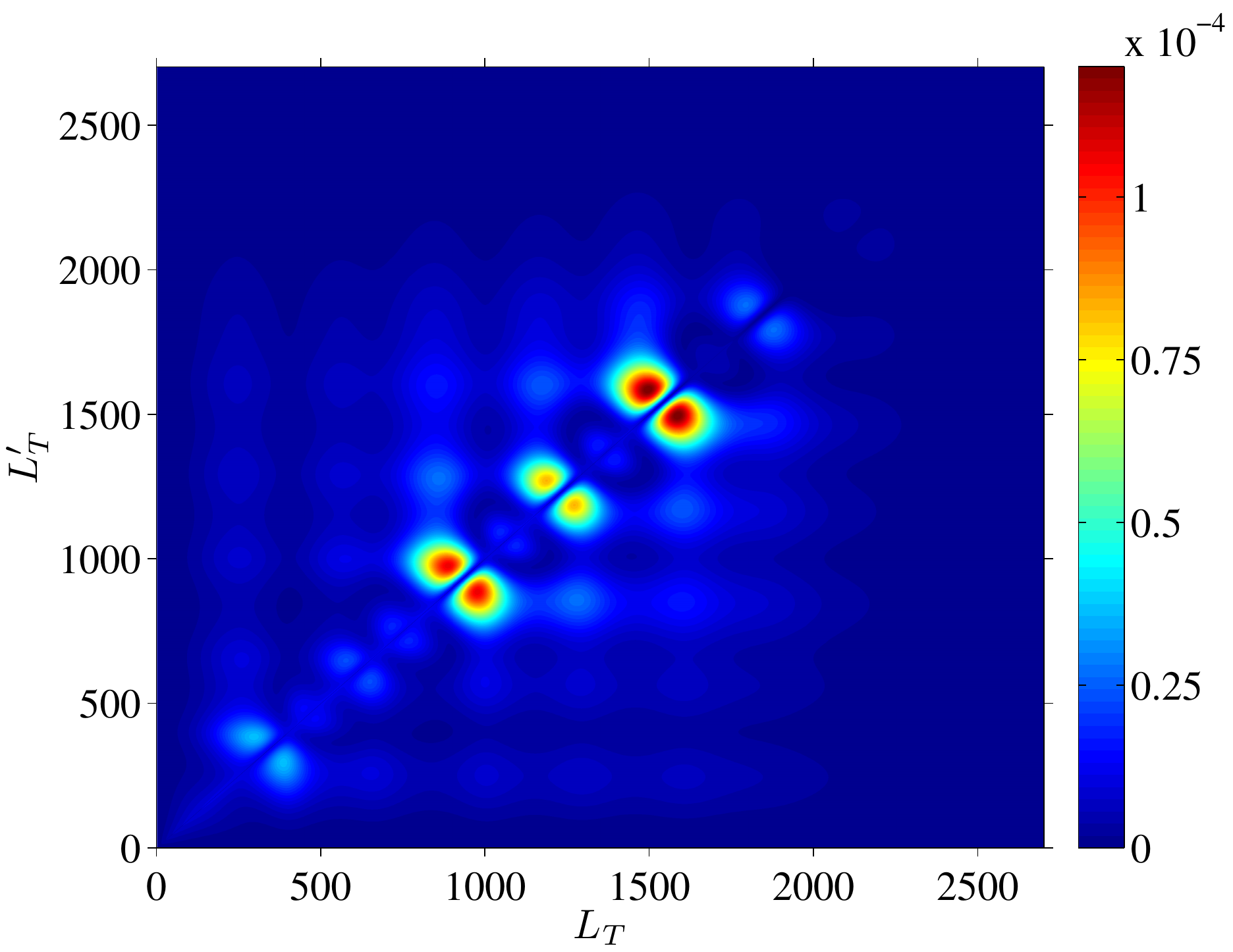}}
\subfloat[][$\mathcal{O}\[(C^{\phi\phi})^2\]$ contribution from \cite{aurelien1205}]{
\includegraphics[width=0.45\textwidth]{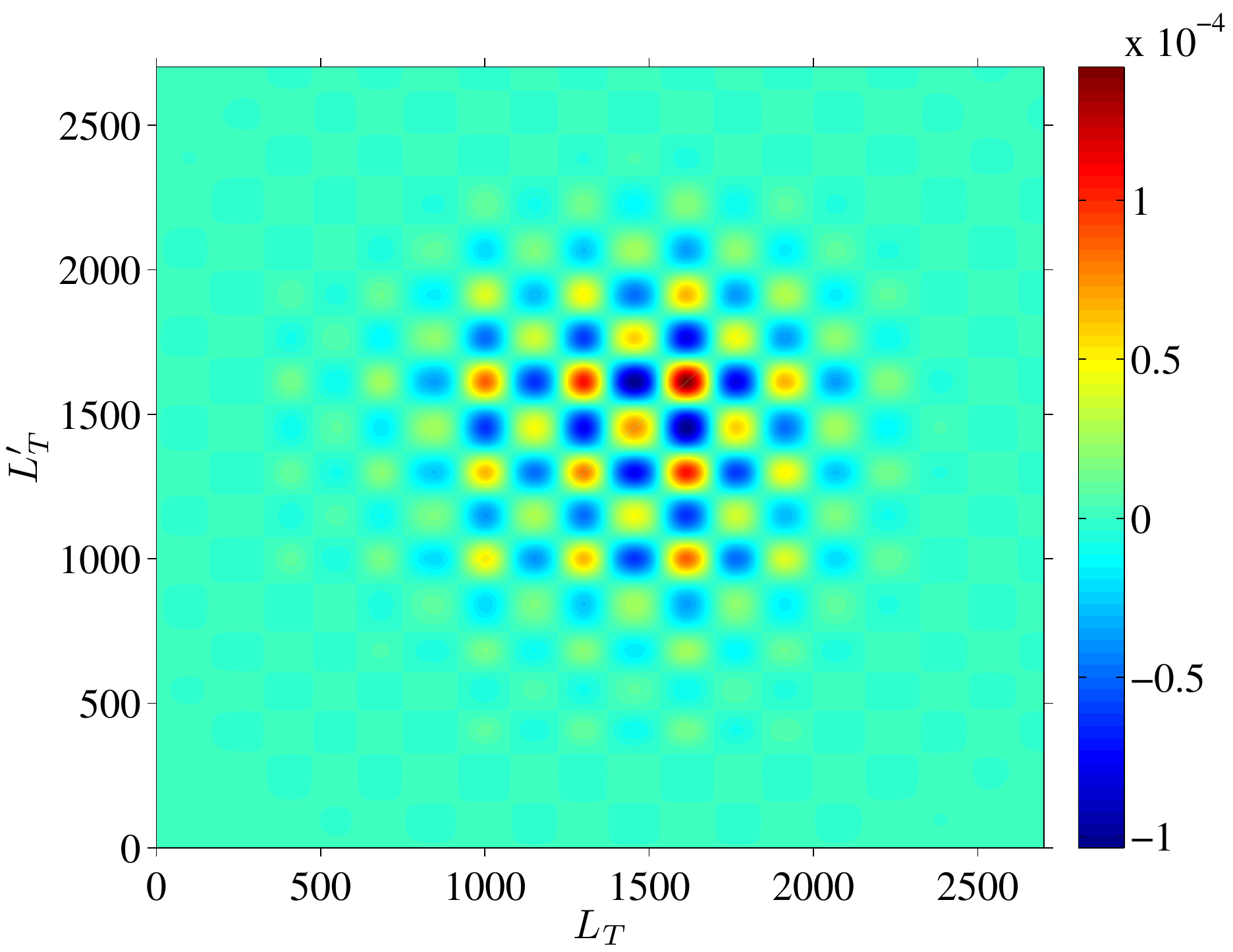}}
\caption{(a) Theoretical off-diagonal part of the correlation
  $\mathrm{correl}(\hat C^{\tilde T\tilde T}_{L_T}, \hat C^{\tilde T\tilde
    T}_{L'_T})$ to first order in $C^{\phi\phi}$, given by the second term on the right of   Eq.~\eqref{eq:CTTexpt-cov} (see
  \cite{li_smith_cooray0607}).  The covariance is converted to a correlation by dividing by $[\mathrm{var}_G(C^{\tilde T\tilde T}_{L_T,\expt})\mathrm{var}_G(C^{\tilde T\tilde T}_{L'_T,\expt})]^{1/2}$. 
(b) $\mathcal{O}\[(C^{\phi\phi})^2\]$ contribution to the correlation identified in~\cite{aurelien1205}. 
}
\label{fig:cov-ttlensed-ttlensed-theo}
\end{figure}

The off-diagonal contributions to the \emph{correlation} between the empirical
$\hat{C}^{\tilde{T}\tilde{T}}_{l,\expt}$ are shown in
\fig{cov-ttlensed-ttlensed-theo} (see \cite{li_smith_cooray0607} and
\cite{aurelien1205} for similar plots). 
 The checkerboard structure of
the $\mathcal{O}[(C^{\phi\phi})^2]$ contribution arises because fluctuations in the lensing power produce changes in the lensed temperature spectra of opposite signs at the acoustic peaks and troughs.
Both of the corrections in Eq.~\eqref{eq:CTTexpt-cov} give correlations that are at most of order $10^{-4}$ and are rather localised in the $(l,l')$ plane. (Note that the correlations are suppressed on small scales where noise dominates the diagonal variance.)
The impact of these
non-diagonal contributions is found to be negligible for all
calculations in this paper, i.e.~we can assume a Gaussian
diagonal auto-covariance of the temperature power spectrum.

\subsection{Lensing reconstruction}
\label{subsec:recon_autocov}

 The auto-correlation of the lensing reconstruction power
spectrum involves the $8$-point function of the lensed
temperature. Hanson \emph{et al.}~\cite{hanson1008} found that the
dominant off-diagonal contributions  on the full sky are given by
disconnected terms that contribute as (see \cite{KCK0302N1} for a
similar calculation on the flat sky)
\begin{eqnarray}
\label{eq:cov-phi-power}
  \mathrm{cov} (\hat C_L^{\hat\phi\hat\phi}, \hat C_{L'}^{\hat\phi\hat\phi})_{\text{non-diag}}^\mathrm{dom}
  & = & 
   \frac{32A_L^2A_{L'}^2}{(2L+1)(2L'+1)} \sum_{l_1} \frac{1}{2l_1+1}
  (C^{\tilde T\tilde T}_{l_1,\mathrm{expt}})^2 
 \[ \sum_{l_2}\tilde g_{l_1l_2}^2(L)
C^{\tilde T\tilde T}_{l_2,\mathrm{expt}} \]
\[ \sum_{l_3}\tilde g_{l_1l_3}^2(L')
C^{\tilde T\tilde T}_{l_3,\mathrm{expt}} \].\;\;\;
\end{eqnarray}
This dominates over more
tightly-coupled terms that involve products of four weights $\tilde g$
that do not factor in the form $\tilde g^2\tilde g^2$ and
therefore enforce a reduced summation volume.  The variance
($L=L'$) on the full sky is predominantly
\begin{equation}
  \label{eq:var-phi-power}
\mathrm{var}(\hat C_L^{\hat\phi\hat\phi})= \frac{2}{2L+1}\langle
C_L^{\hat\phi\hat\phi}\rangle^2 ,
\end{equation}
with small corrections from Eq.~\eqref{eq:cov-phi-power} (for $L=L'$).

As shown in \cite{hanson1008} the off-diagonal reconstruction power
correlation can reach a level of $0.5\%$ and is rather
broad-band.  If the reconstructed
power is binned this can induce correlations of $\mathcal{O}(10\%)$
between different bins.
Physically, these broad-band correlations arise because cosmic-variance fluctuations in the CMB at a given scale
produce fluctuations in $\hat{C}^{\hat\phi\hat\phi}$ that are coherent over a broad range of
scales due to the mode-coupling nature of lens reconstruction (small-scale
CMB fluctuations are used to reconstruct large-scale lenses).
To make this physical interpretation more explicit we note that the
dominant non-diagonal covariance contribution of Eq.~\eqref{eq:cov-phi-power}
can be written as
\begin{equation}
  \label{eq:cov_Cphirec_Cphirec_with_derivative}
    \mathrm{cov} (\hat C_L^{\hat\phi\hat\phi}, \hat C_{L'}^{\hat\phi\hat\phi})_{\text{non-diag}}^\mathrm{dom} =
\sum_{l_1}   
\frac{\partial (2\hat N^{(0)}_L)}{\partial\hat C^{\tilde
    T\tilde T}_{l_1,\expt}}\;
\frac{2}{2l_1+1}(C^{\tilde T\tilde T}_{l_1,\expt})^2\;
\frac{\partial (2\hat N^{(0)}_{L'})}{\partial\hat C^{\tilde
    T\tilde T}_{l_1,\expt}},
\end{equation}
where the (realisation-independent) derivative is given by
\begin{equation}
  \label{eq:Deriv2EmpN0CTT}
  \frac{\partial (2\hat N^{(0)}_L)}{\partial\hat C^{\tilde
    T\tilde T}_{l_1,\expt}}=
\frac{4A_L^2}{2L+1}\sum\limits_{l_2}\tilde g^2_{l_1l_2}(L) C^{\tilde
  T\tilde T}_{l_2,\mathrm{expt}},
\end{equation}
which is non-zero even in the absence of lensing. 
Equation~\eqref{eq:Deriv2EmpN0CTT} describes the change of the
Gaussian reconstruction noise\footnote{When squaring the reconstruction
  $\hat \phi_{LM}$ to form the reconstruction power spectrum, we 
pick up not only the signal power $C_L^{\phi\phi}$ but also the noise of
the reconstruction. Further details on the correspondence between noise terms in
the reconstruction power and the $2\hat N^{(0)}$ expression
are provided in Appendix \ref{se:N0BiasAppendix}. } resulting from
fluctuations of the observed
temperature realisation. In propagating these changes through to the covariance of
$\hat{C}^{\hat{\phi}\hat{\phi}}$, one picks up the sample variance of the total lensed temperature
power spectrum, $\mathrm{var}_G(C^{\tilde{T}\tilde{T}}_{l,\mathrm{expt}})$.

As noted in Sec.~\ref{se:fullSkyLensing}, using the realisation-dependent $\hat N^{(0)}$ bias correction of Eq.~\eqref{eq:empirical_N0} significantly reduces the off-diagonal covariance of the reconstruction power spectrum. To help interpret the $\hat N^{(0)}$ correction, we write it in the form
\begin{equation}
  \label{eq:empN0_subtraction_with_derivatives}
    \hat C_L^{\hat\phi\hat\phi}-2\hat N_L^{(0)} + N_L^{(0)}= 
    \hat C_L^{\hat\phi\hat\phi} - \sum_{l}
  \frac{\partial (2\hat N^{(0)}_L)}{\partial\hat C^{\tilde
    T\tilde T}_{l,\expt}}
 \hat C^{\tilde T\tilde
      T}_{l,\expt} + N_L^{(0)}.
\end{equation}
Therefore, for a given realisation of the lensed temperature, the
empirical $\hat N^{(0)}$ bias correction partly removes the response of the Gaussian
reconstruction noise to changes in the lensed temperature realisation.
To see that this removes the non-diagonal power covariance of the lensing
reconstruction caused by cosmic variance of the lensed temperature, note that both $\mathrm{cov}(\hat
C^{\hat\phi\hat\phi}_L,2\hat N^{(0)}_{L'})$ and $\mathrm{cov}(2\hat
N^{(0)}_L,2\hat N^{(0)}_{L'})$ equal the right hand side of
Eq.~\eqref{eq:cov_Cphirec_Cphirec_with_derivative} at $\mathcal{O}(\phi^0)$. 
The empirical $\hat{N}^{(0)}$ correction leads to a small reduction in the 
\emph{variance} of the binned reconstructed power spectrum, as shown in
Figs.~\ref{fig:fig1-recphi} and~\ref{fig:fig2-recphi}. Any residual covariance after empirical
$\hat{N}^{(0)}$ subtraction is too small
to be detected in our simulations.

\section{Temperature-lensing cross-correlation}
\label{se:temperatureLensingCrossCorrelation}
  For a joint analysis involving the empirical power spectra of the lens reconstruction and the lensed temperature anisotropies, the likelihood should model their cross-correlation to avoid potential double-counting of lensing effects. In this section we calculate the cross-correlation. We recover the two main physical effects introduced in Sec.~\ref{sec:introduction}, i.e.\ a ``noise contribution'' from the cosmic variance of the lensed temperature affecting the noise in the reconstruction over a wide range of scales, and a ``matter cosmic variance'' contribution from cosmic variance of the lenses altering the smoothing of the acoustic peaks in the temperature power spectrum.
We will show in Sec.~\ref{subsec:noise} that the noise
contribution is due to the disconnected part of the lensed temperature
$6$-point function, while in Sec.~\ref{subsec:matter} we show that the matter cosmic variance contribution is
due to the connected part of the $6$-point function.  
Corrections from the lensed temperature trispectrum generally have a
sub-dominant effect on parameter estimation and are discussed in
Appendix~\ref{se:Conn4ptAppendix} (see also \fig{correl_AAprime_plot}
below). 

\subsection{Noise contribution}
\label{subsec:noise}

\subsubsection{Perturbative derivation: Disconnected
  part of lensed temperature 4- and 6-point functions}
\label{se:CovCPhirecCTTexptNoiseContri}
Since the reconstructed lensing potential is quadratic in the lensed
temperature 
the covariance $\mathrm{cov}(\hat C_L^{\hat\phi\hat\phi},
\hat C_{L',\expt}^{\tilde T\tilde T})$ involves the $4$- and $6$-point
functions of the lensed temperature. 
 Using the definition of
$\hat\phi$ in \eqref{eq:phi_esti} we find
\begin{eqnarray}
\nonumber
  \mathrm{cov}(\hat C_L^{\hat\phi\hat\phi},
\hat C_{L',\expt}^{\tilde T\tilde T}) &=& \frac{A_L^2}{(2L+1) (2L'+1)}
\sum\limits_{\underline{l}_1, \underline{l}_2, \underline l_3,
  \underline l_4, M, M'} (-1)^{M+M'}
\left( \begin{matrix} 
l_1 & l_2 & L \\ 
m_1 & m_2 & -M
\end{matrix} \right)
\left( \begin{matrix} 
l_3 & l_4 & L \\ 
m_3 & m_4 & M
\end{matrix} \right)
\tilde g_{l_1l_2}(L)\tilde g_{l_3l_4}(L)\\
  \label{eq:cov_CTT_Cphiphi_rec}
&& \;\times \left[
 \langle \tilde T_{\underline l_1} 
\tilde T_{\underline l_2} \tilde T_{\underline l_3} \tilde
T_{\underline l_4} \tilde T_{L'M'} \tilde T_{L',-M'}\rangle
- \langle \tilde T_{\underline l_1} 
\tilde T_{\underline l_2} \tilde T_{\underline l_3} \tilde
T_{\underline l_4} \rangle\langle \tilde T_{L'M'} \tilde T_{L',-M'}\rangle
\right].
\end{eqnarray}
Since all connected terms vanish  in the absence of lensing we expect the
noise contribution to come from the fully disconnected part. 
If we only keep disconnected terms,
the second line of Eq.~\eqref{eq:cov_CTT_Cphiphi_rec}  can be replaced by
\begin{equation}
  \label{eq:disconnected_contri_to_cov}
  \left[
 \langle \tilde T_{\underline l_1} 
\tilde T_{\underline l_2} \tilde T_{\underline l_3} \tilde
T_{\underline l_4} \tilde T_{L'M'} \tilde T_{L',-M'}\rangle
- \langle \tilde T_{\underline l_1} 
\tilde T_{\underline l_2} \tilde T_{\underline l_3} \tilde
T_{\underline l_4} \rangle\langle \tilde T_{L'M'} \tilde T_{L',-M'}\rangle
\right] \rightarrow
4 \left[ \la \tilde T_{\ul_1}\tilde T_{\ul_3} \ra
  \la \tilde T_{\ul_2}\tilde T_{L'M'} \ra
  \la \tilde T_{\ul_4}\tilde T_{L',-M'} \ra
+ (M' \leftrightarrow -M')\right],
\end{equation}
where we exploited symmetry under relabeling
$(\ul_1\leftrightarrow\ul_2)$ and/or $(\ul_3\leftrightarrow\ul_4)$. We
also used that the contractions $\la \tilde T_{\ul_1}\tilde
T_{\ul_2}\ra$ and $\la \tilde T_{\ul_3}\tilde T_{\ul_4}\ra$ do not
contribute because $\la \hat\phi_{LM}\ra=0$. The fully disconnected
part of Eq.~\eqref{eq:cov_CTT_Cphiphi_rec} is therefore
\begin{equation}
  \label{eq:noisecontri_cov_CTT_Cphiphi_rec}
  \mathrm{cov}(\hat C_L^{\hat\phi\hat\phi},
\hat C_{L',\expt}^{\tilde T\tilde T})_\mathrm{disconn.} =   
\frac{8A_L^2}{(2L+1) (2L'+1)} \left(C_{L',\expt}^{\tilde T\tilde T}
\right)^2 \sum\limits_{l_1} \tilde g^2_{l_1L'}(L) C_{l_1,\expt}^{\tilde
  T\tilde T},
\end{equation}
where the weight $\tilde g_{l_1L'}(L)$ enforces $l_1+L+L'$ to be even.

To interpret this result note that it
can be expressed in terms of the derivative
in Eq.~\eqref{eq:Deriv2EmpN0CTT} as
\begin{equation}
  \label{eq:noise_correlation_heuristic}
  \mathrm{cov}(\hat C^{\hat\phi\hat\phi}_L,\hat C^{\tilde T\tilde
    T}_{L',\expt})_{\mathrm{disconn.}} 
=
  \frac{\partial (2\hat N^{(0)}_L)}{\partial\hat C^{\tilde
    T\tilde T}_{L',\expt}}
\,
 \frac{2}{2L'+1}\left(C^{\tilde T\tilde T}_{L',\expt}\right)^2.
\end{equation}
This part of the covariance is therefore due to the response of the
Gaussian reconstruction noise to changes in the observed temperature
realisation and the resulting covariance with the observed temperature
power.   Based on this
intuition, we anticipate that the covariance can be mitigated by the realisation-dependent
$\hat N^{(0)}$ correction of the reconstruction power bias (see
Sec.~\ref{se:empirical_N0_subtraction_cross_cov} for confirmation).

\subsubsection{Magnitude and structure of the correlation matrix}

\begin{figure}[t]
\centering
\hspace*{-5mm}
\subfloat[][Theoretical noise contribution]{
\includegraphics[width=0.35\textwidth]{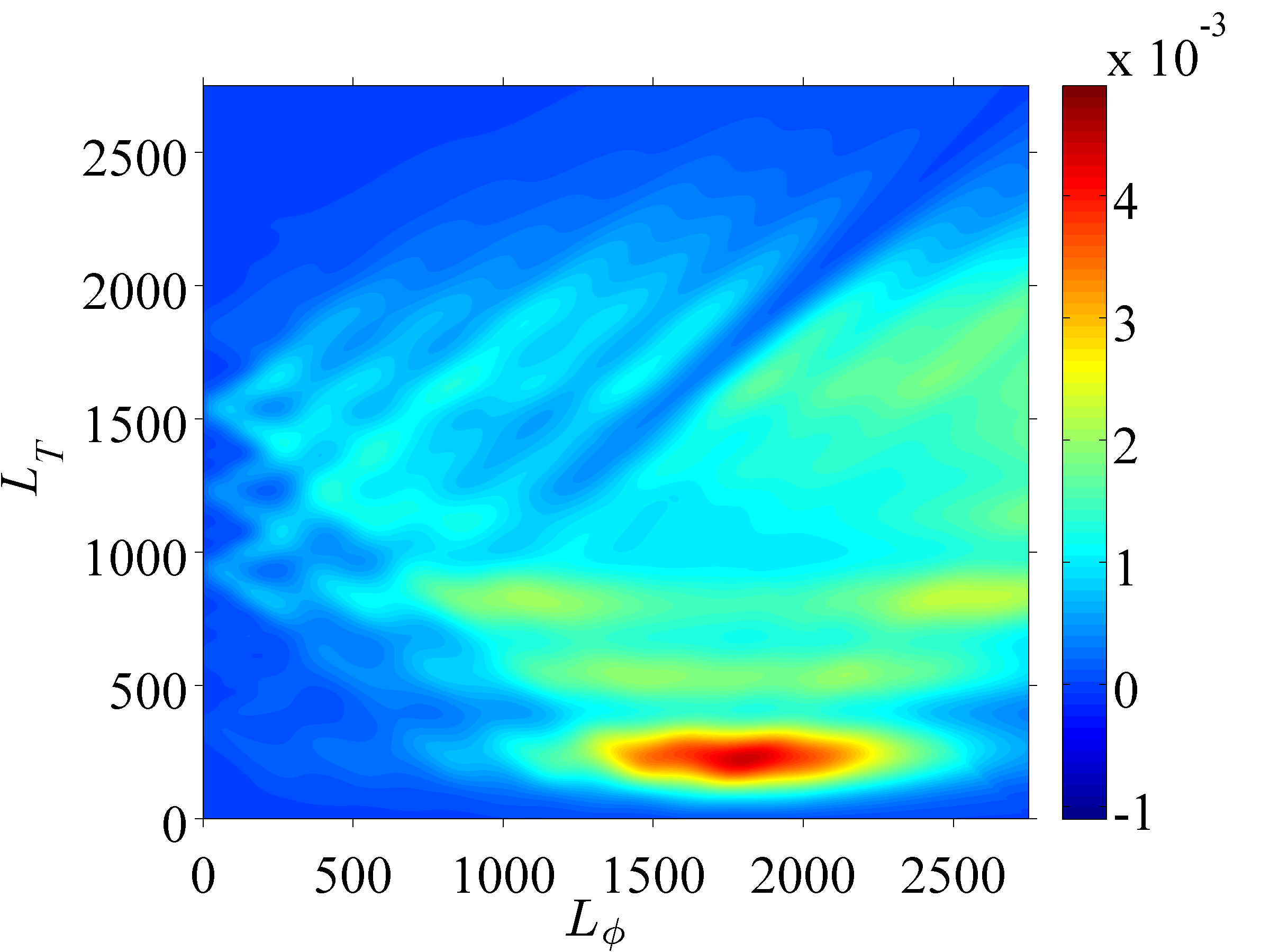}
\label{fig:cov-recphi-t-theory}} 
\hspace*{-5mm}
\subfloat[][Simulations]{
\includegraphics[width=0.35\textwidth]{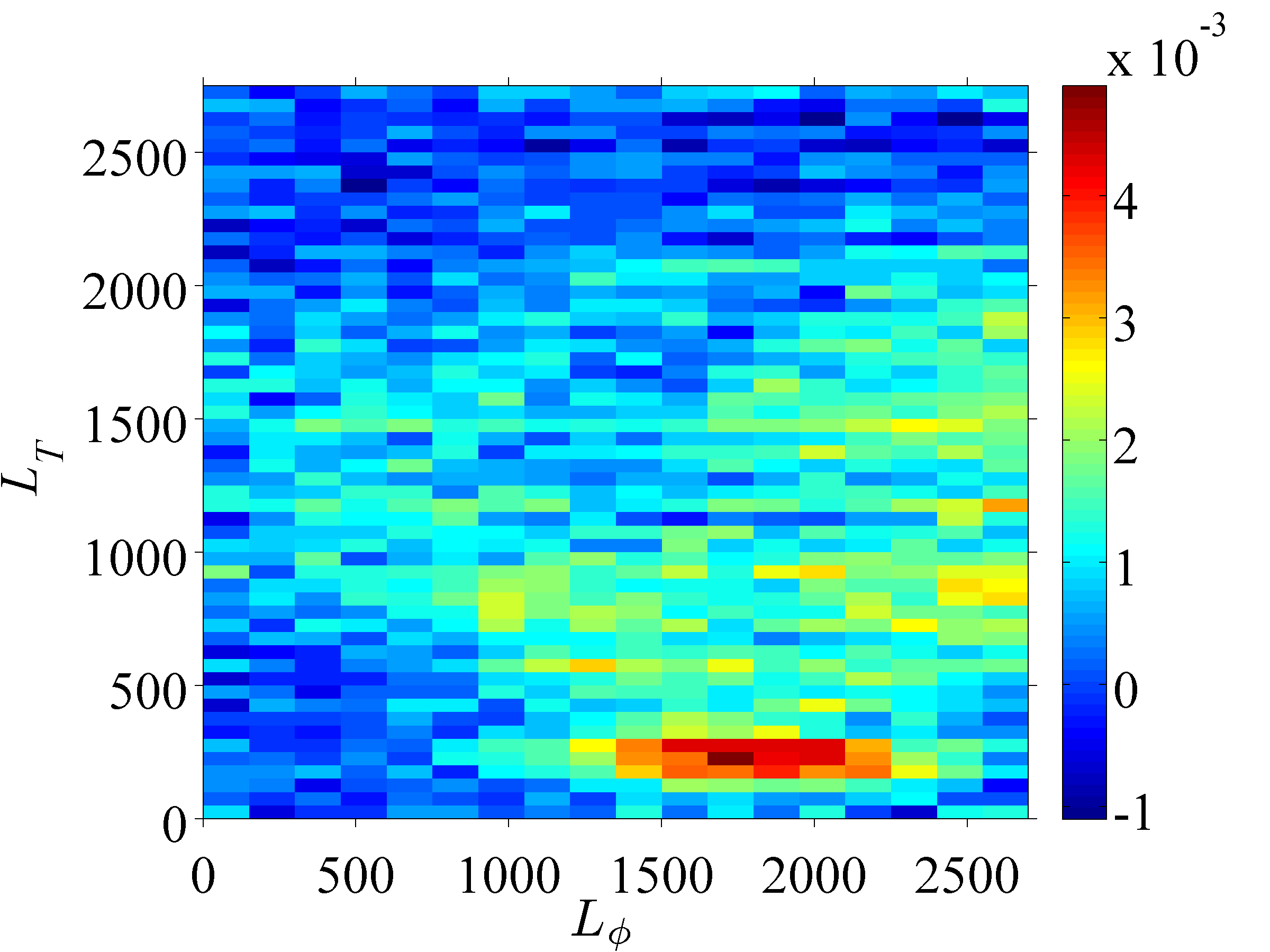}
\label{fig:cov-recphi-t-sims}}
\hspace*{-5mm}
\subfloat[][Simulations with empirical $\hat N^{(0)}$ subtraction]{
\includegraphics[width=0.35\textwidth]{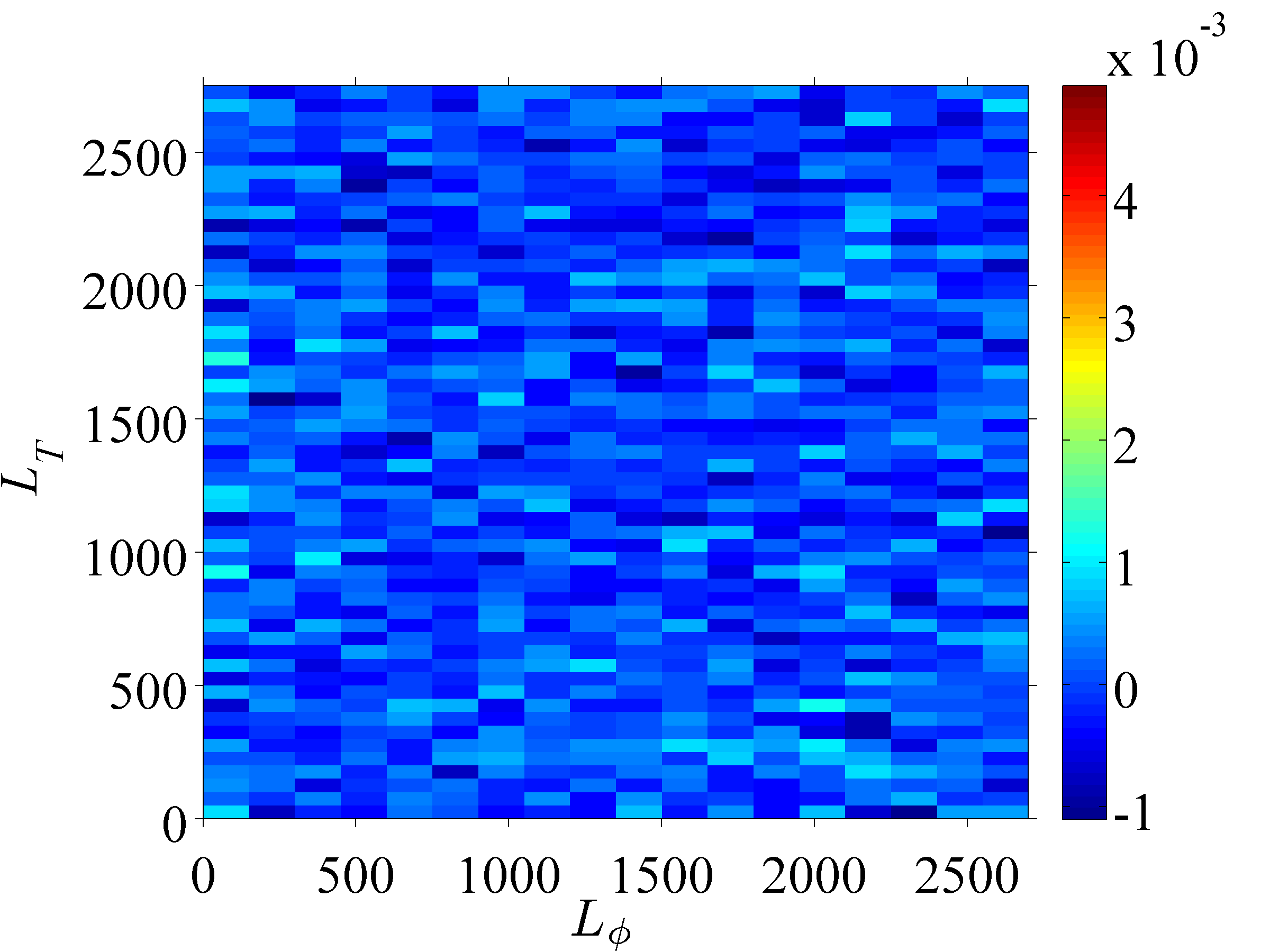}
\label{fig:cov-recphi_SUBempN0-t-sims}}
\caption{(a) Theoretical noise contribution from Eq.~\eqref{eq:noisecontri_cov_CTT_Cphiphi_rec} to the correlation of
  unbinned power spectra of the lensed temperature and the reconstructed
  lensing potential, $\mathrm{correl}
  (\hat C_{L_\phi}^{\hat\phi\hat\phi},\hat C_{L_T,\expt}^{\tilde T\tilde T})$. The acoustic peaks of the
  temperature power spectrum are visible in the vertical direction.
  (b) Estimate of the correlation of unbinned power spectra from
  $1000$ simulations. (c) Same as (b) after subtracting the empirical
  $\hat N^{(0)}$ bias of Eq.~\eqref{eq:empirical_N0} from $\hat C^{\hat\phi\hat\phi}$.  }
\label{fig:cov-recphi-t-all}
\end{figure}

In \fig{cov-recphi-t-theory} we plot the power correlation resulting
from the power covariance in Eq.~\eqref{eq:noisecontri_cov_CTT_Cphiphi_rec} 
(denoting $L_\phi=L$ and $L_T=L'$ for convenience),
\begin{equation}
  \label{eq:correl_noise_contri}
\mathrm{correl}(\hat C^{\hat\phi\hat\phi}_{L_\phi},\hat C^{\tilde T\tilde
  T}_{L_T,\expt})
=
\frac{  \mathrm{cov}(\hat C^{\hat\phi\hat\phi}_{L_\phi},\hat C^{\tilde T\tilde
   T}_{L_T,\expt})}
{\sqrt{\mathrm{var}_G(C^{\phi\phi}_{L_\phi}+N^{(0)}_{L_\phi}+N^{(1)}_{L_\phi})
 \,   \mathrm{var}_G(C^{\tilde T\tilde T}_{L_T,\expt})
}}.
\end{equation}
We plot the correlation of the unbinned spectra.
Note that if the
covariance is broad-band (i.e.~roughly constant over the bin width) the
\emph{correlation} of (sufficiently finely) binned power spectra will
increase roughly proportionally to the square root of the product of
the two bin widths.
The Gaussian variance of $C^{\tilde T\tilde T}_{L_T,\expt}$ in the denominator of
Eq.~\eqref{eq:correl_noise_contri} contains
the beam-deconvolved noisy temperature power spectrum
\eqref{eq:cl-expt}, so that high temperature multipoles are suppressed.

The unbinned power correlation shown in \fig{cov-recphi-t-theory} is
mostly constrained to a cone-like region in the $L_\phi$ vs $L_T$ plane,
with the maximum correlation of $0.5\%$ located at the first
acoustic peak $L_T\sim 200$ and lensing reconstruction multipoles
$1600\lesssim L_\phi\lesssim 1900$. 
To understand the basic structure of the correlation in
\fig{cov-recphi-t-theory} we compute approximations to Eq.~\eqref{eq:noisecontri_cov_CTT_Cphiphi_rec} in the limits $L_T\ll L_\phi$ and
$L_\phi\ll L_T$.

For $L_T\ll L_\phi$, the weights in
Eq.~\eqref{eq:noisecontri_cov_CTT_Cphiphi_rec} restrict the summation from
$l_1=L_\phi-L_T$ to $l_1=L_\phi+L_T$. If we Taylor expand in $l_1$ around $L_\phi$
we get
\begin{equation}
  \label{eq:correl_noise_contri_approx_highPhi_lowTemp_FINAL}
  \mathrm{correl}(\hat C^{\hat\phi\hat\phi}_{L_\phi},\hat C^{\tilde T\tilde
    T}_{L_T,\expt})_\mathrm{disconn.}^{L_T\ll L_\phi}  
\approx 
\frac{[L_\phi(L_\phi+1)]^2A_{L_\phi}}{8\pi} \sqrt{(2L_T+1)(2L_\phi+1)}
\;\frac{L_T(L_T+1)C^{\tilde T\tilde T}_{L_T}}{L_\phi(L_\phi+1)C^{\tilde T\tilde
    T}_{L_\phi,\expt}}. 
\end{equation}
Recalling that $A_{L_\phi} = N_{L_\phi}^{(0)}$ for optimal weights,
we see from \fig{fig1-recphi} that the first term slightly increases with
$L_\phi$. The last term is maximised at the first acoustic peak $L_T\sim 200$ and
at the reconstruction multipole $1600\lesssim L_\phi\lesssim 1900$ where the observed
temperature power $C^{\tilde T\tilde T}_{L_\phi,\expt}$ is minimal (for the
Planck-like noise and beam considered here).   
 In
this region
Eq.~\eqref{eq:correl_noise_contri_approx_highPhi_lowTemp_FINAL} gives
a correlation of around $0.4$--$0.5\%$, which agrees with
\fig{cov-recphi-t-theory}. Equation~\eqref{eq:correl_noise_contri_approx_highPhi_lowTemp_FINAL}
also implies that lower noise in the temperature power spectrum would
move the peak position to higher reconstruction multipoles $L_\phi$.  The
cone structure in \fig{cov-recphi-t-theory}, with apex at $(L_\phi,L_T)\sim
(\text{1600--1900},200)$ and edges $L_\phi-L_T\lesssim (\text{1600--1900}) \lesssim L_\phi+L_T$, encloses the region for which the sum over $l_1$ includes the maximum
of $1/C_{l_1,\text{expt}}^{\tilde{T}\tilde{T}}$ around $l_1=1600$--$1900$.
A similar argument can be applied for the cone patterns in the $L_\phi\lesssim 200$ region.

For high temperature and
low reconstruction multipoles,  $L_\phi\ll L_T$, 
we can Taylor expand Eq.~\eqref{eq:noisecontri_cov_CTT_Cphiphi_rec} 
in $l_1$ around $L_T$ (see \cite{hanson1008} for a similar calculation): 
\begin{eqnarray}
\nonumber
    \mathrm{correl}(\hat C^{\hat\phi\hat\phi}_{L_\phi},\hat C^{\tilde T\tilde
    T}_{L_T,\expt})_\mathrm{disconn.}^{L_\phi\ll L_T}  & \approx &
\frac{[L_\phi(L_\phi+1)]^2A_{L_\phi}^2}{2\pi(A_{L_\phi}+C^{\phi\phi}_{L_\phi})}
 \left( 
\frac{C^{\tilde T\tilde T}_{L_T}}{C^{\tilde T\tilde T}_{L_T,\expt}}
\right)^2
\sqrt{\frac{(2{L_\phi}+1) (2L_T+1)}{4}}
\\
  \label{eq:correl_noise_contri_approx_lowPhi_highTemp}
&& \quad \times \;\frac{1}{4}\left[
\left(\frac{\mathrm{d}\ln (L_T^2 C^{\tilde T\tilde T}_{L_T})}{\mathrm{d}\ln L_T}\right)^2
+ \frac{1}{2}\,\left(\frac{\mathrm{d}\ln C^{\tilde T\tilde T}_{L_T}}{\mathrm{d}\ln L_T}\right)^2
\right],
\end{eqnarray}
where we have neglected $N_{L_\phi}^{(1)}$ and used
$A_{L_\phi}=N_{L_\phi}^{(0)}$.  The quadrature sum of derivatives in the last term
is maximal between acoustic peaks and troughs at $L_T\sim
350$, $625$, $925$, $1225$, $1550$, $1850$ etc., which agrees with the temperature
multipoles where the  full correlation shown in
\fig{cov-recphi-t-theory} is maximal (for $L_\phi\lesssim 200)$.\footnote{The correlation at high $L_T$ is
  suppressed by the noise in $C^{\tilde T\tilde T}_{L_T,\expt}$ which is used to
  normalise the covariance in Eq.~\eqref{eq:correl_noise_contri}.} The maximum value of the 
second line of Eq.~\eqref{eq:correl_noise_contri_approx_lowPhi_highTemp}
is around $36$ (for $L_T\lesssim 2700$).
If we neglect $C^{\phi\phi}_{L_\phi}$
compared to $A_{L_\phi}$, which is roughly acceptable for
$L_\phi\lesssim 10$, then the first term in
Eq.~\eqref{eq:correl_noise_contri_approx_lowPhi_highTemp} is around $2\times
10^{-7}$ (see Fig.~\ref{fig:fig1-recphi}), i.e.~for ${L_\phi}\lesssim 10$,
\begin{equation}
  \label{eq:upper_bound_correl_approx_lowPhi_highTemp}
    \mathrm{correl}(\hat C^{\hat\phi\hat\phi}_{L_\phi},\hat C^{\tilde T\tilde
    T}_{L_T,\expt})_\mathrm{disconn.}^{{L_\phi}\ll L_T}  \lesssim
2\times 10^{-7}\times  \frac{1}{2}\sqrt{(2{L_\phi}+1) (2L_T+1)} \times 
36.
\end{equation}
For example, for ${L_\phi}=10$, $L_T=1500$ the bound is $9\times 10^{-4}$ which
is consistent with the full result shown in
\fig{cov-recphi-t-theory}.

\subsubsection{Comparison with simulations}
\label{sec:noise_sims}

Before assessing the relevance of the noise contribution to the covariance for parameter
estimation we compare the analytic result in Eq.~\eqref{eq:noisecontri_cov_CTT_Cphiphi_rec} with the full covariance estimated from our simulations. For the latter, we use
\begin{equation}
  \label{eq:cov_CTT_Cphiphi_rec_measured_unbinnedcov}
  \widehat{\mathrm{cov}}
(\hat C^{\hat\phi\hat\phi}_{{L}},\hat C^{\tilde
    T\tilde T}_{{L'},\mathrm{expt}}) =
  \frac{1}{N_\mathrm{sims}-1}\sum_{s=1}^{N_\mathrm{sims}} 
(\hat C^{\hat\phi\hat\phi}_{L,s}-\la \hat C^{\hat\phi\hat\phi}_L\ra_\mathrm{sims})
(\hat C^{\tilde T\tilde T}_{L',\expt,s}-\la \hat C^{\tilde T\tilde T}_{L',\expt}\ra_\mathrm{sims}),
\end{equation}
where $s$ labels different realisations and
$\la\cdot\ra_\mathrm{sims}$ denotes the average over $N_\mathrm{sims}$
realisations.  To reduce the noise of the estimates from the finite number of simulations, we average the
measured covariance over a range of $L$ and $L'$ values,\footnote{Note
  that for broad-band covariances this binning procedure does not bias
  the covariance estimate, i.e.,~the binned covariance agrees with the
  unbinned covariance in the limit of averaging over infinitely many
  simulations.  However, for a finite number of simulations, the
  binned covariance is less noisy than the covariance estimate at a
  single $(L,L')$ pair. Note also that binning the estimated covariance is
equivalent to estimating the covariance of the binned spectra.}
\begin{equation}
  \label{eq:cov_CTT_Cphiphi_rec_measured_binnedcov}
  \widehat{\mathrm{cov}}
(\hat C^{\hat\phi\hat\phi}_{\overline{L}_i},\hat C^{\tilde
    T\tilde T}_{\overline{L}'_j,\mathrm{expt}}) =
\frac{1}{\Delta L_i\Delta L'_j}
\frac{1}{[\overline{L}_i (\overline{L}_i+1)]^2\;
\overline{L}'_j (\overline{L}'_j+1)}
\sum_{L=L_i}^{L_{i+1}-1} \sum_{L'=L'_j}^{L'_{j+1}-1}
[L(L+1)]^2L'(L'+1) \widehat{\mathrm{cov}}(\hat C^{\hat\phi\hat\phi}_L,\hat
C^{\tilde T\tilde T}_{L'}),
\end{equation}
where $L_i$ and $L'_j$ are bin boundaries for lensing and temperature
powers, respectively, and $\overline L_i$ and $\overline L'_j$ denote
the corresponding bin centres. The bin widths are 
$\Delta L_i=L_{i+1}-L_i$ and $\Delta L'_j=L'_{j+1}-L'_j$.  We divide
out $L(L+1)$ prefactors to average over relatively slowly varying
quantities. In \fig{cov-recphi-t-sims} we plot the estimate of the
correlation that \emph{unbinned} power spectra would have. This is
obtained by dividing the covariance estimate
of Eq.~\eqref{eq:cov_CTT_Cphiphi_rec_measured_binnedcov} by the theoretical
Gaussian variance of unbinned power spectra as in
Eq.~\eqref{eq:correl_noise_contri} [evaluated at
$(\overline{L}_i,\overline{L'}_j)$].
Within the random scatter from the finite $N_{\text{sims}}$, the estimated correlation agrees
with the theoretical noise contribution of Eq.~\eqref{eq:noisecontri_cov_CTT_Cphiphi_rec}. We can also conclude  that the noise
contribution is the dominant part of the temperature-lensing power
correlation.

\subsubsection{\texorpdfstring{Mitigating the noise contribution with the empirical $\hat
  N^{(0)}$ bias correction}{Mitigating the noise contribution with the empirical \^N(0) bias correction}}
\label{se:empirical_N0_subtraction_cross_cov}

In practice, it is desirable that the temperature and reconstruction
power spectra are uncorrelated so that their respective likelihoods may be simply
combined. Indeed, this is the assumption that has been made in all
joint analyses to date~\cite{2011PhRvL.107b1302S,SPTLensingDetectionVanEngelen1202,2013arXiv1303.5077P}.

Fortunately, the empirical $\hat N^{(0)}$ subtraction, which was originally
proposed in~\cite{hanson1008} to eliminate the non-diagonal
reconstruction power auto-covariance \eqref{eq:cov-phi-power}, also removes the noise
contribution [Eq.~\eqref{eq:noisecontri_cov_CTT_Cphiphi_rec}] to the
temperature-lensing power cross-covariance. To see this note that
if the empirical bias correction
of Eq.~\eqref{eq:empirical_N0} is used, the cross-covariance
changes by
\begin{equation}
  \label{eq:emp_N0_temp_cov}
     -\mathrm{cov}(2\hat N^{(0)}_L,\hat C_{L',\expt}^{\tilde T\tilde
  T}) =-\sum_l
  \frac{\partial (2\hat N^{(0)}_L)}{\partial\hat C^{\tilde
    T\tilde T}_{l,\expt}}
\mathrm{cov}(\hat C^{\tilde
    T\tilde T}_{l,\expt},\hat C^{\tilde T\tilde T}_{L',\expt})
=-\mathrm{cov}(
\hat C^{\hat\phi\hat\phi}_L, \hat  C_{L',\expt}^{\tilde T\tilde
  T})_\mathrm{disconn.}
+\mathcal{O}(\phi^2).
\end{equation}
We establish a more general version of this result in Appendix~\ref{app:edgeworth},
where we show that the generalisation of the $\hat{N}^{(0)}$ correction for anisotropic surveys removes the noise contribution to the covariance with any quadratic estimate (including e.g.\ cross-spectra) of the temperature power spectrum.
We confirm the reduction in the cross-covariance with simulations in
\fig{cov-recphi_SUBempN0-t-sims}. Corrections to Eq.~\eqref{eq:emp_N0_temp_cov} from the non-Gaussian terms in the covariance of the temperature power spectra
[see Eq.~\eqref{eq:CTTexpt-cov}] at $\mathcal{O}(\phi^2)$ and
$\mathcal{O}(\phi^4)$
reach at most $3\times 10^{-5}$ in the correlation,
which is roughly one order of magnitude smaller than the matter cosmic
variance contribution discussed below (see \fig{cov-inphi-t-theory}).
These corrections are too small to be  visible in
\fig{cov-recphi_SUBempN0-t-sims} and we neglect them in the following.

\subsection{Matter cosmic variance contribution}
\label{subsec:matter}

\subsubsection{Warm-up: Power covariance of input lensing potential and
  lensed temperature}

We expect the cosmic variance of the lenses to induce a
power-correlation of the lensed temperature with the lensing
reconstruction since greater lensing power in a given realisation leads to additional smoothing of
the empirical temperature power spectrum.
As a warm-up, we calculate how the same effect gives rise to a covariance between the power spectrum of the temperature and the (unobservable) power spectrum of the lensing potential, as if we were able to measure $\phi$ directly with no noise.
This correlation can be
extracted from simulations simply by measuring the correlation of the power spectra
of the lensed temperature and the input lensing potential without performing
any lensing reconstruction.
  To calculate the covariance
perturbatively, note that for a fixed realisation of the input lensing
potential $\phi_\mathrm{in}$ the lensed temperature power spectrum
obtained by averaging only over the unlensed CMB is given by ($L'\neq
0$)
\begin{eqnarray}
  \nonumber
  \frac{1}{2L'+1}\sum_{M'}\la |\tilde T_{L'M'}|^2\ra_\mathrm{CMB}
&=&C_{L'}^{TT} + \frac{1}{2L'+1}\sum_{l_1l_2}\hat
C_{l_1}^{\phi_\mathrm{in}\phi_\mathrm{in}} C_{l_2}^{TT}F_{L'l_1l_2}^2- \hat R
L'(L'+1)C_{L'}^{TT}\\
\label{eq:TTpower_avg_only_over_CMB}
&=&C_{L'}^{TT} + \sum_{L''}\frac{\partial C^{\tilde T\tilde T}_{L'}}{\partial
  C^{\phi\phi}_{L''}}\hat C^{\phi_\mathrm{in}\phi_\mathrm{in}}_{L''}.
\end{eqnarray}
Here $\hat C^{\phi_\mathrm{in}\phi_\mathrm{in}}$ denotes the empirical
power spectrum of the input lensing potential realisation
$\phi_\mathrm{in}$\footnote{We denote this as
  $\phi_\mathrm{in}$, while the reconstructed lensing potential is
  $\hat\phi$. Recall that $C$ denotes theoretical and $\hat C$ empirical power
  spectra.} and $\hat R$ is defined by replacing $C^{\phi\phi}$ in
Eq.~\eqref{eq:R_defn} by $\hat C^{\phi_\mathrm{in}\phi_\mathrm{in}}$.  The
first line of Eq.~\eqref{eq:TTpower_avg_only_over_CMB} can be derived from
the expansion in Eq.~\eqref{eq:deltaT_expansion} at second order in 
$\phi_\mathrm{in}$.  The term linear in $\phi_\mathrm{in}$ vanishes
because it is proportional to the monopole of $\phi_{\text{in}}$.
In Eq.~\eqref{eq:TTpower_avg_only_over_CMB} we neglected noise in the temperature
measurement since this does not contribute to the covariance that we aim to calculate.
The derivative in
the second line of Eq.~\eqref{eq:TTpower_avg_only_over_CMB} is given by
Eq.~\eqref{eq:CTT_Cphiphi_deriv}. Neglecting the $T$-$\phi$ correlation, we then
have
\begin{align}
  \mathrm{cov}(\hat C^{\phi_\mathrm{in}\phi_\mathrm{in}}_L,
\hat C^{\tilde T\tilde T}_{L',\expt}) &=
\langle \hat C^{\phi_\mathrm{in}\phi_\mathrm{in}}_L\langle \hat C^{\tilde T\tilde T}_{L',\expt}
\rangle_{\text{CMB}}\rangle_{\text{LSS}} - \langle \hat C^{\phi_\mathrm{in}\phi_\mathrm{in}}_L \rangle \langle \hat C^{\tilde T\tilde T}_{L',\expt}\rangle ,\nonumber \\
&= \frac{2}{2L+1}(C^{\phi\phi}_L)^2\,
\frac{\partial C^{\tilde T\tilde T}_{L'}}{\partial
  C^{\phi\phi}_{L}},
  \label{eq:cov_CTT_Cphiphi_IN}
\end{align}
where $\langle \cdot \rangle_{\text{LSS}}$ denotes an average over realisations of
$\phi$ (i.e., over large-scale structure).

\subsubsection{Power covariance of reconstructed lenses and lensed
  temperature from the connected 6-point function}

The power covariance [Eq.~\eqref{eq:cov_CTT_Cphiphi_rec}] of the
\emph{reconstructed} lensing potential and the lensed temperature receives
contributions from the connected $4$- and $6$-point functions, as well as the Gaussian (disconnected) part. The leading-order trispectrum is linear in $C^{\phi\phi}$ and so cannot give rise to the expected matter cosmic variance contribution calculated above. The trispectrum contributions are discussed in detail in Appendix~\ref{se:Conn4ptAppendix}, and are shown to have a sub-dominant effect on parameter estimation (see \fig{correl_AAprime_plot} below).
We therefore focus here on the contribution from
the connected $6$-point function.  Terms independent of $\phi$ do
not contribute to the connected part and terms linear in $\phi$ vanish
when averaging over large-scale structure. Since the $\phi^2$
contribution can be shown to vanish~\cite{KCK0208cumulants} 
as well as $\phi^3$ terms which vanish for Gaussian $\phi$,
the leading-order contribution is of order $\phi^4$.
The $\mathcal{O}(\phi^4)$ contribution from the connected $6$-point
function to the temperature-lensing power covariance is
calculated in detail in Appendix~\ref{se:connected_6point_orderphi4}. The
dominant contribution there [Eq.~\eqref{eq:matter_contri_theo}] is exactly of the
form calculated above  [Eq.~\eqref{eq:cov_CTT_Cphiphi_IN}] for the 
power covariance of the input lensing potential and the
lensed temperature.

\subsubsection{Magnitude and structure of the covariance matrix}

\begin{figure}[t]
\subfloat[][Theoretical matter cosmic variance contribution]{
\includegraphics[width=0.45\textwidth]{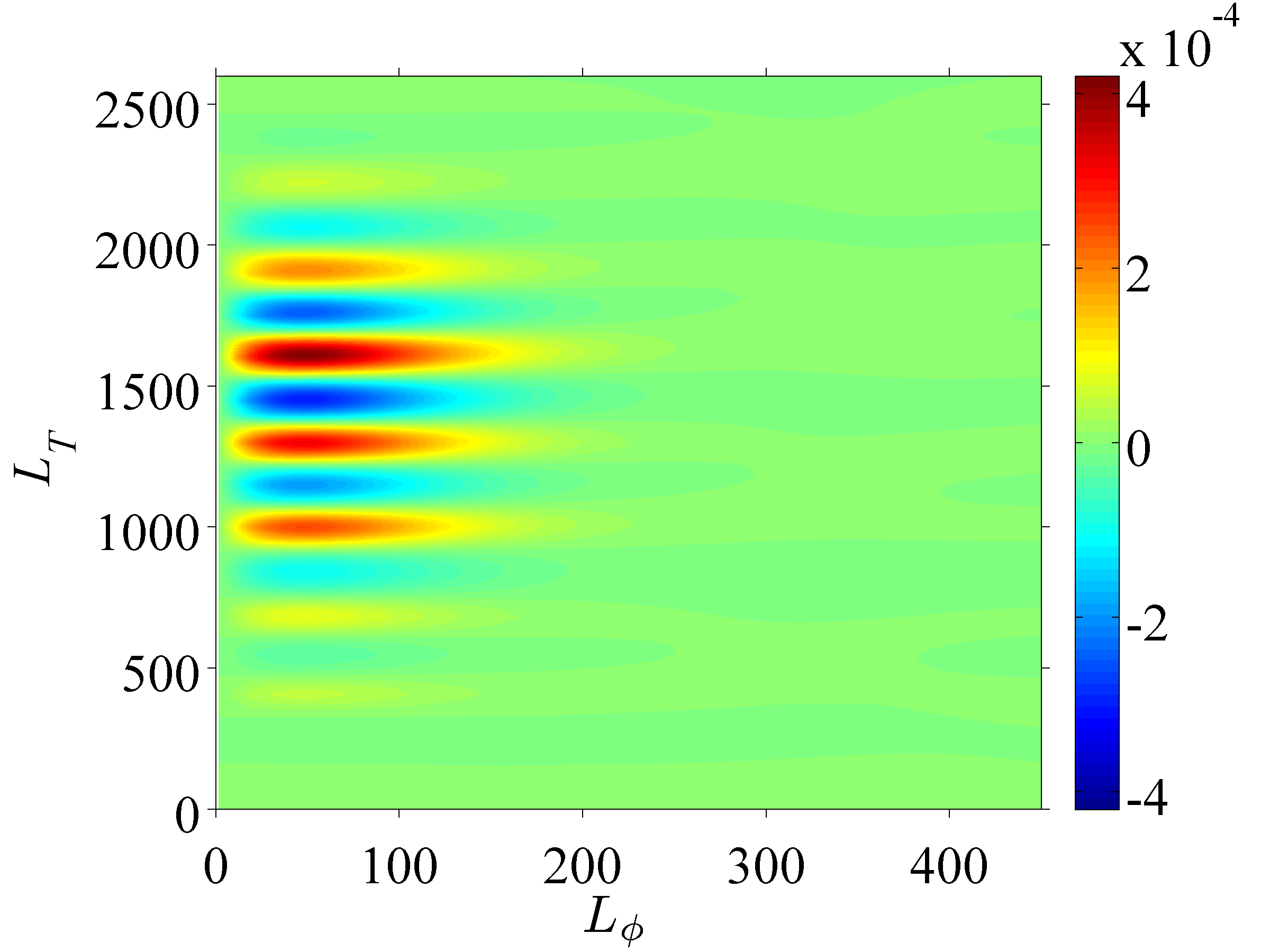}
\label{fig:cov-inphi-t-theory}} 
\hspace{0.05\textwidth}
\subfloat[][Simulations]{
\includegraphics[width=0.45\textwidth]{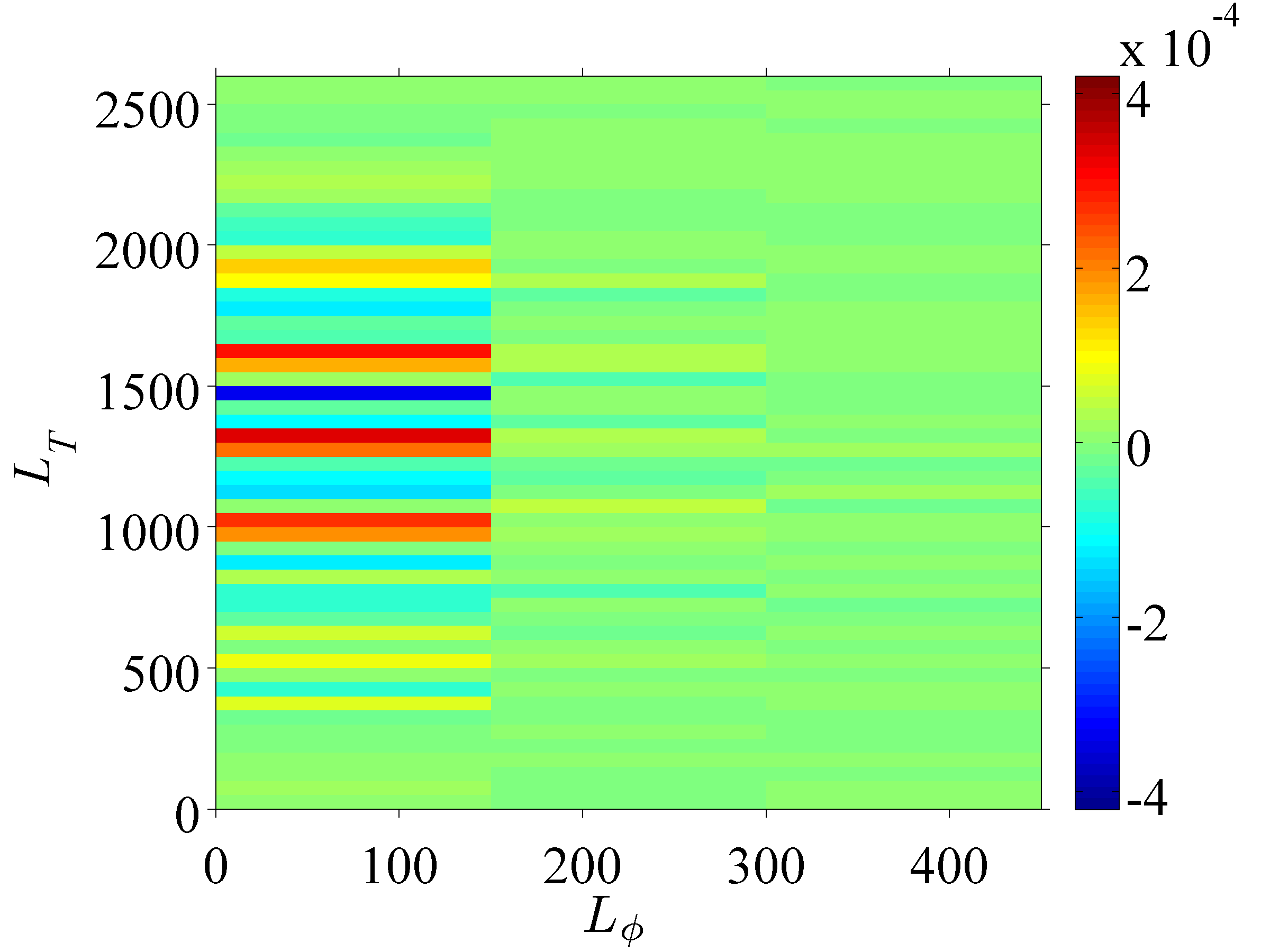}
\label{fig:cov-inphi-t-sims}}
\caption{(a) Theoretical matter cosmic variance contribution, from
  Eq.~\eqref{eq:matter_contri_theo}, to the correlation of the
  unbinned power spectra of the reconstructed lensing potential and the lensed
  temperature anisotropies.  The covariance is
  converted to a correlation following Eq.~\eqref{eq:correl_noise_contri}. 
 (b) Measured correlation
  $\widehat{\mathrm{correl}} (\hat
  C_{L_\phi}^{\phi_\mathrm{in}\phi_\mathrm{in}},\hat C_{L_T}^{\tilde
    T\tilde T}-\hat C^{TT}_{L_T})$ of the power spectrum of the input lensing potential
  and the difference of noise-free lensed and unlensed temperature power spectra
  from $1000$ simulations.  }
\label{fig:cov-inphi-t}
\end{figure}

The \emph{correlation} of unbinned temperature and lensing power spectra due to
Eq.~\eqref{eq:matter_contri_theo} is at most $0.04\%$ (see \fig{cov-inphi-t-theory}).
Since higher lensing power lowers the acoustic peaks and increases the troughs of the temperature
power spectrum, the correlation is
negative at multipoles $L_T$ where the temperature power peaks and positive
where it has a trough. The correlation is small for
$L_\phi\gtrsim 150$ because the
acoustic peak smearing is mainly caused by large-scale lenses
\cite{smith_hu_kaplinghat0607}. 

\begin{figure}[t]
\centering
\includegraphics[width=0.45\textwidth]{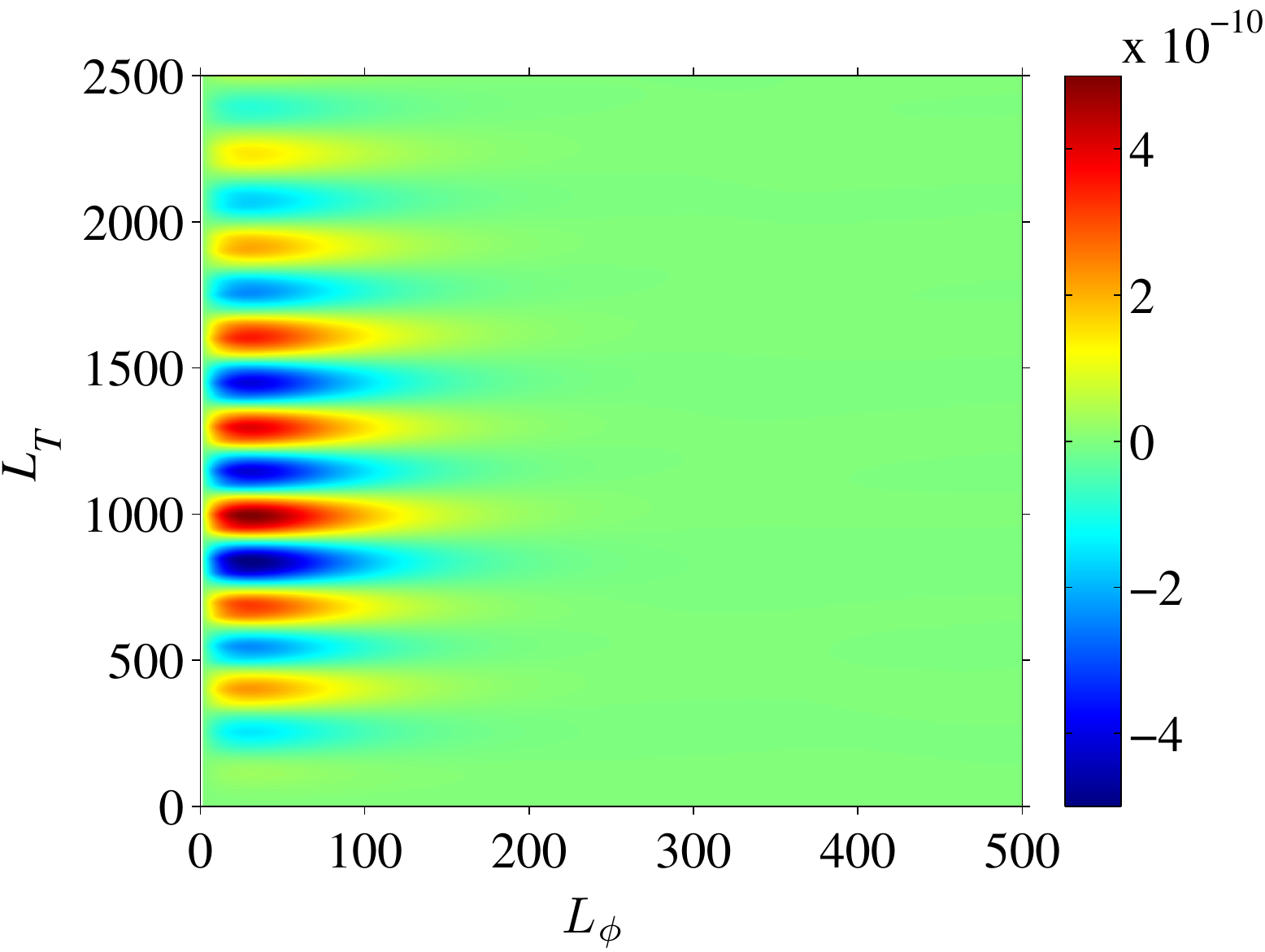}
\hspace{0.05\textwidth}
\includegraphics[width=0.45\textwidth]{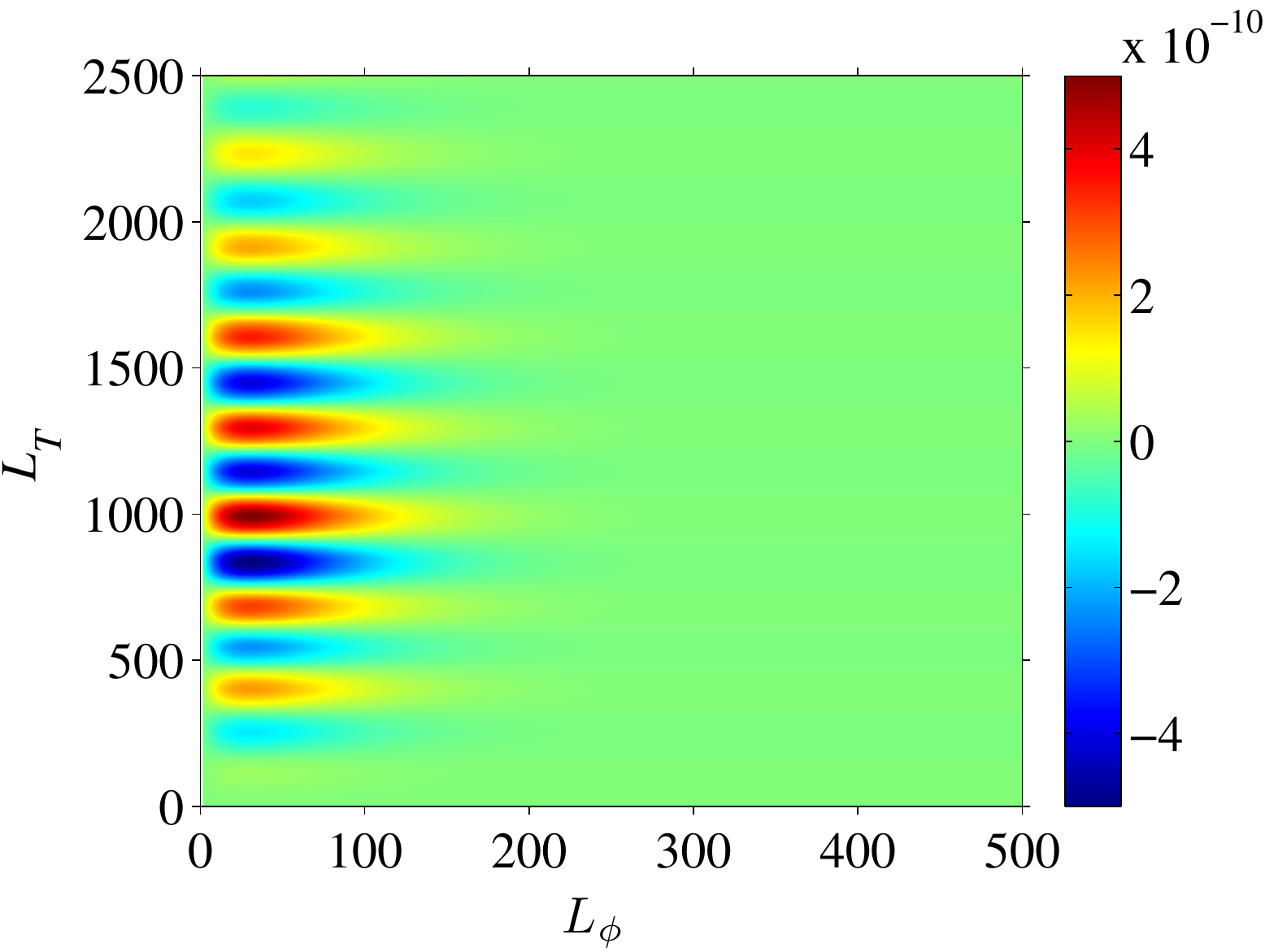}
\caption{\emph{Left}: The approximate contribution to the \emph{covariance} between
$[L_\phi(L_\phi+1)]^2\hat{C}_{L_\phi}^{\hat{\phi}\hat{\phi}}/(2\pi)$
 and
$L_T(L_T+1)\hat{C}_{L_T}^{\tilde{T}\tilde{T}}/(2\pi)$ 
from cosmic variance of the lenses, derived
from Eq.~\eqref{eq:matter_contri_theo}. \emph{Right}: The rank-one
approximation to the matrix on the left from retaining only the
largest singular value.}
\label{fig:cvcov_full_single}
\end{figure}

\begin{figure}[t]
\centering
\includegraphics[width=0.36\textwidth]{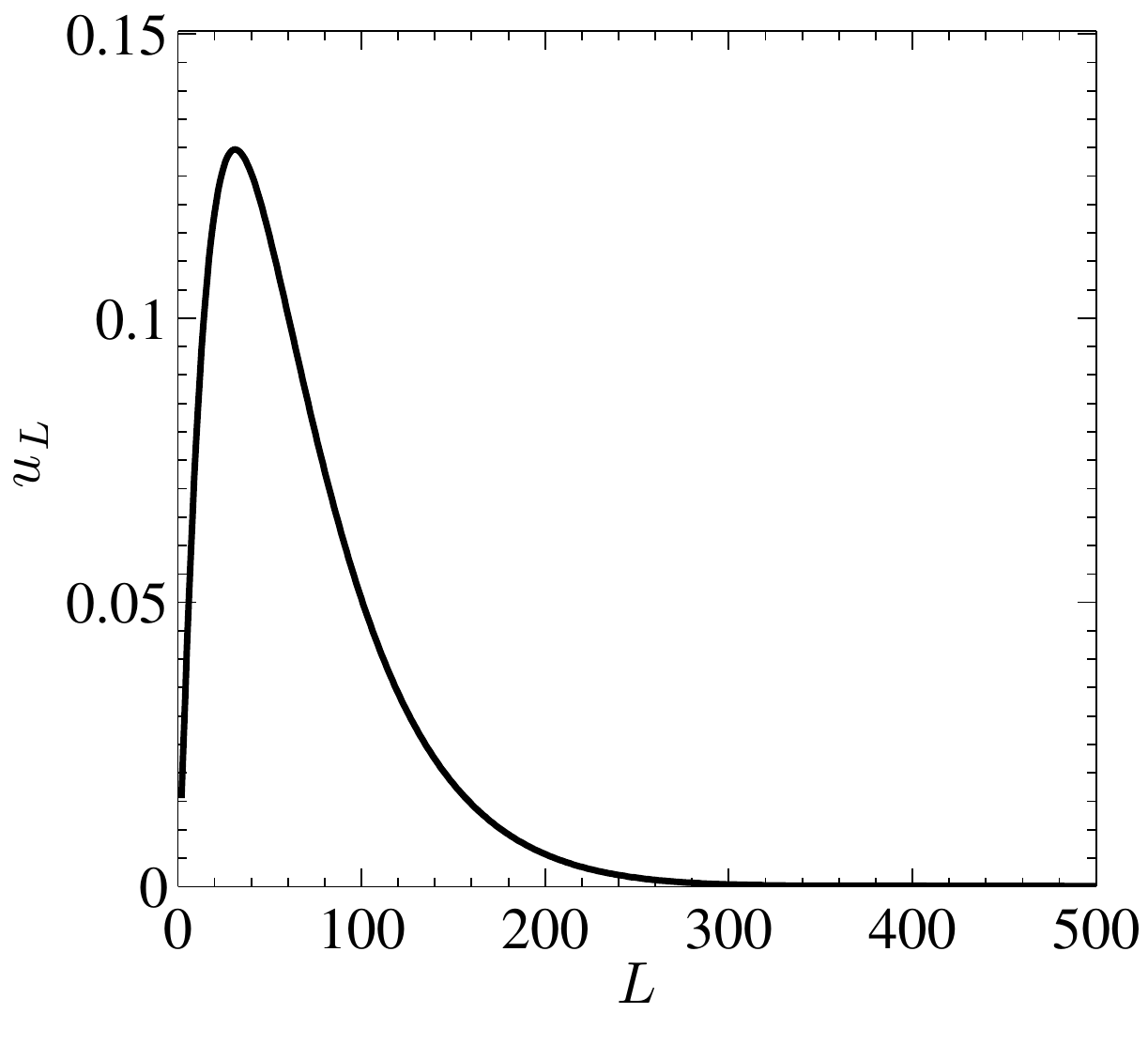}
\hspace{0.05\textwidth}
\includegraphics[width=0.36\textwidth]{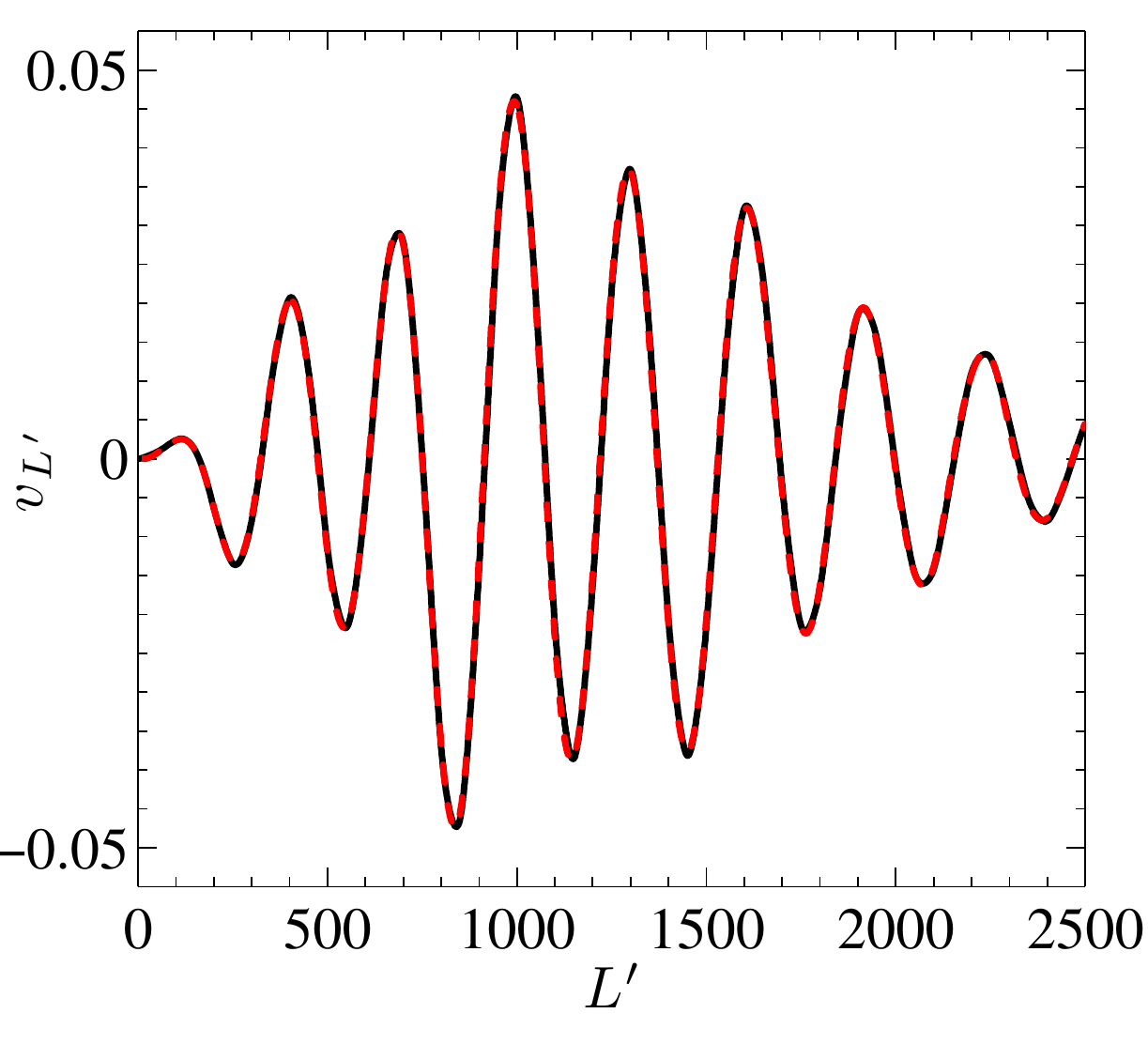}
\caption{Singular vectors $u_L$ (left) and $v_{L'}$ (right, black) of the matter cosmic variance contribution to the covariance between
$[L(L+1)]^2\hat{C}_L^{\hat{\phi}\hat{\phi}}/(2\pi)$ and
$L'(L'+1)\hat{C}_{L'}^{\tilde{T}\tilde{T}}/(2\pi)$ corresponding to the dominant singular value; see Eq.~(\ref{eq:svd}). The approximation of Eq.~\eqref{eq:vLprimeApprox} is shown in red dashed on the right.}
\label{fig:cov_evecs}
\end{figure}

The approximate \emph{covariance} matrix between $[L(L+1)]^2\hat{C}_L^{\hat{\phi}\hat{\phi}}/(2\pi)$ and
$L'(L'+1)\hat{C}_{L'}^{\tilde{T}\tilde{T}}/(2\pi)$, calculated from Eq.~(\ref{eq:matter_contri_theo}), is shown in the left-hand panel of Fig.~\ref{fig:cvcov_full_single}. By perfoming a singular-value decomposition of this matrix, we find that it has a very low-rank structure. For example, retaining the multipoles of the reconstruction to $L=500$ (and up to $L'=2500$ for the temperature), the first singular value is 34 times larger than the second. The covariance matrix can be accurately approximated in rank-one form,
\begin{equation}
\text{cov}\left(\frac{[L(L+1)]^2 \hat{C}_L^{\hat{\phi}\hat{\phi}}}{2\pi},
\frac{L'(L'+1)\hat{C}_{L'}^{\tilde{T}\tilde{T}}}{2\pi}\right)_\eqref{eq:matter_contri_theo} \approx \lambda_1 u_L v_{L'} ,
\label{eq:svd}
\end{equation}
as shown in the right-hand panel of Fig.~\ref{fig:cvcov_full_single}. Here,
the leading singular value $\lambda_1 = 8.3\times 10^{-8} \,\mu \mathrm{K}^2$; the
corresponding (normalised) singular vectors $u_L$ (for the reconstructed lensing power) and $v_{L'}$ (for the temperature power) are plotted in Fig.~\ref{fig:cov_evecs}. The significance of the rank-one covariance is that cosmic variance of the lenses produces correlated changes in the measured lensed temperature power spectrum with definite shape given by $v_{L'}$. Moreover, cosmic-variance fluctuations in $\hat{C}_L^{\phi\phi}$ that are orthogonal to $u_L$ do not influence $\hat{C}_{L'}^{\tilde{T}\tilde{T}}$.

We can understand the low-rank stucture of the covariance by evaluating the derivative
$\partial C_{L'}^{\tilde{T}\tilde{T}}/\partial C_L^{\phi\phi}$ with the flat-sky form of Eq.~(\ref{eq:CTT_expansion}). Following Ref.~\cite{huHarmonicCMBLensing0001303}, we have
\begin{equation}
C_{L'}^{\tilde{T}\tilde{T}} = (1-{L'}^2R)C_{L'}^{TT} + \int \frac{d^2 \vL}{(2\pi)^2}
\left[(\vL'-\vL)\cdot \vL\right]^2 C_L^{\phi\phi} C^{TT}_{|\vL'-\vL|} ,
\label{eq:flatskypower}
\end{equation}
where $R$ is given by Eq.~(\ref{eq:R_defn}). It is clear from Fig.~\ref{fig:cvcov_full_single} that the covariance is only significant between large-scale lens modes ($L < 100$) and intermediate- and small-scale temperature modes (where lensing has a significant effect on the power spectrum). In the limit
$L \ll L'$, and for $L$ small compared to the CMB acoustic scale, we can Taylor expand
$C^{TT}_{|\vL'-\vL|}$ in Eq.~(\ref{eq:flatskypower}) to obtain
\begin{equation}
\frac{\partial C_{L'}^{\tilde{T}\tilde{T}}}{\partial C_L^{\phi\phi}} \approx \frac{L^5}{16 {L'}^2}
\left[5 L'\frac{d}{dL'}\left(\frac{L'(L'+1) C_{L'}^{TT}}{2\pi}\right) + 3 {L'}^2 \frac{d^2}{d{L'}^2}\left(\frac{L'(L'+1) C_{L'}^{TT}}{2\pi}\right)
\right]
\end{equation}
at leading order in $L/L'$. Since this is a rank-one matrix, the same is true of the covariance between $\hat{C}_L^{\hat{\phi}\hat{\phi}}$ and $\hat{C}_{L'}^{\tilde{T}\tilde{T}}$. We can read off that the singular vector $v_{L'}$ in Eq.~(\ref{eq:svd}) is
\begin{equation}
\label{eq:vLprimeApprox}
v_{L'} \propto 5 L'\frac{d}{dL'}\left(\frac{L'(L'+1) C_{L'}^{TT}}{2\pi}\right) + 3 {L'}^2 \frac{d^2}{d{L'}^2}\left(\frac{L'(L'+1) C_{L'}^{TT}}{2\pi}\right) ,
\end{equation}
which agrees well with that determined directly by singular-value decomposition (see Fig.~\ref{fig:cov_evecs}). This singular vector is also similar to the change
in $C_L^{TT}$ due to lensing over the acoustic part of the spectrum, where the dominant
contribution is from large-scale lenses. Finally, we note that the rank-one structure of the derivative $\partial C_{L'}^{\tilde{T}\tilde{T}}/\partial C_L^{\phi\phi}$ is consistent with the findings of Ref.~\cite{smith_hu_kaplinghat0607} where it is shown that the lensed temperature power spectrum is sensitive to essentially a single mode of $C_L^{\phi\phi}$.

\subsubsection{Comparison  with
  simulations}
Before assessing the importance for parameter estimation of the matter cosmic variance
contribution to the cross-covariance, we validate it against
simulations. The measurements of the temperature-lensing power
correlation in \fig{cov-recphi-t-sims} are too noisy to resolve clearly
the matter cosmic variance contribution.
However,
we can test Eq.~\eqref{eq:cov_CTT_Cphiphi_IN} by correlating the lensed temperature
with the empirical power spectrum $\hat C^{\phi_\mathrm{in}\phi_\mathrm{in}}$ of the
realisation of the input lensing potential without performing any
reconstruction. To reduce the noise of the covariance estimate we work with
the empirical power of the lensed temperature without including beam
effects or noise, which do not affect the matter cosmic variance
effect we are looking for.
Since we have so far neglected $T$-$\phi$
correlations in all calculations, we try to eliminate correlations between the
lensing potential and the unlensed temperature in the simulations by calculating
$\widehat{\mathrm{cov}}(\hat C^{\phi_\mathrm{in}\phi_\mathrm{in}}_L,\hat C^{\tilde T\tilde T}_{L'}-\hat C^{TT}_{L'})$.
Here, $\hat C^{\phi_\mathrm{in}\phi_\mathrm{in}}$ is the empirical power of the input lensing
potential and $\hat C^{TT}$ is the empirical power of the unlensed
temperature.  Additionally, subtracting the unlensed from the lensed empirical temperature power
 reduces the noise of the covariance estimate because it
eliminates the scatter due to cosmic variance of the unlensed
temperature. Otherwise, our estimate of the covariance follows the procedure described
in Sec.~\ref{sec:noise_sims}.
As shown in
\fig{cov-inphi-t-sims}, these broad-band estimates are consistent with the 
theoretical expectation from Eq.~\eqref{eq:cov_CTT_Cphiphi_IN}.

\subsubsection{Mitigating the matter cosmic variance contribution}
\label{sec:lensmitigation}

As we shall show in Sec.~\ref{sec:impact}, the impact on parameter errors of ignoring the covariance between the lensed temperature and reconstruction power due to cosmic variance of the lenses is small for an experiment like Planck. Indeed, this is why the covariance is not accounted for in the current Planck likelihood~\cite{2013arXiv1303.5077P}. However, the covariance is simple to model using Eq.~(\ref{eq:matter_contri_theo}) and could easily be included in a joint analysis. Equivalently, the covariance could be diagonalised by appropriate modifications of the empiricial temperature or lensing power spectra. A symmetric way to do this, mirroring the empirical $N^{(0)}$ correction that we advocate applying to the empirical lensing reconstruction power, is to modify the measured temperature power spectrum by
\begin{equation}
  \label{eq:matter_CV_mitigation}
  \hat C^{\tilde T\tilde T}_{l,\expt} \to  \hat C^{\tilde T\tilde
    T}_{l,\expt}- 
\sum_{l'} A_{ll'}(\hat{C}_{l'}^{\hat{\phi}\hat{\phi}}-2\hat{N}_{l'}^{(0)}) + 
\left\la \sum_{l'}
A_{ll'}(\hat{C}_{l'}^{\hat{\phi}\hat{\phi}}-2\hat{N}_{l'}^{(0)}) \right\ra,
\end{equation}
where 
\begin{equation}
  \label{eq:matter_CV_mitigation_anthony_Amatrix}
  A_{ll'} = \frac{\partial C_{l}^{\tilde T\tilde T}}{\partial C_{l'}^{\phi\phi}}
\left(\frac{C_{l'}^{\phi\phi}}{\langle
\hat{C}_{l'}^{\hat{\phi}\hat{\phi}}\rangle}\right)^2 .
\end{equation}
The expectation value is introduced in Eq.~(\ref{eq:matter_CV_mitigation}) to preserve
the mean value of $\hat C^{\tilde{T}\tilde{T}}_{l,\expt}$. The construct $(C_{l'}^{\phi\phi}/\langle
\hat{C}_{l'}^{\hat{\phi}\hat{\phi}}\rangle)^2$ acts like a Wiener filter on the 
empirical spectrum $\hat{C}_{l'}^{\hat{\phi}\hat{\phi}}-2\hat{N}_{l'}^{(0)}$.
The effect is to `delens' the lensed temperature power spectrum with the (CMB-averaged) lensing effect from any scales in the reconstructed potential power spectrum where the signal-to-noise on the reconstruction is high.
In practice, high signal-to-noise lens reconstructions are never achieved with reconstructions based only on the temperature.

An alternative procedure is suggested by Eq.~(\ref{eq:svd}): project out from $\hat{C}_L^{\hat{\phi}\hat{\phi}}$ the dominant singular vector $u_L$. Although this is lossy, it does have the virtue that the $\tilde{T}\tilde{T}$ part of the likelihood is left unchanged. We discuss the effect of such a lossy projection on the errors of amplitude estimates in Sec.~\ref{sec:impact_amp}.

\subsection{Towards a complete model for power covariances}

The power covariances of Eqs.~\eqref{eq:cov_Cphirec_Cphirec_with_derivative},
\eqref{eq:noise_correlation_heuristic} and
\eqref{eq:matter_contri_theo} can be regarded as a natural extension of the model
for temperature and polarization power covariances found in
\cite{aurelien1205}. We can
summarise the covariances in the  unified
form
\begin{eqnarray}
\nonumber
  \mathrm{cov}(\hat C^{XY}_L, \hat C^{ZW}_{L'}) &=&
  \delta_{LL'} \text{cov}_G(\hat C^{XY}_L, \hat C^{ZW}_{L'})
  +
  \sum_{ll'}
  \frac{\partial C^{XY}_L}{\partial C^{UU}_l}
  \mathrm{cov}(\hat C^{UU}_l,\hat C^{UU}_{l'})
  \frac{\partial  C^{ZW}_{L'}}{\partial C^{UU}_{l'}}
\\
  \label{eq:Cphi_correls_summary}
&&\;
+\sum_{ll'}
  \frac{\partial  C^{XY}_L}{\partial  C^{VV}_l}
  \mathrm{cov}(\hat C^{VV}_l,\hat C^{VV}_{l'})
  \frac{\partial C^{ZW}_{L'}}{\partial  C^{VV}_{l'}},
\end{eqnarray}
where $U$ and $V$ are
listed in Table~\ref{tab:complete_covariance_model} for different
combinations of $XY$ and $ZW$. 
In this context we make the 
 identifications
\begin{equation}
  \label{eq:derivative1UnifiedFormula}
\frac{\partial C^{\hat\phi\hat\phi}_L}{\partial C^{\phi\phi}_l}
\equiv
\delta_{Ll},
\qquad
\frac{\partial C^{\hat\phi\hat\phi}_L}{\partial C^{\tilde T\tilde
    T}_{l}}
\equiv
  \frac{\partial (2\hat N^{(0)}_L)}{\partial\hat C^{\tilde
    T\tilde T}_{l,\expt}} .
\end{equation}
The first term on the right of Eq.~\eqref{eq:Cphi_correls_summary} is the Gaussian covariance
that would arise for Gaussian fields $X$, $Y$, $Z$ and $W$.

\begin{table}[th]
\renewcommand{\arraystretch}{1.1}
\begin{tabular}{@{}lllllllll@{}}
\hline
$XY$, $ZW$ & \phantom{a} & $UU$ & \phantom{a} & $VV$ & \phantom{aa} & Equation here & \phantom{a} & Reference  \\ 	\hline\hline
Any combination of $\tilde T\tilde T$, $\tilde T\tilde E$, $\tilde E\tilde
E$ && $\phi\phi$ && \emph{-- --} && \eqref{eq:CTTexpt-cov} && \cite{aurelien1205} \\
$\tilde B\tilde B$, $\tilde B\tilde B$ && $\phi\phi$ && $EE$ &&
 && \cite{aurelien1205} \\
$\hat\phi\hat\phi$, $\hat\phi\hat\phi$ && \emph{-- --} && $\tilde T\tilde T$ 
&&  \eqref{eq:var-phi-power}, \eqref{eq:cov_Cphirec_Cphirec_with_derivative} && \cite{KCK0302N1,hanson1008} and this work \\
$\hat\phi\hat\phi-2\hat{N}^{(0)}+N^{(0)}$, $\hat\phi\hat\phi-2\hat{N}^{(0)}+N^{(0)}$
&& \emph{-- --} && \emph{-- --}
&& \eqref{eq:var-phi-power}
&& \cite{hanson1008} \\
$\hat\phi\hat\phi$, $\tilde T\tilde T$ && 
$\phi\phi$ && $\tilde T\tilde T$ &&
\eqref{eq:matter_contri_theo},
\eqref{eq:noise_correlation_heuristic} && this work\\
$\hat\phi\hat\phi-2\hat N^{(0)}+N^{(0)}$, $\tilde T\tilde T$ &&
$\phi\phi$ && \emph{-- --} &&\eqref{eq:matter_contri_theo}
 && this work\\
\hline
\end{tabular}
\caption{Covariance types $UU$ and $VV$ that are picked up by power
  covariances $\mathrm{cov}(\hat C^{XY}_L, \hat C^{ZW}_{L'})$ according to
  Eq.~\eqref{eq:Cphi_correls_summary}. Lensed power spectra are denoted
  with tildes and for $\tilde T\tilde T$ the $VV$ covariance should be understood to correspond to
  $C^{\tilde T\tilde T}_{l,\expt}$ including noise.
  The notation $\hat\phi\hat\phi-2\hat N^{(0)}+N^{(0)}$ stands for
  $\hat C_L^{XY}=\hat
  C_L^{\hat\phi\hat\phi}-2 \hat N_L^{(0)}+N_L^{(0)}$, and in this case the Gaussian
variance in the first term of Eq.~\eqref{eq:Cphi_correls_summary} involves
$\langle \hat{C}_L^{\hat{\phi}\hat{\phi}}\rangle^2$, while the Gaussian covariance with
$\hat{C}_{L'}^{\tilde{T}\tilde{T}}$ involves $(C_L^{T\phi})^2$.
The symbol \emph{-- --} means that the corresponding
  contribution is cancelled or can be neglected.  For the
  temperature-lensing covariances in the bottom two rows the leading
  trispectrum contribution of Eq.~\eqref{eq:4ptConnBprimaryContriToTempLensingPowerCov} should be
  added to Eq.~\eqref{eq:Cphi_correls_summary}.
}   
\label{tab:complete_covariance_model}
\end{table}

The general formula of Eq.~\eqref{eq:Cphi_correls_summary} does not include
trispectrum contributions to the temperature-lensing covariance, but
the dominant correction has a simple form [Eq.~\eqref{eq:4ptConnBprimaryContriToTempLensingPowerCov}], which can be
added straightforwardly to the covariance.  Generally, terms involving $C^{T\phi}$ in
Eq.~\eqref{eq:Cphi_correls_summary} can be neglected. 
While we evaluated
derivatives perturbatively in $\phi$, non-perturbative corrections can
be included from numerical derivatives of accurate lensed power spectra
\cite{aurelien1205,camb}. However we do not expect these corrections
to be significant here.  All combinations listed in Table
\ref{tab:complete_covariance_model} have been verified with
simulations in \cite{aurelien1205,hanson1008} or in this work.
Extending the covariance model to polarization-based lensing reconstructions
would be interesting but is beyond the scope of this paper.

\subsection{Impact of correlations on parameter estimation}
\label{sec:impact}

\subsubsection{Lensing amplitude estimates}
\label{sec:impact_amp}

As a first step in assessing the impact of covariances between the temperature and lens-reconstruction power spectra on parameter estimation, we consider constraining 
an overall amplitude parameter $A$ of a fiducial lensing power
spectrum, $C_l^{\phi\phi}=AC_l^{\phi\phi}|_\mathrm{fid}$
\cite{calabrese0803LensingAmplitude} with all other parameters fixed.
The value of $A=1$ corresponds to lensing at the level expected in the fiducial model, while $A=0$ corresponds to no lensing.

The lensing amplitude can be estimated from the
reconstructed lensing potential with 
\begin{equation}
  \label{eq:Aphiphi_estimator}
  \hat A=
\frac{\sum_{l,l'}
C_l^{\phi\phi}\left(\mathrm{cov}_{\hat\phi\hat\phi}^{-1}\right)_{ll'}\left(\hat C_{l'}^{\hat\phi\hat\phi}-N^{(0)}_{l'}-N^{(1)}_{l'}\right)
}{
\sum_{L,L'}C_L^{\phi\phi}\left(\mathrm{cov}_{\hat\phi\hat\phi}^{-1}\right)_{LL'}
 C_{L'}^{\phi\phi}},
\end{equation}
where $\mathrm{cov}_{\hat\phi\hat\phi}$ denotes the auto-covariance of the
reconstructed lensing power given by Eqs.~\eqref{eq:cov-phi-power} and
\eqref{eq:var-phi-power}, evaluated for  $A=1$. 
Equation~(\ref{eq:Aphiphi_estimator}) is the maximum-likelihood estimator for the lensing amplitude if
the likelihood is modeled as a multi-variate Gaussian in the empirical
power spectrum of the lensing reconstruction. This form of the
likelihood will be motivated later.
Alternatively, the lensing amplitude can be extracted directly from the
lensed temperature power spectrum without
invoking lensing reconstruction by
\begin{equation}
  \label{eq:ATT_estimator} 
  \hat A' = \frac{\sum_l \left( \hat C^{\tilde T\tilde
        T}_{l,\mathrm{expt}} - C^{TT}_{l,\mathrm{expt}} \right)
  \left( \mathrm{cov}_{\tilde T\tilde T,\mathrm{expt}}^{-1}  \right)_{ll}
\left(  C^{\tilde T\tilde T}_{l} - C^{TT}_{l} \right)
}
{\sum_{l'} \left(  C^{\tilde T\tilde
        T}_{l'} - C^{TT}_{l'} \right)^2
  \left( \mathrm{cov}_{\tilde T\tilde T,\mathrm{expt}}^{-1}  \right)_{l'l'}},
\end{equation}
where the auto-covariance of
the temperature power is approximated by its leading-order diagonal piece
[see Eq.~\eqref{eq:CTTexpt-cov}].

Since the reconstruction-based amplitude $\hat A$ is linear in the
empirical reconstruction power and the 
temperature-based amplitude $\hat A'$ is linear in the empirical
lensed temperature power, the covariance of $\hat A$ and $\hat A'$ 
involves the temperature-lensing power covariance that we computed
earlier,
\begin{equation}
  \label{eq:cov_AAprime}
  \mathrm{cov}(\hat A,\hat A') =
\sigma_A^2\sigma_{A'}^2
\sum_{l,l',l''}
C_l^{\phi\phi}\left(\mathrm{cov}_{\hat\phi\hat\phi}^{-1}\right)_{ll'}
\mathrm{cov}(\hat C^{\hat\phi\hat\phi}_{l'},\hat C_{l'',\mathrm{expt}}^{\tilde T\tilde T})
\left(\mathrm{cov}^{-1}_{\tilde T\tilde T,\mathrm{expt}}\right)_{l''l''}
\left(C^{\tilde T\tilde T}_{l''}-C^{TT}_{l''}\right),
\end{equation}
where the standard deviations are\footnote{If no empirical $\hat
  N^{(0)}$ subtraction is used we evaluate $\sigma_A$ with
  non-diagonal reconstruction power auto-covariance,
which gives $\sigma_A\approx
  2.7\%$ if $l^\phi_\mathrm{max}\ge 500$ for our noise and beam
  specifications.  The estimated sample standard deviation of $\hat A$ from simulations
  is larger by a factor of up to $1.07$ compared to the theoretical
  expectation. 
If the $\hat
  N^{(0)}$ subtraction is used we evaluate $\sigma_A$ with diagonal
  reconstruction power auto-covariance,
which yields $\sigma_A\approx 2.5$--$2.6\%$ for
  $l^\phi_\mathrm{max}\ge 500$. The estimated sample standard
  deviation from simulations is larger by a factor of at most $1.05$.
  The modest reduction in $\sigma_A$ with empirical $\hat{N}^{(0)}$ subtraction is expected
   given the origin of this estimator as the approximate maximum-likelihood estimator for the  trispectrum (see Appendix \ref{app:edgeworth}). 
For the lensing amplitude $A'$ estimated from the temperature power spectrum, we find $\sigma_{A'}\approx 3.9\%$ for
$l_\mathrm{max}^{T}=2002$ and our Planck-like noise model. The
estimated sample standard deviation from simulations is larger by a factor of $1.01$.
} 
\begin{eqnarray}
  \label{eq:sigma_Aphiphi}
  \sigma_{A}=\left[
\sum_{l,l'}
C_l^{\phi\phi}\left(\mathrm{cov_{\hat\phi\hat\phi}^{-1}}\right)_{ll'}
C_{l'}^{\phi\phi}
\right]^{-1/2},
\quad
  \sigma_{A'}=\left[
\sum_{l}
\left(C_{l}^{\tilde T\tilde T}-C_{l}^{TT}\right)^2
\left(\mathrm{cov}_{\tilde T\tilde T,\mathrm{expt}}^{-1}\right)_{ll}
\right]^{-1/2}.
\end{eqnarray}

The covariance of $\hat A$ and $\hat A'$ can be measured in $N_\mathrm{sims}$ simulations
with
\begin{equation}
  \label{eq:cov_AAprime_esti}
  \widehat{\mathrm{cov}}(\hat A,\hat A')=
\frac{1}{N_\mathrm{sims}-1}\sum_{s=1}^{N_\mathrm{sims}}
(\hat A_{s}-\la \hat  A\ra_\mathrm{sims}) 
(\hat A'_{s}-\la \hat A'\ra_\mathrm{sims}),
\end{equation}
where $s$ labels different realisations. We get the correlation by
dividing by the theoretical standard deviations of Eq.~\eqref{eq:sigma_Aphiphi}. 
 Approximating the spectra as Gaussian variables, the variance of the estimated covariance is $\sigma_A^2 \sigma_{A'}^2 + [\text{cov}(\hat{A},\hat{A}')]^2$, i.e.~the variance of $(\hat{A}-1)(\hat{A}'-1)$, divided by
$N_\mathrm{sims}-1$.\footnote{As a product of two approximately normally distributed variables the
random variable $(\hat A_s-\la \hat A\ra_\mathrm{sims}) (\hat A'_s-\la \hat
  A'\ra_\mathrm{sims})$ is not normally distributed. 
  However, the average over $N_\mathrm{sims}=1000$ realisations is
  approximately normally distributed due to the central limit theorem.
} If we ignore the (small) correlation between $\hat{A}$ and $\hat{A}'$,
the theoretical standard error of the measured covariance is therefore
$\sigma_A\sigma_{A'}/\sqrt{N_\mathrm{sims}-1}$, i.e.~the theoretical
error of the estimated correlation is roughly $1/\sqrt{N_\mathrm{sims}-1}\approx
3.2\%$ for $1000$ simulations.

\begin{figure}[t]
\subfloat[][No $\hat N^{(0)}$ subtraction]{
\includegraphics[width=0.5\textwidth]{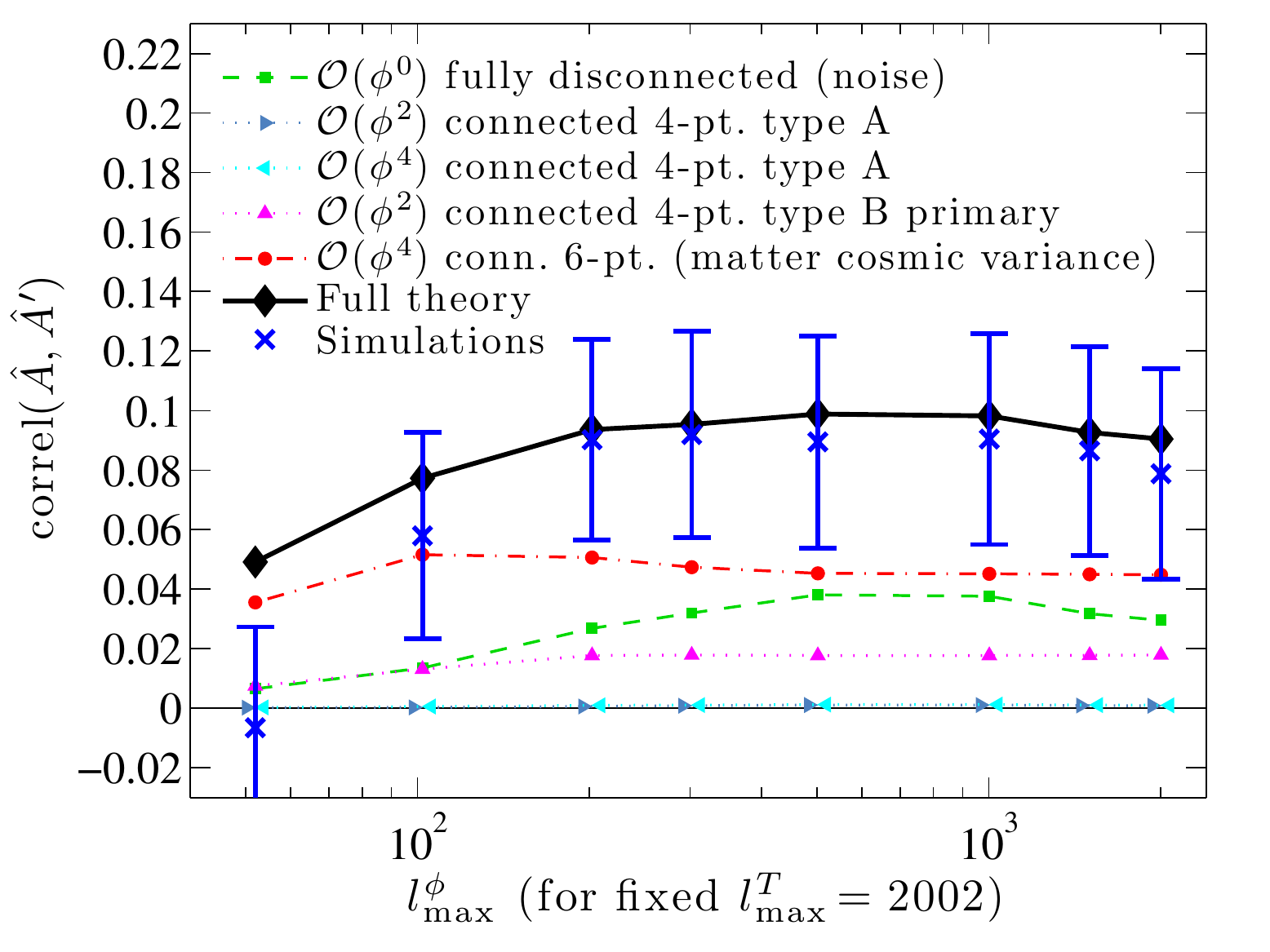}
\label{fig:correl_AAprime_no_emp_N0}}
\subfloat[][With $\hat N^{(0)}$ subtraction]{
\includegraphics[width=0.5\textwidth]{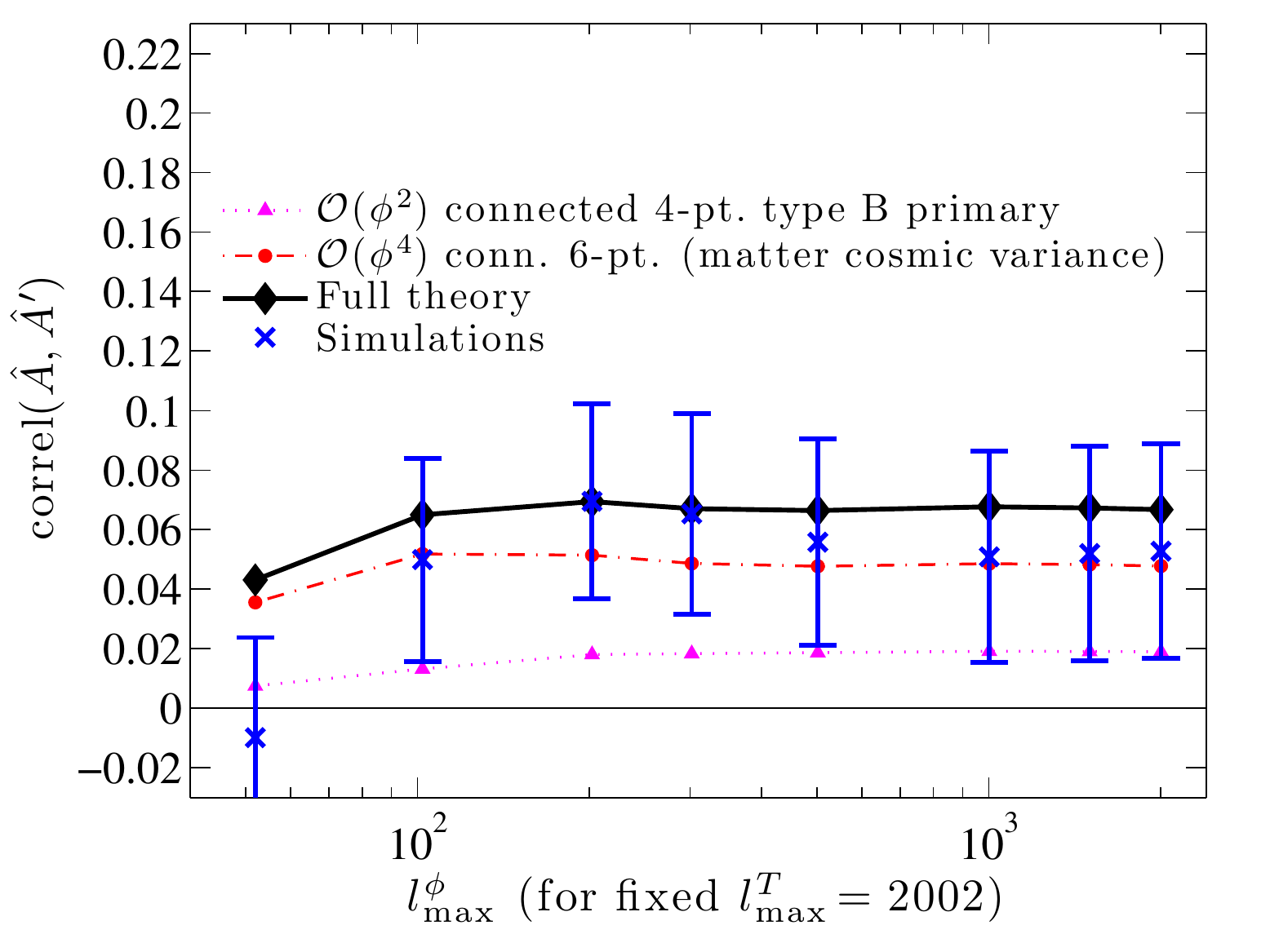}
\label{fig:correl_AAprime_with_emp_N0}}
\caption{(a) Correlation $\mathrm{\widehat{correl}}(\hat A,\hat
  A')=\widehat{\mathrm{cov}}(\hat A,\hat A')/(\sigma_A\sigma_{A'})$ of
  the lensing amplitude estimates of Eqs.~\eqref{eq:Aphiphi_estimator} and
  \eqref{eq:ATT_estimator} measured in $1000$ simulations (blue crosses), if no empirical $\hat
  N^{(0)}$ subtraction is applied. The
  theoretical correlation from Eq.~\eqref{eq:cov_AAprime} (black) includes
  the noise contribution of Eq.~\eqref{eq:noisecontri_cov_CTT_Cphiphi_rec},
  the matter cosmic variance contribution
  of Eq.~\eqref{eq:matter_contri_theo} and the dominant connected
  $4$-point contributions from Eqs.~\eqref{eq:4ptConnAcontriToTempLensingPowerCov} and
  \eqref{eq:4ptConnBprimaryContriToTempLensingPowerCov}. In the
  simulations the lensing potential is
  reconstructed from the lensed temperature power spectrum up to
  $l_\mathrm{max}^\mathrm{rec}=2750$. Then, $\hat A$ is estimated from
  $\hat C_l^{\hat\phi\hat\phi}$ from $l^\phi_\mathrm{min}=2$ up to the
  multipole  
  $l^\phi_\mathrm{max}$, which is varied along the horizontal
  axis. The amplitude $\hat A'$ is estimated using $\hat C^{\tilde T\tilde
    T}_{l,\mathrm{expt}}$ up to $l_\mathrm{max}^T=2002$. Theoretical
  expressions are evaluated with the same cut-offs (non-diagonal covariance
  matrices are cut off before inverting them).   Error bars show standard
  errors on the measured correlation. The errors are very correlated because
  they all involve overlapping low-$l$ reconstruction modes. 
  (b) Same as (a) but with empirical $\hat N^{(0)}$ subtraction.
  Since this removes some covariance contributions, the theoretical
  covariance is given by
  Eqs.~\eqref{eq:4ptConnBprimaryContriToTempLensingPowerCov} and
  \eqref{eq:matter_contri_theo} only.  }
\label{fig:correl_AAprime_plot}
\end{figure}

Figure~\ref{fig:correl_AAprime_plot} shows that the lensing amplitude
correlation measured in our simulations agrees well with the
theoretical correlation of Eq.~\eqref{eq:cov_AAprime} if all contributions to
the temperature-lensing power covariance are taken into account. As
one of the main results of this paper we find that the correlation is
at most $7\%$ if the realisation-dependent $\hat N^{(0)}$
subtraction [Eq.~\eqref{eq:empirical_N0}] is used.
Without this $\hat N^{(0)}$ subtraction the correlation can reach up
to $10\%$ because the disconnected noise contribution
of Eq.~\eqref{eq:noisecontri_cov_CTT_Cphiphi_rec} is not cancelled.
 A plausibility argument for the relatively small level of
amplitude correlations is presented in Appendix
\ref{se:plausibilityOfSmallCorrel}. Briefly, the disconnected noise
contribution is small since the temperature modes that bring most
information to the reconstruction are in between acoustic peaks and
troughs, but the temperature modes that influence $\hat A'$ most strongly are at the peaks and troughs. Since these disjoint modes vary independently, the amplitude correlation is suppressed. The matter cosmic variance
contribution to the amplitude correlation is small since the errors in the measurements of $\hat A$ and $\hat A'$ are dominated by cosmic variance of the temperature, not the lenses.

Figure~\ref{fig:correl_AAprime_plot} also illustrates the relative
importance of the
individual covariance contributions derived above.
 The dominant effect comes from the matter cosmic
 variance contribution [Eq.~\eqref{eq:matter_contri_theo}] which induces an
 amplitude correlation of around $4$--$5\%$ for any
 $l^\phi_\mathrm{max}\gtrsim 100$. The disconnected noise contribution
 [Eq.~\eqref{eq:noisecontri_cov_CTT_Cphiphi_rec}] implies a slightly smaller amplitude
 correlation of $3$--$4\%$ for $l^{\phi}_\mathrm{max}\gtrsim
 300$.\footnote{Although for the \emph{power spectrum} cross-correlation the maximal noise
   contribution is about an 
   order of magnitude larger than the 
maximal matter cosmic
  variance contribution, the latter
  can be more relevant for the correlation of amplitude estimates
  because of the phase argument given in the text.}
 An additional contribution to the temperature-lensing power covariance
 comes from the lensed temperature trispectrum discussed in
 Appendix~\ref{se:Conn4ptAppendix}. The dominant term
 [Eq.~\eqref{eq:4ptConnBprimaryContriToTempLensingPowerCov}] gives rise to a 
 $2\%$ lensing amplitude correlation.
The agreement between simulations and
theory in \fig{correl_AAprime_plot} gives us confidence that the power
correlations modeled here include all relevant contributions for amplitude measurements.

Instead of fixing $l^\phi_\mathrm{min}=2$ and varying
$l^\phi_\mathrm{max}$ it is worthwhile to consider the disjoint
reconstruction bins
$[l^\phi_\mathrm{min},l^\phi_\mathrm{max}]=[40,84]$, $[85,129]$,
$[130,174]$, $\dots$, $[355,400]$ used for the Planck analyses in
\cite{2013arXiv1303.5076P,2013arXiv1303.5077P}. If the
realisation-dependent $\hat N^{(0)}$ is used, the theoretical
correlation of the lensing amplitude estimated from one of these bins
alone with the lensing amplitude estimated from the temperature power
(for $l^T_\mathrm{max}=2002$) is $5\%$, $3.6\%$ and $2\%$ for the
first three reconstruction bins, and decreases further for
the remaining higher-$l$ bins. This is consistent with the
correlations estimated from our $1000$ simulations.  In particular,
this result shows that the lensing amplitude estimated from the
temperature power spectrum and the reconstruction amplitudes used in the Planck
lensing likelihood~\cite{2013arXiv1303.5077P}
are nearly uncorrelated, which justifies neglecting this correlation in
the likelihood.

For experiments with superior noise and beam characteristics the
matter cosmic variance contribution to the temperature-lensing power
covariance does not change, but the lensing amplitude errors
decrease. We therefore expect the corresponding amplitude
correlation to increase.  For example,  for a full-sky experiment with SPT-like noise and beam
specificiations, $\sigma_N=18\, \mu\mathrm{K}\,\mathrm{arcmin}$ and
$\sigma_\mathrm{FWHM}=1\,\mathrm{arcmin}$, the amplitude
correlation from the matter cosmic variance contribution alone is around
$10$--$11\%$  for $l_\mathrm{max}^T=
2002$ and $l_\mathrm{max}^\phi=500$--$1000$.

\subsubsection*{Combined lensing amplitude estimate}
We have presented two estimators of the lensing amplitude: $\hat A$ is
linear in the reconstruction power and 
$\hat A'$ is
linear in the CMB power. These two estimates can be combined with inverse variance
weighting,
\begin{equation}
  \label{eq:combined_lensing_amplitude}
  \hat A_C=\frac{1}{\sigma_A^{-2}+\sigma_{A'}^{-2}} \left(
\frac{\hat A}{\sigma_A^2}+ \frac{\hat A'}{\sigma_{A'}^2}
\right).
\end{equation}
This combined estimator is the maximum-likelihood estimator for the
lensing amplitude if $\hat A$ and $\hat A'$ are assumed to be
uncorrelated.
If there is a correlation between $\hat A$ and $\hat A'$ this does not change the expectation value of $\hat{A}_C$, but it does change its
variance, which  is then given by\footnote{To first order in the correlation, this sampling variance of the combined $\hat{A}_C$ is the same as the sampling variance of the optimal combined estimate that takes account of the correlations.}
\begin{equation}
  \label{eq:A_combined_variance}
  \mathrm{var}(\hat A_C) = 
\sigma_A\sigma_{A'}\left(
\frac{\sigma_{A}}{\sigma_{A'}}+\frac{\sigma_{A'}}{\sigma_{A}}
\right)^{-2}
\left[
\left(
\frac{\sigma_{A}}{\sigma_{A'}}+\frac{\sigma_{A'}}{\sigma_{A}}
\right)
+2\,\mathrm{correl}(\hat A,\hat A')
\right].
\end{equation}
A correlation between $\hat A$ and $\hat A'$ therefore increases the
$1\sigma$ error of the combined estimator
\eqref{eq:combined_lensing_amplitude} by a factor of
\begin{equation}
  \label{eq:A_combined_stddev_change_due_to_correlAAprime}
  \frac{\sigma_{A_C}|_{\mathrm{cov}(A,A')\neq
      0}}{\sigma_{A_C}|_{\mathrm{cov}(A,A')=0}}
=  \sqrt{1+
\frac{2(\sigma_A/\sigma_{A'})}{1+(\sigma_A/\sigma_{A'})^2}
\mathrm{correl}(\hat A,\hat A')
}
\leq \sqrt{1+\mathrm{correl}(\hat A,\hat A')}
\end{equation}
compared to the error if $\hat A$ and $\hat A'$ were
uncorrelated. Since correlations between $\hat A$ and $\hat A'$ were
found to be at most $7\%$ if the empirical $\hat N^{(0)}$
subtraction is used, the $1\sigma$ error of the combined lensing
estimate changes by at most $3.5\%$ for Planck
($5.5\%$ for the full-sky SPT-like experiment mentioned above). Noting that this is the error on the error bar, the
correlations between $\hat A$ and $\hat A'$ found above can be safely
neglected when combining these two estimates of the lensing amplitude.

We briefly introduced a projection technique in Sec.~\ref{sec:lensmitigation} to remove the covariance between the reconstructed lensing power and the lensed temperature power spectrum due to the cosmic variance of the lenses.
A simple way to perform the projection is to modify the covariance matrix $(\text{cov}_{\hat\phi\hat\phi})_{ll'}$ in Eq.~\eqref{eq:Aphiphi_estimator} by adding $\lambda' u_l u_{l'}$, where $u_l$ is the dominant singular vector in Eq.~(\ref{eq:svd}), and taking $\lambda'$ to infinity. To the extent that the covariance between the lensing and temperature power spectra is really rank-one, this procedure removes the correlation between $\hat{A}$ and $\hat{A}'$ exactly. However, the variance of $\hat{A}$ is increased by projection: it is still given by Eq.~(\ref{eq:sigma_Aphiphi}) but with the modified $(\text{cov}_{\hat\phi\hat\phi})_{ll'}$. For $l^\phi_{\mathrm{max}}=500$, we find that $\sigma_A$ is increased from $0.025$ to
$0.042$, i.e.\ a 70\% increase. This falls to 60\% for  $l^\phi_{\mathrm{max}}=1000$. The reason for the large increase is that $u_l$ is rather similar in shape to the signal whose amplitude we are trying to reconstruct. Given the large increase in the error on $\hat{A}$, and that ignoring the effect of the covariance between the lensing reconstruction and temperature power spectra is relatively harmless, we do not advocate the use of projection to remove the correlations.

\subsubsection{Cosmological parameters}

We naively expect the impact of power correlations on cosmological
parameters to be smaller than for the lensing amplitude, because the
latter is directly related to the lensing potential on all scales and can
therefore accumulate contributions from the full power covariances. We
confirm this with a simple Fisher analysis. The covariance matrix for the joint data
vector $\underline{\hat{C}}=(\hat C^{\tilde T\tilde T}_{\expt}, \hat
C^{\hat\phi\hat\phi}-2\hat N^{(0)}+N^{(0)})$ is
\begin{equation}
  \label{eq:jointCovMat}
  \mathrm{cov}_{{LL'},\,\mathrm{joint}} \equiv \mathrm{cov}(\underline{\hat{C}}_L,\underline{\hat{C}}_{L'}) = 
\left( \begin{matrix} 
\delta_{LL'}\mathrm{var}_G(C^{\tilde T\tilde T}_{L,\expt}) & 
\mathrm{cov}(\hat C^{\tilde T\tilde T}_{L,\expt},\hat C^{\hat\phi\hat\phi}_{L'}-2\hat N^{(0)}_{L'})
 \\ 
\mathrm{cov}(\hat C^{\hat\phi\hat\phi}_{L}-2\hat N^{(0)}_{L}, \hat C^{\tilde T\tilde T}_{L',\expt})
 & \delta_{LL'}\mathrm{var}_G(\la \hat C_L^{\hat\phi\hat\phi}\ra)
\end{matrix} \right).
\end{equation}
Fisher errors are obtained by taking the square root of the diagonal
entries of the inverse of the Fisher matrix
\begin{equation}
  \label{eq:jointFisherMatrix}
  F_{ij} =\sum_{LL'} \frac{\partial \underline{C}_L}{\partial
    p_i}
(\mathrm{cov}_\mathrm{joint}^{-1})_{LL'}\frac{\partial \underline{C}_{L'}}{\partial
    p_j}
\end{equation}
for cosmological parameters $\mathbf{p}= (\Omega_bh^2, \Omega_ch^2, h,
\tau, A_s, n_s, \Omega_\nu h^2, \Omega_K)$ and theoretical power
spectra $\underline C= (C^{\tilde T\tilde T},C^{\phi\phi})$
(assuming cosmology-independent $ N^{(0)}$ for simplicity). Including the off-diagonal temperature-lensing
covariances of
Eqs.~\eqref{eq:4ptConnBprimaryContriToTempLensingPowerCov} and \eqref{eq:matter_contri_theo}
 in
Eq.~\eqref{eq:jointCovMat} increases the Fisher errors for these
parameters by at most $0.7\%$ ($0.5\%$ if only
Eq.~\eqref{eq:matter_contri_theo} is used) compared to a completely
diagonal covariance matrix.\footnote{To obtain accurate derivatives
  for the Fisher matrix we assumed a fiducial cosmology with massive
  neutrinos, $\mathbf{p}_\mathrm{fid} = (0.0226, 0.1123, 0.704, 0.087,
  2.167\times 10^{-9}, 0.963, 0.005, 0)$, which differs slightly from
  the cosmology used throughout the rest of the paper. The Fisher errors
  were computed for $l_\mathrm{max}^{T}=2002$ and
  $l_\mathrm{max}^{\phi}=1002$. } The
off-diagonal part of the joint covariance matrix
can therefore safely be neglected for
cosmological parameter estimation with a Planck-like experiment.

\section{Towards a lensing likelihood}

\label{se:towardsLensingLikelihood}

As argued in Sec.~\ref{sec:introduction}, dealing with the exact likelihood for the lensed CMB temperature is generally computationally prohibitive. For this reason, we have focussed on a form of data compression whereby the non-Gaussian lensed CMB is represented by its 2- and 4-point functions (the latter via the lensing reconstrucion power spectrum). In computing the correlations between these spectra, we have implicitly been assuming that the likelihood takes the form of a multi-variate Gaussian in the spectra. In this section, we test the accuracy of this assumption in simple parameter-estimation exercises.

\subsection{Lensing amplitude from lensing reconstruction}

As a toy model, we first aim to constrain the lensing amplitude $A$
from the lensing reconstruction alone. 
Considering an isotropic CMB survey, with Planck-like
noise as described earlier, we consider two simple models for the likelihood, both of which depend only on the empirical power spectrum of the reconstruction. The first is the usual isotropic likelihood for a Gaussian field:
\begin{equation}
\label{eq:likelihood1}
  -2\ln \mathcal{L}_1(\hat\phi | A )  =   \sum_{l} (2l+1) \left(
  \frac{\hat{C}_l^{\hat\phi\hat\phi} }{ AC_l^{\phi\phi}+N_l
  } 
+ \ln \big\vert AC_l^{\phi\phi}+N_l \vert
\right) + (\text{const.}) \, .
\end{equation}
This would be correct if $\hat{\phi}$ were a Gaussian field. However, 
since the reconstruction is manifestly non-Gaussian, we do not expect this likelihood to perform well. The second is Gaussian in the empirical power spectrum
$\hat C^{\hat\phi\hat\phi}$:
\begin{equation}
 \label{eq:likelihood2}
-2\ln\mathcal{L}_2 (\hat{C}^{\hat\phi\hat\phi} | A) =
  \sum_{l,l'}
 \left[\hat{C}_l^{\hat\phi\hat\phi}-(AC_l^{\phi\phi}+N_l)\right] (\mathrm{cov}_{\hat\phi\hat\phi}^{-1})_{ll'} 
 \left[\hat{C}_{l'}^{\hat\phi\hat\phi}-(AC_{l'}^{\phi\phi}+N_{l'})\right] + (\text{const.}) \, .
\end{equation}
The theoretical reconstruction power auto-covariance
$\mathrm{cov}_{\hat\phi\hat\phi}$, given by Eqs.~\eqref{eq:cov-phi-power} and \eqref{eq:var-phi-power},
and the bias of the reconstructed lensing power, $N=N^{(0)}+N^{(1)}$,
 are evaluated for the fiducial
amplitude $A=1$.
The empirical $\hat N^{(0)}$ subtraction
is obtained by replacing 
$N$ with $\hat
N=2\hat N^{(0)}-N^{(0)}+N^{(1)}$.
For $\mathcal{L}_1$, the maximum-likelihood estimate for $A$ (given a realisation of the lensing
reconstruction) is found numerically by direct evaluation of $\mathcal{L}_1$ for
various $A$. The maximum likelihood estimator for $A$ based on the
second likelihood $\mathcal{L}_2$ is given by
Eq.~\eqref{eq:Aphiphi_estimator}.

We compute estimates $\hat A$ of the lensing amplitude for $1000$
realisations of the lensed CMB. The sample mean of $\hat A$ should be
 unity and the sample variance of $\hat A$, i.e.~the scatter of the
best-fit amplitude over different realisations, should agree with the
typical width of the likelihood evaluated for a single
realisation. Checking these two properties provides a non-trivial test
of the likelihood $\mathcal{L}_1$. In contrast, for $\mathcal{L}_2$,
rather than testing the
accuracy of $\mathcal{L}_2$, 
the sample
mean and sample variance of $\hat A$ just  test our understanding of
the mean and covariance of $\hat C^{\hat\phi\hat\phi}$.\footnote{This is because the estimated lensing amplitude [Eq.~\eqref{eq:Aphiphi_estimator}] is 
linear in $\hat C^{\hat\phi\hat\phi}$. For example, if the true likelihood
depends on 
the third power of $\hat C^{\hat\phi\hat\phi}$,  it would be
possible that this only shows up in
the skewness of $\hat A$. This issue will be addressed later by
considering the tilt of the lensing power spectrum, which depends
non-linearly on the reconstruction power.}
This test is still useful to
check for residual biases and the accuracy of our model for the reconstruction power
covariance. 

\begin{figure}[tp]
\subfloat[][$\mathcal{L}_1$ without $N^{(1)}$: biased]{
\includegraphics[width=0.28\textwidth]{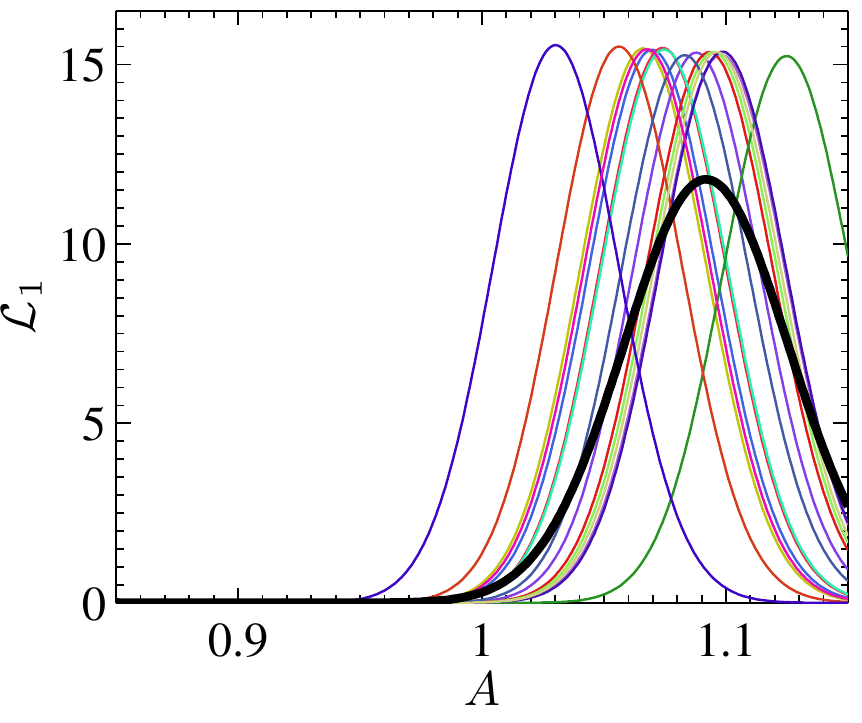}
\label{fig:likeli1d_LP1_noNL1}}
\qquad
\subfloat[][$\mathcal{L}_1$ with $N^{(1)}$: unbiased but underestimates variance]{
\includegraphics[width=0.28\textwidth]{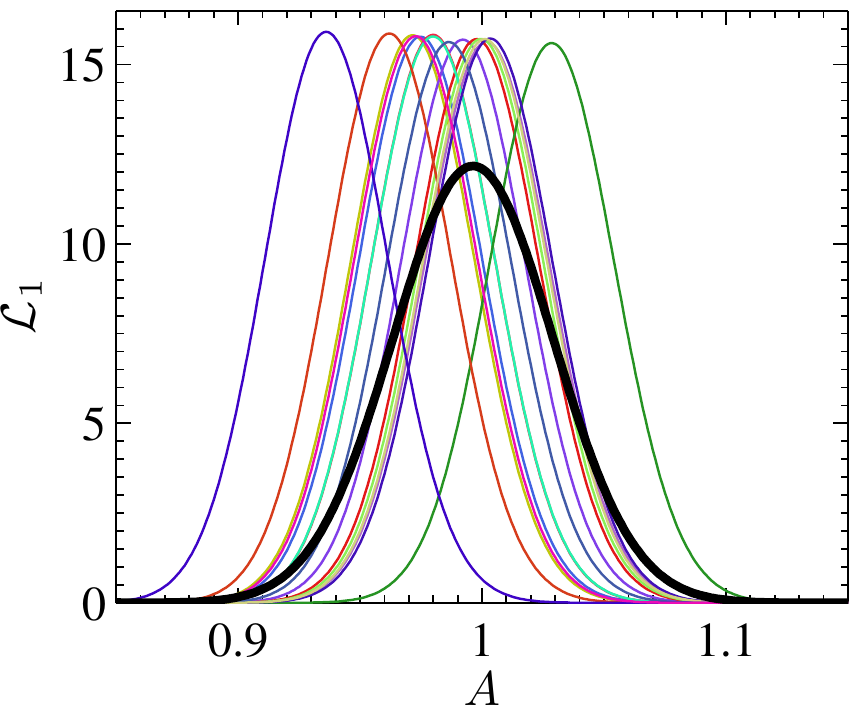}
\label{fig:likeli1d_LP1_withNL1}}
\qquad
\subfloat[][$\mathcal{L}_1$ for Gaussian mock $\hat\phi_\mathrm{G}$ 
  with power $C^{\phi\phi}+N^{(0)}+N^{(1)}$: good]{
\includegraphics[width=0.28\textwidth]{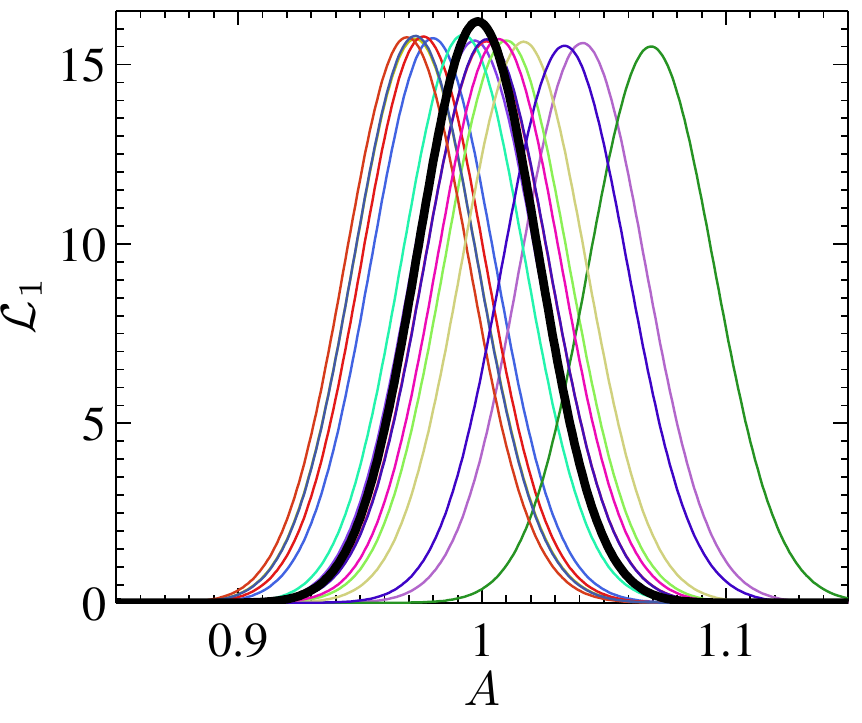}
\label{fig:likeli1d_LP4}}
\\
\subfloat[][$\mathcal{L}_2$ with diagonal covariance: underestimates
variance]{
\includegraphics[width=0.28\textwidth]{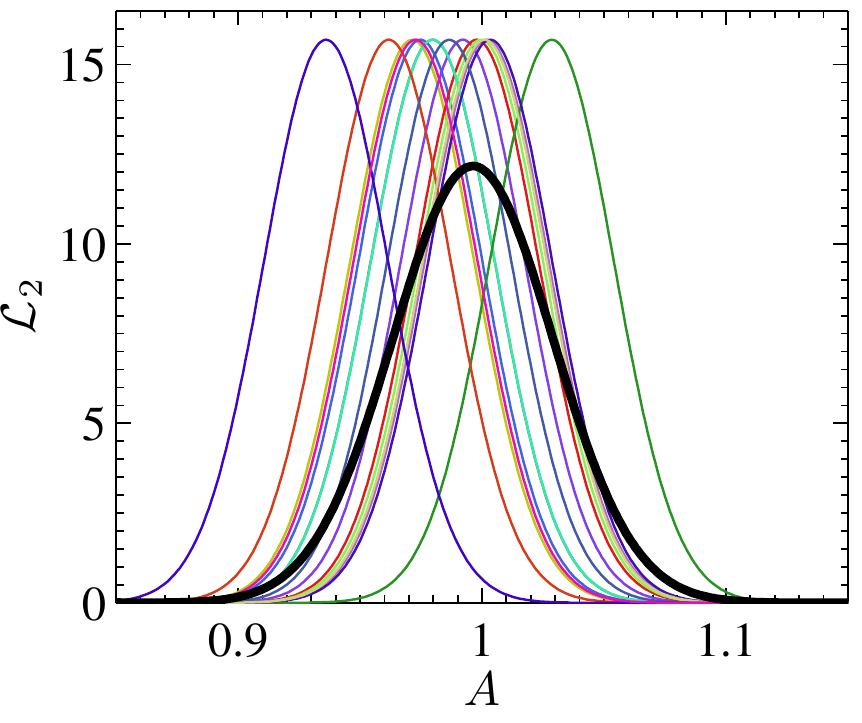}
\label{fig:likeli1d_LP13}}
\qquad
\subfloat[][$\mathcal{L}_2$ with non-diagonal,
non-Gaussian covariance: good]{
\includegraphics[width=0.28\textwidth]{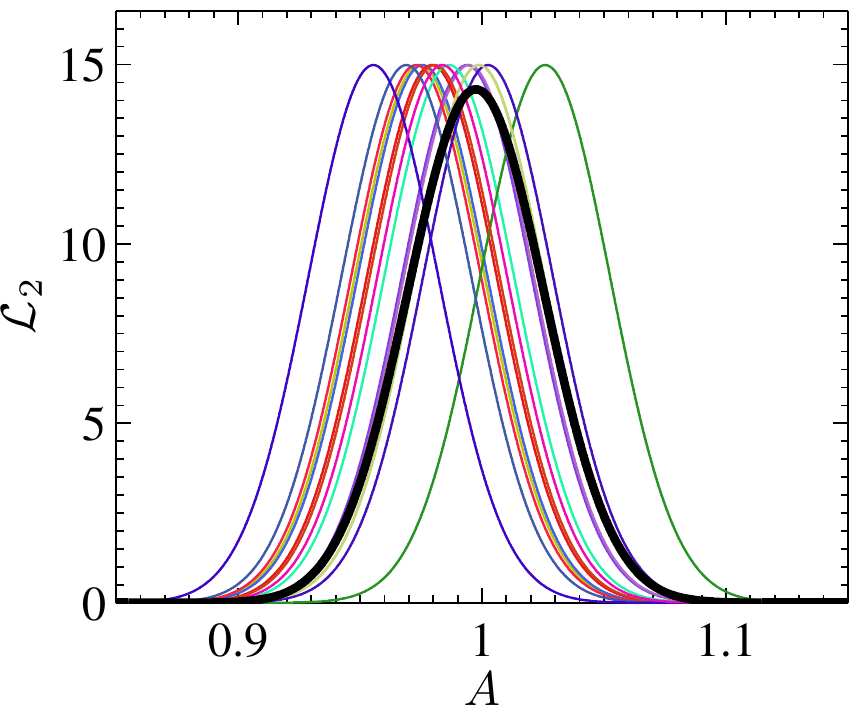}
\label{fig:likeli1d_LP20}} 
\qquad
\subfloat[][$\mathcal{L}_2$ with empirical $\hat N^{(0)}$ subtraction
  and diagonal
  covariance: good]{
\includegraphics[width=0.28\textwidth]{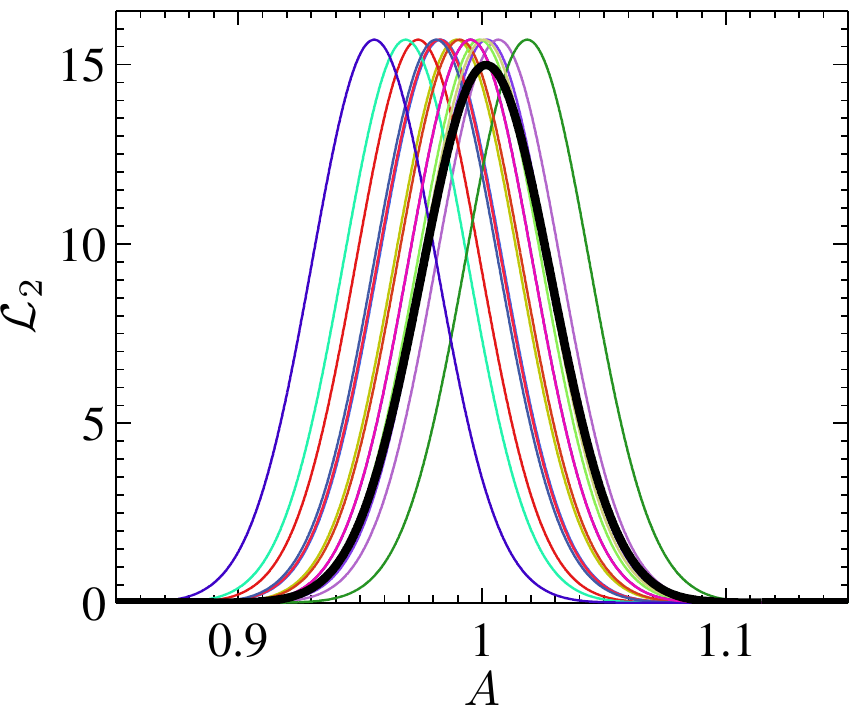}
\label{fig:likeli1d_LP15}}
 \caption{Likelihoods for the lensing amplitude for $15$ different realisations (thin, coloured lines), using the lensing reconstruction up to
   $l_\mathrm{max}^{\phi}=2650$, compared with the scatter in the best-fit amplitude obtained over 1000 simulations. Each likelihood
peaks at some best-fit parameter $\hat A$, and the mean and scatter of these best-fitting amplitudes over all 1000 simulations correspond to the Gaussian curves (thick black).
The upper panels show $\mathcal{L}_1$ from
Eq.~\eqref{eq:likelihood1}, while the lower panels show $\mathcal{L}_2$
given by Eq.~\eqref{eq:likelihood2}. 
All
curves are normalised such that their integral over $A$ is unity.}
\label{fig:likeli1d_all}
\end{figure}

Figure~\ref{fig:likeli1d_all} compares the likelihood evaluated for several
individual realisations (coloured) with a Gaussian (black) with mean
and standard deviation given by sample mean and standard deviation of
$\hat A$ averaged over all $1000$ realisations (for
$l_\mathrm{max}^\phi=2650$ and our Planck-like noise model).
Including the $N^{(1)}$ bias is important at high multipoles for both
likelihoods: e.g.~without it, $\mathcal{L}_1$ overestimates the
lensing amplitude by $9\%$; see \fig{likeli1d_LP1_noNL1}. Including
the $N^{(1)}$ bias in $\mathcal{L}_1$ yields the correct lensing
amplitude $A=1$ in the mean, but the scatter of $\hat{A}$ over realisations is
more than $30\%$ larger than the typical width of $\mathcal{L}_1$ in a
single realisation; see \fig{likeli1d_LP1_withNL1}. The likelihood $\mathcal{L}_1$
underpredicts the error of $A$ because of the non-Gaussianity of
$\hat\phi$. This is demonstrated in \fig{likeli1d_LP4}, for which we replace the reconstructions with Gaussian simulations of a field with power spectrum
$C_l^{\phi\phi} + N_l^{(0)} + N_l^{(1)}$. In this case, $\mathcal{L}_1$ should be exact and
the scatter does indeed match the widths of individual realisations.

 In
contrast, $\mathcal{L}_2$ can partly model the non-Gaussianity of
$\hat\phi$ through the non-diagonal reconstruction power
auto-covariance.  
We compute
$\hat A$ based on $\mathcal{L}_2$ for 
different forms of the lensing covariance.  Neglecting off-diagonal
contributions to $\mathrm{cov}_{\hat\phi\hat\phi}$ gives likelihood-based errors for
$A$ less than $80\%$ of the scatter of $\hat A$ across the
simulations; see \fig{likeli1d_LP13}.  If we include non-Gaussian, off-diagonal contributions given by Eq.~\eqref{eq:cov-phi-power}, we find that $\mathcal{L}_2$ predicts the scatter
in $A$ to better than $5\%$; see \fig{likeli1d_LP20}.  Similar results are achieved with the
empirical $\hat{N}^{(0)}$ correction
of Eq.~\eqref{eq:empirical_N0} and diagonal (Gaussian) reconstruction power
covariance; see \fig{likeli1d_LP15}.

\subsection{Two-parameter likelihood tests with lensing amplitude and
  lensing tilt}

\begin{figure}[tp]
\centering
\hspace*{-6mm}
\subfloat[][$\mathcal{L}_2$ with diagonal covariance]{
\includegraphics[width=0.48\textwidth]{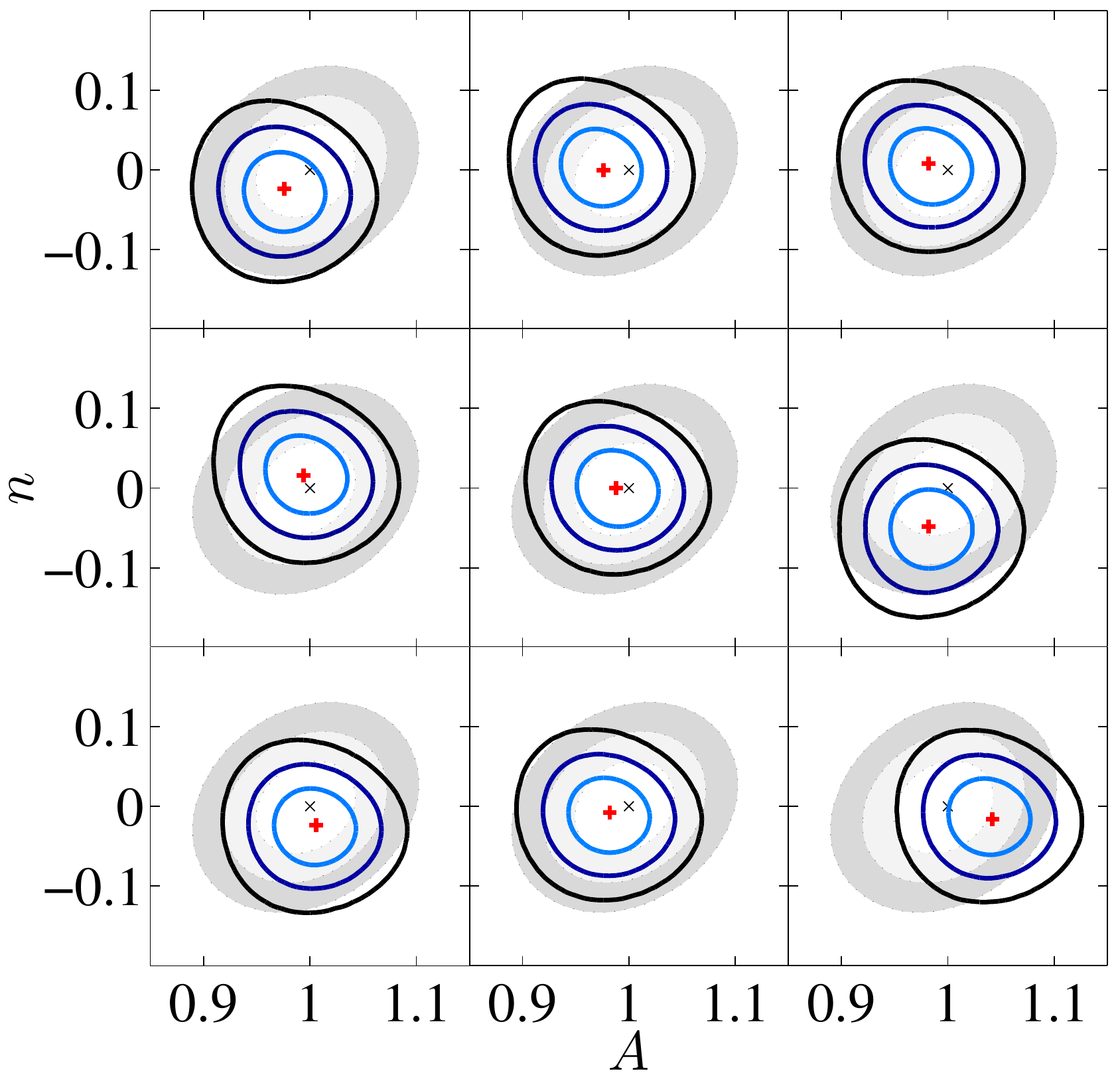}
\label{fig:contour_binned_tmp1b}}
\hspace*{3mm}
\subfloat[][$\mathcal{L}_2$ with non-diagonal covariance]{
\includegraphics[width=0.48\textwidth]{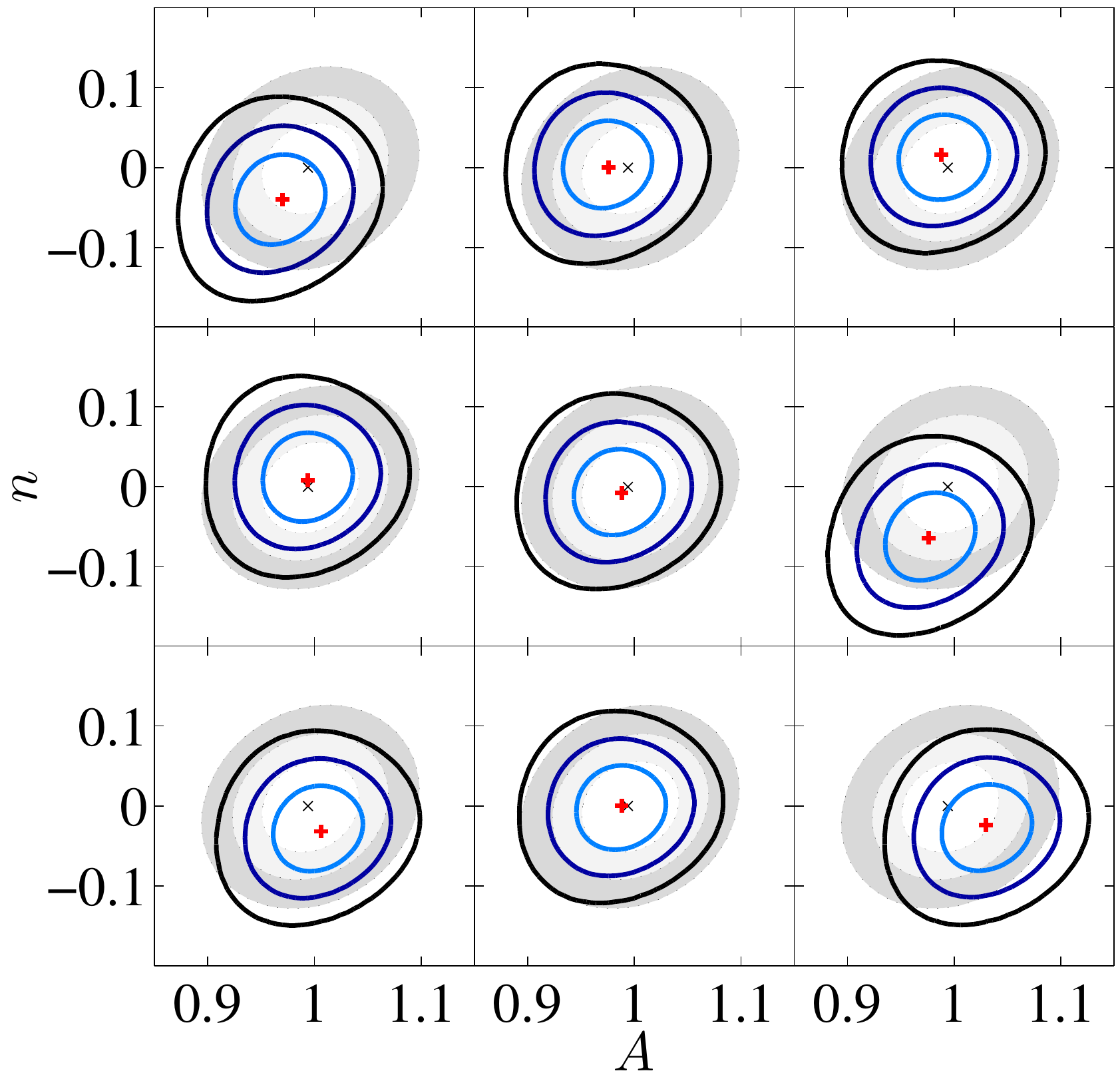}
\label{fig:contour_binned_tmp2b}}
\caption{Likelihood test if lensing amplitude $A$
  (horizontal axis) and tilt $n$ (vertical axis) are varied. 
 \emph{Thick lines:} Contours enclosing $68\%$, $95\%$ and $99.7\%$ of the probability for the likelihood $\mathcal{L}_2$ \eqref{eq:likelihood2} evaluated for nine
  realisations. \emph{Gray filled ellipses:} 
  Contours of a Gaussian with central point and covariance matrix
  estimated from the scatter of the best-fit parameters over 1000 realisations.
 The lensing power
  auto-covariance $\mathrm{cov}_{\hat\phi\hat\phi}$ is assumed to be diagonal
  in (a), while (b) also includes the non-diagonal contribution in Eq.~\eqref{eq:cov-phi-power}.
The reconstruction power is used up to
  $l_\mathrm{max}^\phi=1002$ and is binned as described in the text. 
}
\label{fig:contour_binned_2plots_maintext}
\end{figure}

To test the likelihood approximation
$\mathcal{L}_2$ we use the lensing reconstruction additionally to constrain the lensing
 tilt $n$ defined by   
\begin{align}
  \label{eq:lensing_tilt}
  C_l^{\phi\phi}=A \left(\frac{l}{l_*}\right)^n C_l^{\phi\phi}|_\mathrm{fid}.
\end{align}
The pivot multipole $l_*=124$ is chosen such that the Fisher matrix
associated with $\mathcal{L}_2$ is
diagonal [for $l_\mathrm{max}^\phi=\mathcal{O}(10^3)$], implying that
the parameters $A$ and $n$ are approximately uncorrelated.
The
likelihoods for nine realisations are compared with the scatter of the
best-fit parameters over $1000$ realisations in
\fig{contour_binned_2plots_maintext}.  If the non-diagonal
lensing power covariance of Eq.~\eqref{eq:cov-phi-power} is included we find
good agreement (without empiricial $\hat{N}^{(0)}$ subtraction).
Note that we have binned  the reconstruction power in bins with
boundaries at
\begin{equation}
  \label{eq:planck-binning}
l=2, 13, 35, 75,
115, 155, 195, \dots \quad \mbox{(increasing by $40$ above $l=35$)}.
\end{equation}
Similar results for the unbinned case will be summarised in
\fig{frac_contourareas_all} below. 

\begin{figure}[tp]
\centering
\hspace*{-3mm}
\subfloat[][$\mathcal{L}_2$, diagonal covariance]{
\includegraphics[width=0.33\textwidth]{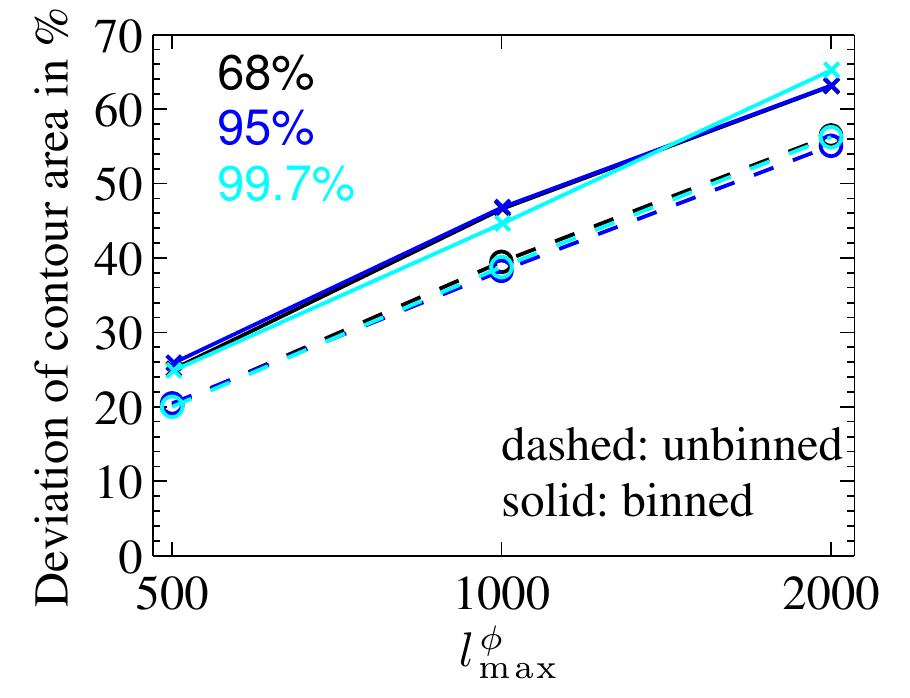}
\label{fig:frac_contour_areas_LP13}}
\subfloat[][$\mathcal{L}_2$, non-diagonal covariance]{
\includegraphics[width=0.33\textwidth]{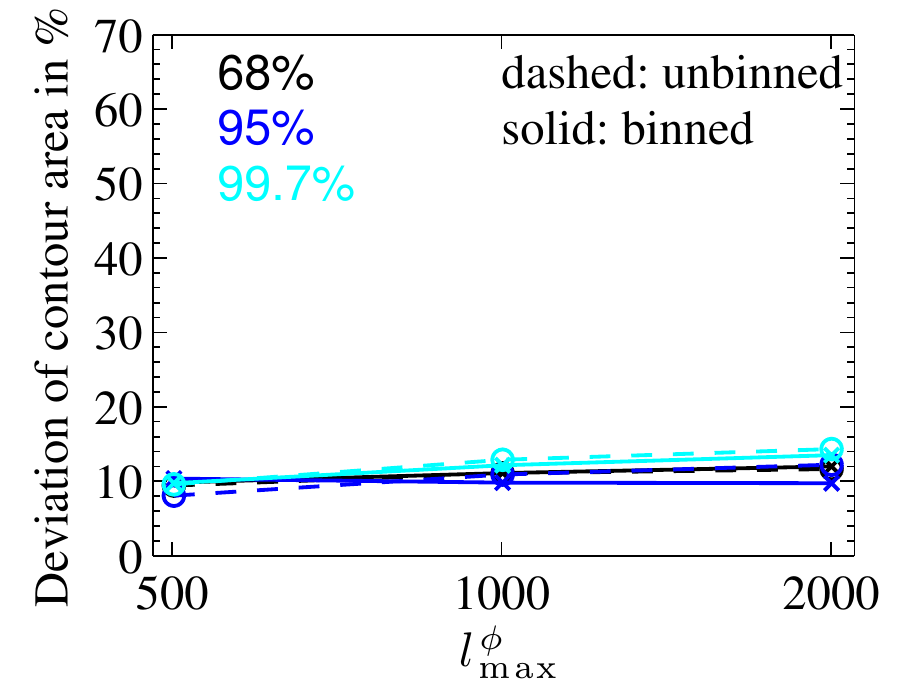}
\label{fig:frac_contour_areas_LP20}}
\subfloat[][$\mathcal{L}_2$, empirical $\hat N^{(0)}$ correction, diagonal covariance]{
\includegraphics[width=0.33\textwidth]{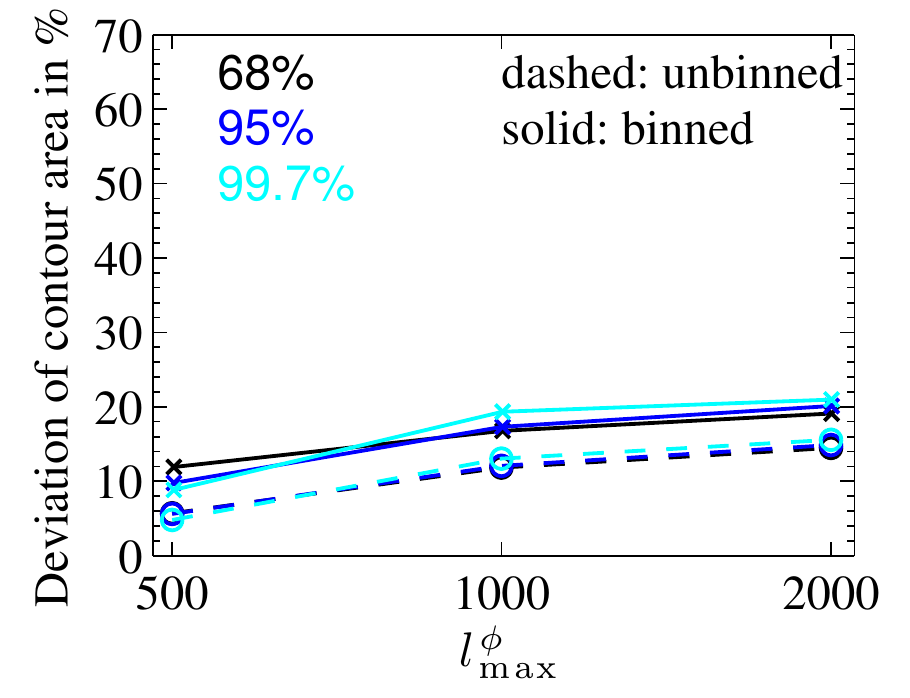}
\label{fig:frac_contour_areas_LP15}}
\caption{Quantitative comparison of areas in the $A$ vs.~$n$ plane enclosed by contours of the $\mathcal{L}_2$ likelihood.
We show the fractional deviation of the
  areas of Gaussians (with sample mean and sample covariance derived from the scatter of the best
  fit-parameters over 1000 realisations) from the average area in a single
  realisation. Positive values mean that the width of the likelihood typically
 underestimates the scatter of its peak across realisations. Results are shown for unbinned power
  spectra (dashed lines) and power spectra binned according to the scheme of
  Eq.~\eqref{eq:planck-binning} (solid lines); and for $68\%$ (black), $95\%$ (blue) and $99.7\%$ (cyan) confidence levels.  The
  maximum multipole of the reconstruction power is varied along
  the horizontal axis (only for three values which are
  connected by straight lines to guide the eye).  The lensing power auto-covariance
  $\mathrm{cov}_{\hat\phi\hat\phi}$ in $\mathcal{L}_2$ is assumed to be
  diagonal in (a), while (b) also includes the non-diagonal
  contribution of Eq.~\eqref{eq:cov-phi-power}.  Panel (c) is for empirical $\hat
  N^{(0)}$ bias correction and diagonal covariance.  The contours in
  \fig{contour_binned_2plots_maintext} correspond to the crosses at
  $l^\phi_\mathrm{max}=1002$ in (a) and (b). }
\label{fig:frac_contourareas_all}
\end{figure}

To quantify the level of agreement between the likelihoods for individual
realisations and the scatter of their best-fit parameters, we compare the 
areas of the confidence contours shown in Fig.~\ref{fig:contour_binned_2plots_maintext}.
We show in  \fig{frac_contourareas_all} the fractional deviation of the
areas of the Gaussian, with sample mean and sample covariance matched to the scatter of the best-fit parameters over realisations, from the average area of the likelihoods for individual realisations
(i.e.~the fractional deviation of gray background areas from average areas enclosed by the solid lines in \fig{contour_binned_2plots_maintext}).

Neglecting the off-diagonal
contribution to the lensing power covariance, which is largest at
high reconstruction multipoles \cite{hanson1008}, gives narrow misshapen likelihoods that underestimate the scatter across simulations. This is particularly so for $l_\mathrm{max}^\phi\gtrsim
1000$ where the confidence areas disagree by around $40$--$65\%$.
Binning does not help because it does not reduce the
broad-band correlations of the reconstruction power.
  The agreement is better when the non-diagonal 
reconstruction power covariance is used (the disagreement of
confidence areas is at most $14\%$). 
 Alternatively, if the empirical $\hat N^{(0)}$
bias correction and the diagonal covariance is used, the confidence
areas deviate by at most $22\%$. 
If we assume circular contours the fractional deviation of the contour
radius is $\sqrt{1+d}-1 < d/2$
 if $d$ is the fractional deviation of the
contour areas. Taking this as the approximate fractional error of the
marginalised error bars of $A$ or $n$ shows that the error on the
error bars is smaller than $11\%$ if the non-diagonal reconstruction covariance or empirical $\hat
N^{(0)}$ subtraction are used in $\mathcal{L}_2$. Therefore these two cases
provide a reasonably accurate model for the lensing likelihood in this test. 
If diagonal reconstruction
power covariance is assumed, and no empirical $\hat N^{(0)}$ subtraction performed, the error on the error bars can reach $30\%$ even if binning is used. 
It is also worth noting that in this last case only the
confidence areas \emph{increase} with $l^\phi_\mathrm{max}$, i.e.~the analysis
is clearly non-optimal. 

Note that in the above, ideally, we should use a histogram of the best-fit parameters instead of fitting a Gaussian to their scatter.
This would test the tails of the distribution much better
because it would include possible skewness etc. However we find that histograms
from $1000$ simulations are too noisy to be useful for this purpose, giving results
that scatter significantly with changes in histogram binning widths $\Delta A$ and
$\Delta n$.

\section{Conclusions}
\label{se:conclusions}

To include the CMB lensing reconstruction power spectrum in a joint likelihood analysis with the power spectrum of the temperature anisotropies requires knowledge of the cross-covariance of the two spectra. We computed this cross-covariance between the CMB 4-point and 2-point functions perturbatively, identifying
two physical contributions. The disconnected part of the 
$6$-point function of the lensed temperature leads to a noise contribution which can
be interpreted as the response of the statistical noise in the lens reconstruction to fluctuations in the underlying CMB temperature field.
The connected
$\mathcal{O}(\phi^4)$ piece of the $6$-point
function gives rise to a second contribution attributable to the
cosmic variance of the lenses, which causes the power spectrum of the lens reconstruction and the smoothing effect in the anisotropy power spectrum to covary.
The temperature-lensing power covariance can therefore be written as
\begin{equation}
  \label{eq:conclusion_temperatureLensingPowerCov}
  \mathrm{cov}(\hat C^{\hat\phi\hat\phi}_L,\hat C^{\tilde T\tilde
    T}_{L',\expt}) \;= \;
  \frac{\partial (2\hat N^{(0)}_L)}{\partial\hat C^{\tilde
    T\tilde T}_{L',\expt}}\;
 \frac{2}{2L'+1}(C^{\tilde T\tilde T}_{L',\expt})^2
\left[1+2\frac{C^{\phi\phi}_L}{A_L}\right]
\;+\;
\frac{2}{2L+1}(C^{\phi\phi}_L)^2\;
\frac{\partial C^{\tilde T\tilde T}_{L'}}{\partial
  C^{\phi\phi}_{L}},
\end{equation}
where perturbative expressions for the derivatives are given in Eqs.~\eqref{eq:CTT_Cphiphi_deriv} and~\eqref{eq:Deriv2EmpN0CTT}.  Both
contributions were confirmed with simulations. The second term in the
square brackets represents the leading correction from the connected 4-point
function; see
Eq.~\eqref{eq:4ptConnBprimaryContriToTempLensingPowerCov}.  The
$C^{T\phi}$ correlation gives a diagonal contribution to the temperature-lensing power correlation, which is less than $5\%$ for Planck-like specifications and falls rapidly with
$L$. This generally has a negligible impact on parameter constraints derived jointly from the CMB 2- and 4-point functions.

We showed that correcting for the Gaussian $N^{(0)}$ bias in the reconstruction power with the data-dependent $\hat{N}^{(0)}$, advocated by~\cite{hanson1008} to remove auto-covariances of the lensing reconstruction power spectrum,
also removes the noise
contribution to the temperature-lensing power correlation and we provided
an intuitive interpretation of this result.

For Planck-like specifications, estimates of the lensing amplitude $A$ based on
the lensing reconstruction or the peak smearing of the lensed
temperature power spectrum can be correlated at around the $10\%$ level
due to the power correlations. If the correlations are ignored, this gives a mis-estimate of the error on a joint amplitude estimate of only $5\%$, which should be negligible.
The data-dependent $\hat N^{(0)}$ bias correction reduces the amplitude correlation
further to $7\%$ and the error of the error to $3.5\%$.
Intuitively, we can understand the smallness of the correlation (found
perturbatively and with simulations) by noting that: (i)
covariance of the amplitude estimates due to cosmic variance of the lenses is limited by the small number of modes of $C^{\phi\phi}$ that influence the acoustic region of the temperature power spectrum, and is diluted significantly by CMB cosmic variance (and noise); and (ii) roughly disjoint scales in the CMB contribute to the amplitude determination from peak smearing and to the lens reconstruction limiting the
correlation due to CMB cosmic variance. (See Appendix \ref{se:plausibilityOfSmallCorrel} for further details of these arguments.)
For a joint analysis of the power spectrum of a temperature-based CMB lensing reconstruction and the power spectrum of the temperature anisotropies themselves,
the likelihoods for these two observables can therefore be simply combined for a Planck-like experiment (as was the case for the 2013 Planck analysis~\cite{2013arXiv1303.5077P}).

Non-Gaussianity of the lensing reconstruction complicates the construction of a likelihood. We showed that the usual likelihood for isotropic Gaussian fields does not
perform well for lens reconstruction in simple parameter tests, significantly underestimating the scatter seen in the best-fitting parameters across simulations.
We obtained better results with simple likelihoods that are Gaussian in the measured spectra (with fiducial covariance matrix) provided that power spectrum covariances were properly modeled or data-dependent $N^{(0)}$ subtraction included. In two-parameter tests based on the amplitude and tilt of a fiducial lensing power spectrum, the widths of these Gaussian likelihoods reproduce the scatter in parameters across simulations at the $10\%$ level.

With polarization-based reconstructions becoming feasible with current observations, 
it will be  important to extend the analysis presented here to polarization (see \cite{ZahnEtAl2013} for work in this direction). While we expect that many of our results can be simply applied to reconstructions based on the temperature \emph{and} polarization, the correlations are likely to be much more significant and particularly so for the most powerful $EB$-based reconstructions.
We leave this to future work.

\section*{Acknowledgements}

\noindent  
We would like to thank Uros Seljak for illuminating
discussions, particularly related to Appendix
\ref{se:plausibilityOfSmallCorrelLensCV}.
We also thank 
Helge Gruetjen, Simon Su and Oliver Zahn for helpful
discussions.
The numerical calculations for this paper were performed
on the COSMOS supercomputer, part of the DiRAC HPC Facility
jointly funded by STFC and the Large Facilities Capital Fund of
BIS. We are grateful to Andrey Kaliazin for computational support.  We
acknowledge use of LensPix \cite{lenspix} and HEALPix \cite{healpix}.
MMS was supported by STFC, DAMTP Cambridge and St John's College
Cambridge, and he thanks the University of Zurich, Institut d'astrophysique de Paris and
Berkeley Center for Cosmological Physics for hospitality and the
opportunity to present this work.

\appendix

\section{Why are the lensing amplitude cross-correlations so small?}
\label{se:plausibilityOfSmallCorrel}
The calculations in this paper give a rigorous derivation of the
cross-correlation of the power spectra of the lensing reconstruction and the
temperature anisotropies. In this appendix we present simple physical arguments for why the correlation of the lensing amplitudes estimated from the reconstruction power and the anisotropy power are so small.
We present these arguments
first for the correlation due to CMB cosmic variance and then for the
correlation due to cosmic variance of the lenses.

\subsection{Cosmic variance of the CMB}
\label{se:plausibilityOfSmallCorrelCMBCV}

\begin{figure}[t]
\centering   
\subfloat[][$l^\phi_\mathrm{max}=10$]{
 \includegraphics[width=0.48\textwidth]{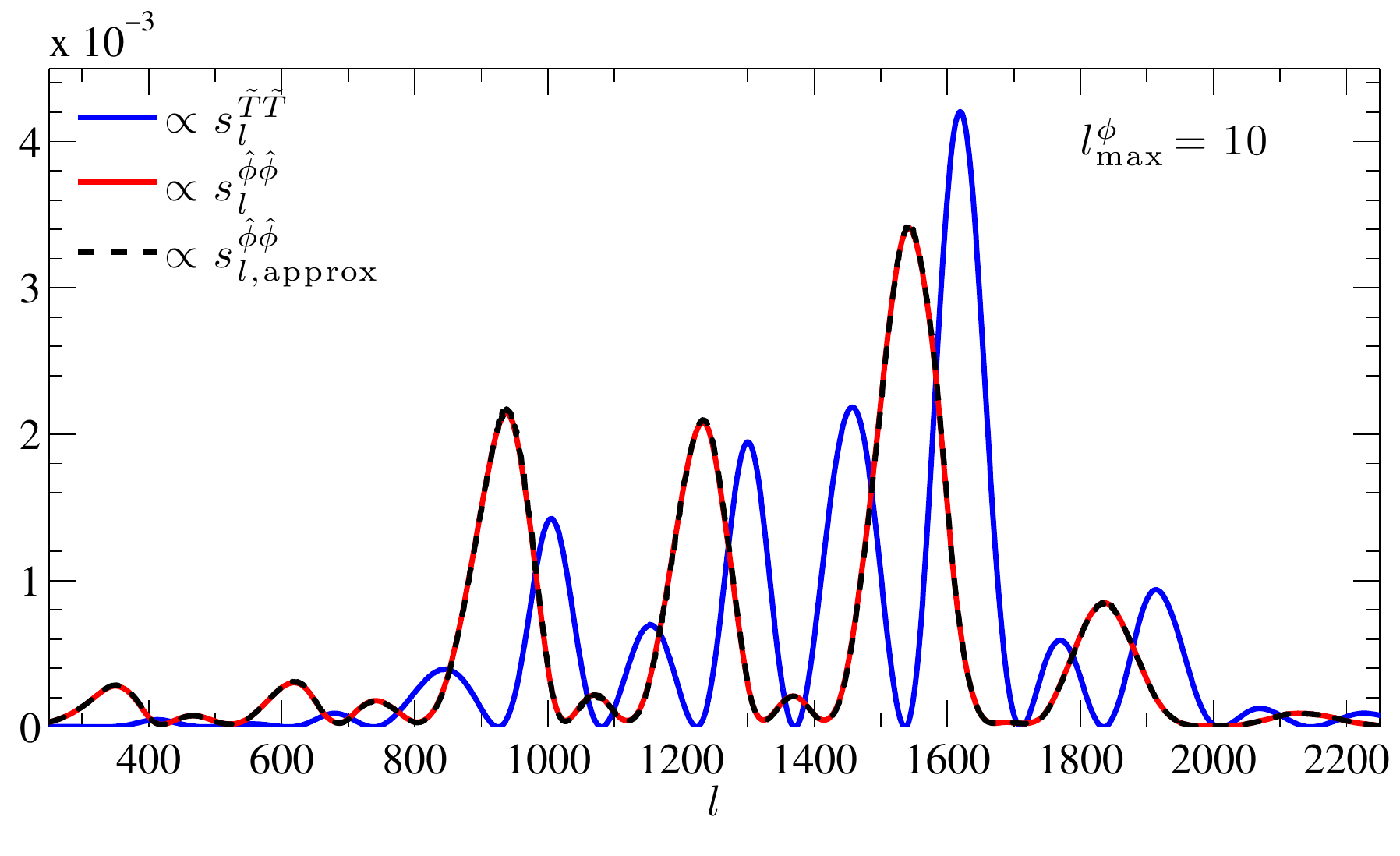}
\label{fig:intuitiveSmallnessNoiseContriLmax10}}
\hspace{0.01\textwidth}
\subfloat[][$l^\phi_\mathrm{max}=500$]{
  \includegraphics[width=0.48\textwidth]{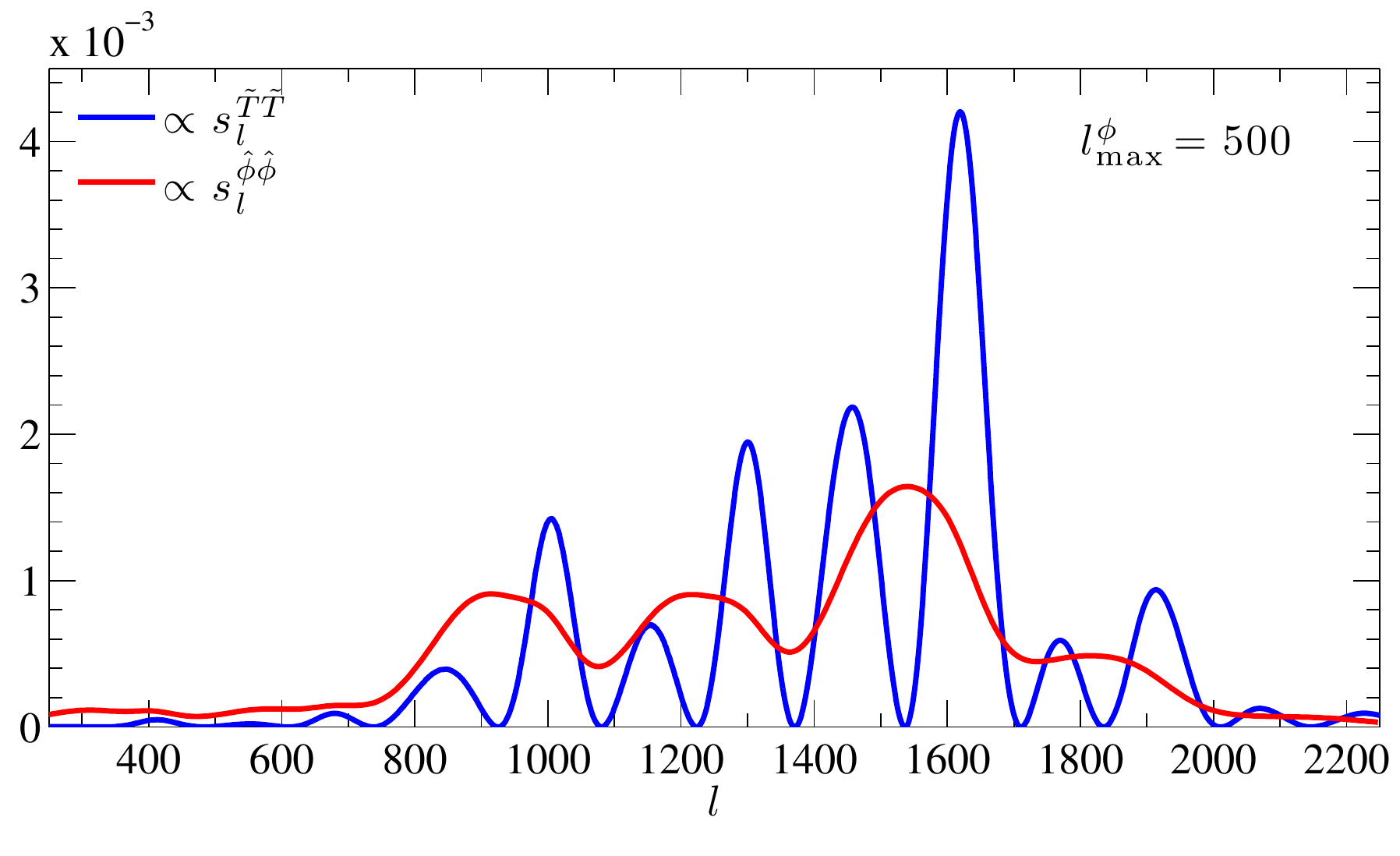}
\label{fig:intuitiveSmallnessNoiseContriLmax500}}
  \caption{Contribution of different CMB multipoles $l$ to the $(S/N)^2$
    of the lensing amplitude for estimators based on the
    CMB power spectrum [blue; Eq.~\eqref{eq:SNsquaredTT}] and on the
    lensing reconstruction [red; Eq.~\eqref{eq:CMBCVImpactOnAPhi}]
    for $l^{\phi}_\mathrm{max}=10$ (left) and $l^\phi_\mathrm{max}=500$
    (right).  The approximation
    in Eq.~\eqref{eq:CMBCVImpactOnAPhi_LargeScaleLenses} for large-scale
    lenses is shown in black dashed in the left plot. 
   Since we are not interested in the total $S/N$ but only in its distribution over different CMB scales, all
curves are normalised such their integral over $l$ is unity.  }
\label{fig:intuitiveSmallnessNoiseContriAllPlots}
\end{figure}

Due to the smoothing effect of lensing, most of the constraint on the
lensing amplitude $\hat A'$ estimated from the CMB power spectrum
comes from the CMB on scales of the acoustic peaks and troughs.
This can be seen directly from the contribution $s_l^{\tilde T\tilde
  T}$ to the total signal-to-noise squared [$(S/N)^2$] associated with the CMB
cosmic variance at multipole $l$ (see the blue curve in
\fig{intuitiveSmallnessNoiseContriAllPlots}):
\begin{equation}
  \label{eq:SNsquaredTT}
  \left(\frac{S}{N}\right)_{A'}^2 = \frac{1}{\sigma_{A'}^2} = 
\sum_{l=l_\mathrm{min}^{T}}^{l_\mathrm{max}^{T}}
\underbrace{\frac{(C_l^{\tilde T\tilde T}-C_l^{TT})^2}{
\text{var}_G(C^{\tilde T\tilde T}_{l,\expt})}}_{s_l^{\tilde T\tilde T}}.
\end{equation}
  In contrast,
  in the limit of very large-scale lenses, and as argued in more detail
  below, the reconstruction combines local convergence and shear measurements,
  for which scales in the CMB where the power spectrum changes rapidly are
  most informative. For large-scale lenses, the $(S/N)^2$ on the reconstruction-based amplitude estimate
  $\hat A$ is thus expected to be dominated by CMB modes \emph{between} acoustic peaks and troughs.
Therefore, the lensing amplitude estimates $\hat A$ and $\hat A'$ are
determined by rather disjoint CMB modes with independent CMB cosmic-variance
fluctuations.
We therefore expect the amplitude correlation due to CMB cosmic
variance to be suppressed (in case this correlation is not mitigated
by the empirical $\hat N^{(0)}$ subtraction anyway).

To make this point more quantitative, note that the reconstruction
power $\hat C^{\hat\phi\hat\phi}$ is affected by CMB cosmic variance
through the disconnected CMB $4$-point contribution $N^{(0)}$. Keeping
the estimator normalisation $A_L$ and weights $\tilde g$ fixed in
Eq.~\eqref{eq:N0}, the contribution from the CMB at multipole $l$ to
the $(S/N)^2$ of the reconstruction-based amplitude
estimate $\hat A$ is monitored by
\begin{equation}
  \label{eq:CMBCVImpactOnAPhi}
s_l^{\hat\phi\hat\phi}=  
\left| \frac{\delta \sigma_A^{-2}}{\delta \ln C^{\tilde T\tilde T}_{l,\expt}} \right|_{\text{fix}\,A_L,\tilde{g}} 
=  \sum_{L=l_\mathrm{min}^{\phi}}^{l_\mathrm{max}^{\phi}}  
\Bigg|\frac{-2(C_L^{\phi\phi})^2}{\frac{2}{2L+1}\big(
N_L^{(0)} + C_L^{\phi\phi} + N_L^{(1)}\big)^3}
\,
\frac{4A_L^2}{2L+1}\sum_{l_1}\tilde g^2_{l_1l}(L)C^{\tilde T\tilde T}_{l_1,\expt}
\Bigg|
C^{\tilde T\tilde T}_{l,\expt},
\end{equation}
where we used $\sigma_A$ from Eq.~\eqref{eq:sigma_Aphiphi} and kept only
the dominant diagonal part of the reconstruction power
auto-covariance.  As shown in
Fig.~\ref{fig:intuitiveSmallnessNoiseContriAllPlots}, $s_l^{\hat\phi\hat\phi}$
(red) is out of phase compared to $s_l^{\tilde T\tilde T}$
(blue). To understand this structure, we separate the sums over $L$
and $l_1$ in Eq.~\eqref{eq:CMBCVImpactOnAPhi}
by restricting ourselves to very large-scale lenses,
$l^\phi_\mathrm{max}\ll l$, and using the large scale
approximation for $N^{(0)}$ derived in \cite{hanson1008} [see their Eq.~(19)],
to find
\begin{equation}
  \label{eq:CMBCVImpactOnAPhi_LargeScaleLenses}
  s_l^{\hat\phi\hat\phi} \approx   s^{\hat\phi\hat\phi}_{l,\text{approx}}=
  f(l^\phi_\mathrm{min},l^\phi_\mathrm{max})
\frac{1}{\text{var}_G(C^{\tilde T\tilde T}_{l,\expt})}\left[
\left( C^{\tilde T\tilde T}_{l}\frac{\drm \ln(l^2C^{\tilde
      T\tilde T}_l)}{\drm\ln l}\right)^2 + 
\frac{1}{2}\left( C^{\tilde T\tilde T}_{l}\frac{\drm \ln C^{\tilde
      T\tilde T}_l}{\drm\ln l}\right)^2
\right],
\quad [l^{\phi}_\mathrm{max}\lesssim \mathcal{O}(10)].
\end{equation}
Here, the prefactor $f$ depends on the minimum and
maximum reconstruction multipole but not on the CMB multipole $l$.
The terms in square brackets have the form of the quadrature sum of the
information in convergence and shear. Convergence changes locally the angular scale of the CMB anisotropies and so would contribute nothing to the $(S/N)^2$ for 
a scale-invariant spectrum, $l^2 C_l = \text{const.}$, while shear contributes nothing for a white-noise spectrum, $C_l = \text{const}$.\footnote{The relation between large-scale lenses and the
  induced local convergence and shear is discussed in detail by
  \cite{BucherCarvalhoEtAl1004}, who also find agreement between the
  $(S/N)^2$ of a combined convergence and shear estimate
  with the large-scale limit
of the $(S/N)^2$ of the trispectrum reconstruction. This correspondence has
  also been used to approximate the squeezed limit of the ISW-lensing
  bispectrum \cite{lewis1101}.} 
Thus, for large-scale lenses, the $(S/N)^2$ for the
reconstruction-based amplitude gets most contributions from CMB scales where
the gradient of the CMB power spectrum is maximal, i.e.~between
acoustic peaks and troughs (see red and black curves in
\fig{intuitiveSmallnessNoiseContriLmax10}).  

In reality, temperature multipoles that are not precisely at peaks or
troughs and not precisely in between them will affect both amplitude
estimates, which implies a small amplitude correlation. Intermediate-
and small-scale lenses can mix CMB modes over multipole ranges
comparable to the acoustic peak separation so that they are affected by wider
ranges of CMB multipoles than argued above (see red curve in
\fig{intuitiveSmallnessNoiseContriLmax500}), which implies
a somewhat larger amplitude correlation. However, since the CMB scales
that are most important for the reconstruction still have negligible
impact on the amplitude estimated from the temperature power, we
expect the correlation of the amplitudes to stay rather small.

\subsection{Cosmic variance of the lenses}
\label{se:plausibilityOfSmallCorrelLensCV}
We now consider the contribution of cosmic variance of the lenses to the covariance of the lensing amplitude estimates given in Eq.~\eqref{eq:cov_AAprime}. It is instructive to consider a toy-model where the reconstruction ``noise''  power is proportional to
$C^{\phi\phi}$, i.e.~$N_l^{(0)} + N_l^{(1)} = \lambda C_l^{\phi\phi}$. Taking the limit
$\lambda \rightarrow 0$ is equivalent to being able to observe $\phi$ directly with no measurement error, while $\lambda \rightarrow \infty$ corresponds to there being no information in the reconstruction. With $N_l^{(0)} + N_l^{(1)} = \lambda C_l^{\phi\phi}$,
the weighting of the reconstruction power spectrum in $\hat{A}$ is the same as for an ideal reconstruction (i.e.~one with no $N^{(0)}$ and $N^{(1)}$ noise). Provided we then determine $\hat{A}$ from all those $\phi$ modes that influence the temperature power spectrum,
the contribution to the amplitude \emph{covariance} from cosmic variance of the lenses simplifies significantly to give\footnote{Note that $\sum_{l=1}^{l^\phi_\text{max}} (\partial C_{l'}^{\tilde T\tilde T}/\partial C^{\phi\phi}_{l}) C^{\phi\phi}_l=C^{\tilde T\tilde T}_{l'}-C^{TT}_{l'}$ for sufficiently large $l^\phi_\text{max}$.}
\begin{equation}
\text{cov}(\hat{A},\hat{A}') = \sigma_{A,\text{ideal}}^2 .
\label{eq:appcv1}
\end{equation}
Here, $\sigma_{A,\text{ideal}}^2 = [\sum_l (l+1/2)]^{-1} \approx 2 / (l_{\text{max}}^{\phi})^2$ is the variance of the reconstruction-based amplitude
in the ideal limit $\lambda \rightarrow 0$ using all modes up to $l_{\text{max}}^{\phi}$. Since only a few (large-scale) lensing modes affect $\hat A'$, including more lensing modes in the reconstruction dilutes the covariation of $\hat A$ and $\hat A'$ over different realisations of the lenses, because there are increasingly more lensing modes in $\hat A$ whose fluctuations do not enter $\hat A'$.
The amplitude covariance falls inversely as the number of modes in the reconstruction
since the weight in $\hat{A}$ given to those (few) modes of $C^{\phi\phi}$ that influence the temperature power spectrum falls as the total number of modes.
Note that the covariance is independent of the weighting of the measured temperature power spectrum in $\hat{A}'$, provided $\hat{A}'$ is appropriately normalised, and it is also independent of additional contributions to the reconstruction noise (e.g.~from CMB cosmic variance, for fixed $l^\phi_\text{max}$). 
The \emph{variance} of $\hat{A}$ does depend on the reconstruction noise level, with 
\begin{eqnarray}
\sigma_A^{-2} &\approx&  \sum_l \frac{2l+1}{2} \left(\frac{C_l^{\phi\phi}}{C_l^{\phi\phi}
+ N_l^{(0)} + N_l^{(1)}}\right)^2  \nonumber \\
&=& \frac{1}{(1+\lambda)^2} \sum_l \frac{2l+1}{2}
= \frac{1}{(1+\lambda)^2} \sigma_{A,\text{ideal}}^{-2} .
\end{eqnarray}

The result $\text{cov}(\hat{A},\hat{A}') = \sigma_{A,\text{ideal}}^2$ for ideal weighting is necessary to ensure that the lensed CMB spectrum adds no further information on the lensing amplitude when combined with an ideal measurement of $\phi$ itself on all scales that are relevant for peak smearing of the temperature power spectrum.
To see this, note that we can combine the amplitude estimates $\hat{A}$ and $\hat{A}'$ optimally into a single estimate $\hat{A}_{\text{opt.}}$, properly taking account of their correlation.
If we do this, the inverse variance of the optimal estimate is given by contracting the
inverse covariance matrix of the estimates:
\begin{equation}
\sigma_{A,\text{opt.}}^{-2} = \text{cov}^{-1}(\hat{A},\hat{A}) +  2\, \text{cov}^{-1}(\hat{A},\hat{A}') +  \text{cov}^{-1}(\hat{A}',\hat{A}') .
\end{equation}
This evaluates to
\begin{equation}
\sigma_{A,\text{opt.}}^{-2} = \frac{\sigma_A^2 + \sigma^2_{A'} - 2 \sigma^2_{A,\text{ideal}}}{\sigma_A^2 \sigma^2_{A'} - \sigma_{A,\text{ideal}}^4} ,
\end{equation}
on using Eq.~\eqref{eq:appcv1} for the covariance.
In the ideal case, taking the limit $\lambda \rightarrow 0$, we have
$\sigma_A = \sigma_{A,\text{ideal}}$ so that $\sigma_{A,\text{opt}.} = \sigma_{A,\text{ideal}}$.
This as it must be -- the observation of the peak smearing in the power spectrum adds no new information to that obtained from the ideal measurement of $\phi$.
In the opposite limit, $\lambda \rightarrow \infty$, we have
$\sigma_A \rightarrow \infty$ and $\sigma_{A,\text{opt.}} = \sigma_{A'}$ and all information is coming from the temperature power spectrum.

The \emph{correlation} induced by matter cosmic variance,
\begin{equation}
\label{eq:intuitiveMCVamplitudeCorrel}
\text{correl}(\hat{A},\hat{A}') = \frac{\sigma^2_{A,\text{ideal}}}{\sigma_A \sigma_{A'}}
= \frac{\sigma_A}{(1+\lambda)^2 \sigma_{A'}},
\end{equation}
reaches its maximal value of $\sigma_{A,\text{ideal}}/\sigma_{A'}$ if the variance of the reconstruction power spectrum
is only due to matter cosmic variance, $\lambda=(N_L^{(0)}+N_L^{(1)})/C_L^{\phi\phi}\rightarrow 0$; and it falls monotonically 
with increasing $\lambda$, tending to zero as $\lambda\rightarrow\infty$ (when CMB cosmic variance dominates the reconstruction uncertainty). This is expected since we assume matter and CMB fluctuations to be independent.
 More generally,
$\sigma_{A}$ is determined by the number of high $S/N$ modes in the
reconstruction, but $\sigma_{A'}$ depends not only on the number of CMB modes but also the fractional size of the power spectrum corrections from lensing relative to the
total spectrum, $C_{l,\text{expt}}^{\tilde{T}\tilde{T}}$. 
The result is that both factors $\sigma_{A,\text{ideal}}/\sigma_A$ and $\sigma_{A,\text{ideal}}/\sigma_{A'}$ in Eq.~\eqref{eq:intuitiveMCVamplitudeCorrel} are less than $1$, diluting the amplitude correlation.

For our Planck-like parameters, the
power spectrum corrections from lensing are only ever a few percent of the total spectrum and so cosmic variance of the CMB limits $\sigma_{A'} \approx 0.04$.
Statistical noise in the lens reconstruction limits
$\sigma_A \approx 0.025$. It is clear from Fig.~\ref{fig:fig1-recphi} that a constant $\lambda$ is not a good approximation for lens reconstruction, but we can crudely limit $\lambda \gtrsim 2$ in which case we expect the amplitude correlation to be
less than $(0.025/0.04)/3^2 \sim 0.06$ which is close to the value plotted in Fig.~\ref{fig:correl_AAprime_plot}.

To summarise, the correlation of the lensing amplitudes due to the cosmic variance of the lenses is generally small since there are a limited number of modes of $C^{\phi\phi}$ that influence the acoustic part of the temperature power spectrum (so the covariance for an ideal reconstruction scales inversely as the number of reconstruction modes), and the small covariance [less than $O(10^{-4})$] is diluted
by cosmic variance of the CMB (and noise), which dominates the error on $\hat{A}'$ and contributes significantly to the error on $\hat{A}$. We emphasise that these conclusions assume that the temperature power spectrum at multipoles $l \gtrsim 3000$, where the lensing-induced power from small-scale lenses dominates the unlensed power, does not influence the amplitude estimate (i.e.\ the spectrum is limited by noise or foregrounds there).

\section{\texorpdfstring{Optimal trispectrum estimation and $N^{(0)}$
    subtraction}{Optimal trispectrum estimation and N(0) subtraction}}
\label{app:edgeworth}

In this appendix we show that the lensing power spectrum estimator advocated in Sec.~\ref{se:fullSkyLensing}, which includes a data-dependent $N^{(0)}$ subtraction to remove the (disconnected) Gaussian bias, follows naturally from optimal trispectrum estimation (e.g.~\cite{2010PhRvD..82b3520R}). Moreover, we show that in the Gaussian limit, this estimator is uncorrelated with any quadratic estimate of the temperature power spectrum, generalising the result established in Sec.~\ref{se:empirical_N0_subtraction_cross_cov}.

We start by considering the Gram-Charlier expansion for the probability density function (PDF) of a weakly non-Gaussian, zero-mean temperature field $T$\footnote{In this appendix, to avoid undue clutter in our expressions, we shall suppress the tildes that are used to denote the lensed temperature in the body of this paper.}.
Writing the field in an arbitrary basis (e.g.\ a pixelised map or multipole coefficients) as $T_i$, with covariance $\langle T_i T_j \rangle = C_{ij}$, the expansion truncated at the trispectrum level can be written in the form~\cite{2010PhRvD..82b3520R}
\begin{eqnarray}
P(T)&=&\frac{e^{-T_i C_{ij}^{-1} T_j/2}}{\sqrt{\mathrm{det}(2\pi C)}} \biggl(1 +
\frac{1}{24} \langle T_i T_j T_k T_l \rangle_c \bigl(
\bar{T}_i \bar{T}_j \bar{T}_k \bar{T}_l - [ C_{ij}^{-1} \bar{T}_k \bar{T}_l
+C^{-1}_{kl} \bar{T}_i \bar{T}_j - C^{-1}_{ij} C^{-1}_{kl} ] \nonumber \\
&&\mbox{} \hspace{0.2\textwidth} - [ C_{ik}^{-1} \bar{T}_j \bar{T}_l
+C^{-1}_{jl} \bar{T}_i \bar{T}_k - C^{-1}_{ik} C^{-1}_{jl} ]
- [ C_{il}^{-1} \bar{T}_j \bar{T}_k
+C^{-1}_{jk} \bar{T}_i \bar{T}_l - C^{-1}_{jk} C^{-1}_{il} ]
\bigr)\biggr).
\label{eq:app1}
\end{eqnarray}
Here, $\bar{T}_i \equiv C_{ij}^{-1} T_j$ is the field filtered by the inverse covariance and summation over repeated indices is implicit. Note that $\langle \bar{T}_i \bar{T}_j \rangle = C_{ij}^{-1}$. We have ignored bispectrum contributions (due to the lensing-ISW correlation) which is a good approximation on intermediate and small scales. Using Wick's theorem, it is straightforward to show from Eq.~(\ref{eq:app1}) that the mean and bispectrum of $T$ vanish, whilst the covariance is $C_{ij}$ and the trispectrum is $\langle T_i T_j T_k T_l \rangle_c $, as intended. That the covariance is $C_{ij}$ implies that the data combination associated with the trispectrum term in Eq.~(\ref{eq:app1}) is orthogonal to any quadratic function of $T$ with respect to the Gaussian weight function. In other words, if $T$ were Gaussian, this data combination is uncorrelated with any quadratic function of $T$.

We can form an estimator for $C_L^{\phi\phi}$ by maximising the PDF with respect to
$C_L^{\phi\phi}$. In practice, we can approximate this by one step of a Newton-Raphson
scheme starting at $C_L^{\phi\phi}=0$. The estimator is then proportional to $\partial \ln P(T) / \partial C_L^{\phi\phi}$ evaluated at $C_L^{\phi\phi}=0$. The lensing power spectrum enters the PDF explicitly through the connected 4-point function, and also implicitly through the lensed $\tilde{C}_l^{TT}$ in the covariance $C_{ij}$. The contribution to the $C_L^{\phi\phi}$ estimate from the 4-point function is of the form
\begin{eqnarray}
\hat{C}_L^{\phi\phi} &\sim& \frac{\partial}{\partial C_L^{\phi\phi}}
\langle T_i T_j T_k T_l \rangle_c \bigl(
\bar{T}_i \bar{T}_j \bar{T}_k \bar{T}_l - [ C_{ij}^{-1} \bar{T}_k \bar{T}_l
+C^{-1}_{kl} \bar{T}_i \bar{T}_j - C^{-1}_{ij} C^{-1}_{kl} ] \nonumber \\
&&\mbox{} \hspace{3cm}
- [ C_{ik}^{-1} \bar{T}_j \bar{T}_l
+C^{-1}_{jl} \bar{T}_i \bar{T}_k - C^{-1}_{ik} C^{-1}_{jl} ]
- [ C_{il}^{-1} \bar{T}_j \bar{T}_k
+C^{-1}_{jk} \bar{T}_i \bar{T}_l - C^{-1}_{jk} C^{-1}_{il} ]
\bigr).
\label{eq:app2}
\end{eqnarray}
The terms involving the data in this equation are symmetric under permutations and so we need retain only primary couplings of the trispectrum, i.e., the first term in Eq.~(\ref{eq:4pt}). The primary coupling has the form
\begin{equation}
\frac{\partial}{\partial C_L^{\phi\phi}} \langle T_i T_j T_k T_l \rangle_c^{\mathrm{primary}} 
\propto \sum_M (-1)^M X_{LM}^{ij} X_{L-M}^{kl} ,
\label{eq:app3}
\end{equation}
where, in multipole space, 
\begin{equation}
X_{LM}^{l_1 m_1 \, l_2 m_2} = \tilde{f}_{l_1 L l_2} \left( \begin{matrix} 
l_1 & l_2 & L \\ 
m_1 & m_2 & M
\end{matrix} \right) .
\label{eq:app4}
\end{equation}

The power spectrum estimator in Eq.~(\ref{eq:app2}) generalises that introduced in
Sec.~\ref{se:fullSkyLensing}. It reduces to the simple form given there for an isotropic survey. To see this, note that for an isotropic survey, $\bar{T}_{l_1 m_1} = T_{l_1 m_1}^* / C_{l_1,\,\expt}^{\tilde{T}\tilde{T}}$, and $X_{LM}^{l_1 m_1 \, l_2 m_2} \bar{T}_{l_1 m_1} \bar{T}_{l_2 m_2}$ is proportional to the estimator $\hat{\phi}_{LM}$ for the lensing potential in Eq.~(\ref{eq:phi_esti}). The first set of data-dependent terms in Eq.~(\ref{eq:app2}) can be rewritten as
\begin{equation}
\bar{T}_i \bar{T}_j \bar{T}_k \bar{T}_l - [ C_{ij}^{-1} \bar{T}_k \bar{T}_l
+C^{-1}_{kl} \bar{T}_i \bar{T}_j - C^{-1}_{ij} C^{-1}_{kl} ] = 
(\bar{T}_i \bar{T}_j - \langle \bar{T}_i \bar{T}_j\rangle)
(\bar{T}_k \bar{T}_l - \langle \bar{T}_k \bar{T}_l\rangle) .
\end{equation}
Each factor on the right combines with an $X_{LM}$ to give an unnormalised $\hat{\phi}_{LM} - \langle \hat{\phi}_{LM} \rangle$. 
The ``mean-field'' term $\langle \hat{\phi}_{LM} \rangle$ vanishes (except for $L=0$) for a full-sky survey with homogeneous noise. More generally, in the presence of beam asymmetry, realistic noise, and masking, it can require careful subtraction on large scales
(see e.g.,~\cite{2009MNRAS.400.2169H,2012arXiv1209.0091N,2013arXiv1303.5077P}). The sum of the remaining terms in Eq.~(\ref{eq:app2}) has an expectation value that, for an isotropic survey, is simply $-N_L^{(0)}$ [see Eq.~\ref{eq:N0}]. However, rather than removing the Gaussian (disconnected) bias from
$\hat{C}_L^{\phi\phi}$ with the data-independent $N_L^{(0)}$, the optimal trispectrum estimator debiases with additional data-dependent terms that are equivalent to the $\hat{N}_L^{(0)}$ of Eq.~(\ref{eq:empirical_N0}). For example,
\begin{align}
\sum_M (-1)^M X^{ij}_{LM} X^{kl}_{L-M} C^{-1}_{ik} \bar{T}_j \bar{T}_l &= 4 \sum_M (-1)^{m_1 + M} 
\tilde{g}_{l_1 l_2}(L) \tilde{g}_{l_1 l_4}(L)
\left( \begin{matrix} 
l_1 & l_2 & L \\ 
m_1 & m_2 & M
\end{matrix} \right) 
\left( \begin{matrix} 
l_1 & l_4 & L \\ 
-m_1 & m_4 & -M
\end{matrix} \right) C_{l_1,\,\expt}^{\tilde{T}\tilde{T}} T^*_{l_2m_2} T^*_{l_4 m_4} \nonumber \\
&= 4 \sum_{l_1 l_2} \tilde{g}^2_{l_1 l_2}(L) C_{l_1,\,\expt}^{\tilde{T}\tilde{T}} \left(\frac{1}{2l_2+1}
\sum_{m_2} T_{l_2 m_2} T^*_{l_2 m_2} \right)\nonumber \\
&= 4 \sum_{l_1 l_2} \tilde{g}^2_{l_1 l_2}(L) C_{l_1,\,\expt}^{\tilde{T}\tilde{T}} \hat{C}_{l_2,\,\expt}^{\tilde{T}\tilde{T}} ,
\end{align}
where summation over $(l_1, m_1)$, $(l_2,m_2)$ and $(l_4, m_4)$ is implicit in the first line.

We showed in Sec.~\ref{se:empirical_N0_subtraction_cross_cov} that, for an isotropic survey, the data-dependent $N^{(0)}$ subtraction removes the disconnected covariance between $\hat{C}_L^{\hat{\phi}\hat{\phi}}$ and the empirical power spectrum of the measured temperature anisotropies, $\hat{C}_{l,\,\expt}^{\tilde{T}\tilde{T}}$. 
For the power spectrum estimator in Eq.~(\ref{eq:app2}), this result generalises to anisotropic surveys, and also to arbitrary quadratic power spectrum estimates of the
temperature anisotropies. The latter are often estimated with combinations of cross-spectra, each of which are quadratic forms constructed from maps $T^{(1)}$ and
$T^{(2)}$ for which the noise is independent. Each cross-spectrum therefore involves terms like $T^{(1)}_p T^{(2)}_q$, where $p$ and $q$ are pixel (or multipole) indices, and
the fields $T^{(1)}$ and $T^{(2)}$ are correlated with the field that enters the lensing power spectrum estimate at least through the common temperature anisotropies.
By using Wick's theorem, it is straightforward to show that the complete set of data-dependent terms in Eq.~(\ref{eq:app2}) has vanishing covariance with $T^{(1)}_p T^{(2)}_q$ in the Gaussian limit.

\section{\texorpdfstring{$\hat N^{(0)}$ subtraction as efficient mitigation
    of chance alignments}{N(0) subtraction as efficient mitigation of chance alignments}}
\label{se:N0BiasAppendix}

In this appendix we provide further details to understand why the empirical
$\hat N^{(0)}$ bias subtraction cancels not only the non-diagonal
reconstruction power auto-covariance, but also the noise contribution
to the temperature-lensing power cross-covariance.
After identifying chance-aligned terms in the empirical
reconstruction power spectrum in Eq.~\eqref{eq:CphiphihatNoise}
below (see also \cite{sherwinDasPowerWOBias1011}), we will show that avoiding these terms naturally cancels
the noise contribution to the auto- and cross-covariance.
We then show that empirical $\hat N^{(0)}$ subtraction achieves the same
but in a faster way (due to the specific multipole couplings that are relevant
for the covariances).

Given a CMB realisation $\tilde T_{lm}$, we split the 
\emph{empirical} reconstruction power spectrum into two contributions,
\begin{equation}
  \label{eq:Cphiphihat}
\hat C^{\hat\phi\hat\phi}_L= 
\frac{A_L^2}{2L+1}\sum_{\underline {l}_1\cdots
\underline {l}_4M}
(-1)^M
\left( \begin{matrix} 
l_1 & l_2 & L \\ 
m_1 & m_2 & -M
\end{matrix} \right)
\left( \begin{matrix} 
l_3 & l_4 & L \\ 
m_3 & m_4 & M
\end{matrix} \right)
\tilde g_{l_1l_2}(L)
\tilde g_{l_3l_4}(L)
\tilde T_{\underline{l}_1}\tilde T_{\underline{l}_2}\tilde
T_{\underline{l}_3}\tilde T_{\underline{l}_4}
=
\hat  C^{\hat\phi\hat\phi}_{L,\mathrm{noise}} + 
\hat   C^{\hat\phi\hat\phi}_{L,\mathrm{rest}} ,
\end{equation}
 where the noise term contains the part of the sum over
 $l_i$ where CMB multipoles are chance-aligned,\footnote{It may
   be possible to modify the  specific splitting of
   Eq.~\eqref{eq:Cphiphihat} chosen here without significantly
   changing the conclusions below (e.g.~by coupling $m_3$ and $m_4$ to
   $m_1$ and $m_2$). Our choice allows for a relatively simple
   analytical assessment of the mean and covariance properties, and is sufficiently accurate for the discussion here.}
\begin{equation}
  \label{eq:CphiphihatNoise}
\hat C^{\hat\phi\hat\phi}_{L,\mathrm{noise}} = 
\frac{2A_L^2}{2L+1}\sum_{\underline{l}_1\underline{l}_2m_3m_4M}
(-1)^M
\left( \begin{matrix} 
l_1 & l_2 & L \\ 
m_1 & m_2 & -M
\end{matrix} \right)
\left( \begin{matrix} 
l_1 & l_2 & L \\ 
m_3 & m_4 & M
\end{matrix} \right)
\tilde g^2_{l_1l_2}(L)
s_{l_1l_2}
\tilde T_{l_1m_1}\tilde T_{l_2m_2}
\tilde T_{l_1m_3}\tilde T_{l_2m_4},
\end{equation}
taking only half of the terms for $l_1=l_2$ by defining 
 $s_{l_1l_2}= 1-\delta_{l_1l_2}/2$. The expectation values are as
 desired, at sub-percent level accuracy\footnote{The disconnected part is
$\la \hat C^{\hat\phi\hat\phi}_{L,\mathrm{noise}}\ra_\mathrm{disconn.} = N^{(0)}_L$. 
Using Eq.~\eqref{eq:lensing_trispectrum}
and simplifying the product of a $3j$- and a $6j$-symbol \cite{NISTWigner},
we have
\begin{eqnarray}
\nonumber
  \la \hat C^{\hat\phi\hat\phi}_{L,\mathrm{noise}}\ra_\mathrm{conn.} &=&
  \frac{2A_L^2}{(2L+1)^2}\sum_{l_1l_2} 
s_{l_1l_2}\tilde g^2_{l_1l_2}(L)\tilde f^2_{l_1Ll_2}C_L^{\phi\phi}
+  \frac{A_L^2}{16(2L+1)}
\sum_{m'l_1l_2L'} (-1)^{m'}C^{\phi\phi}_{L'}
\frac{s_{l_1l_2}}{(C^{\tilde T\tilde T}_{l_1,\expt} C^{\tilde T\tilde T}_{l_2,\expt})^2}\\
  \label{eq:CphiphihatNoiseExpConn}
&&
\;\;
\times 
\left[
(\tilde f^\mathrm{(u)}_{l_1Ll_2})^2
\tilde f^\mathrm{(u)}_{l_1L'l_2}
\tilde f^\mathrm{no3j}_{l_1L'l_2}
\mathcal{G}^{l_1Ll_2}_{0-m'm'}
\mathcal{G}^{l_1Ll_2}_{-m'm'0}
\mathcal{G}^{l_1L'l_2}_{m'0-m'}
 +
\tilde f^\mathrm{no3j}_{l_1Ll_2}\tilde f^\mathrm{(u)}_{l_1Ll_2}
\tilde f^\mathrm{(u)}_{l_1L'l_1}\tilde f^\mathrm{(u)}_{l_2L'l_2}
\mathcal{G}^{l_1Ll_2}_{-m'0m'}\mathcal{G}^{l_1L'l_1}_{0-m'm'}
\mathcal{G}^{l_2L'l_2}_{-m'm'0}
\right],
\quad\quad
\end{eqnarray}
where we defined the Gaunt coefficients, $\tilde f^\mathrm{(u)}$ with unsymmetric terms and
$\tilde f^\mathrm{no3j}$ with no $3j$-symbols  by
\begin{eqnarray}
  \label{eq:gauntCoeffs}
  \mathcal{G}^{l_1l_2l_3}_{m_1m_2m_3}&\equiv &
\sqrt{\frac{(2l_1+1) (2l_2+1) (2l_3+1)}{4\pi}}
 \left( \begin{matrix} 
 l_1 & l_2 & l_3 \\ 
 0&0&0
 \end{matrix} \right)
 \left( \begin{matrix} 
 l_1 & l_2 & l_3 \\ 
 m_1 & m_2 & m_3
 \end{matrix} \right),
\\
  \label{eq:fNoWigner}
  \tilde f^\mathrm{(u)}_{l_1Ll_2}&\equiv &
\left[L (L+1)-l_1(l_1+1)+l_2(l_2+1)\right]C^{\tilde T\tilde T}_{l_2}
+ (l_1\leftrightarrow l_2),
\qquad
\tilde f^\mathrm{no3j}_{l_1Ll_2}\equiv \sqrt{\frac{(2l_1+1) (2L+1)
    (2l_2+1)}{16\pi}}  \tilde f^\mathrm{(u)}_{l_1Ll_2}.
\end{eqnarray}
 We find that
\eqref{eq:CphiphihatNoiseExpConn} is smaller than
$C^{\phi\phi}_L+N^{(1)}_L$ by a factor of more than $400$ for any $L$,
 so that [with Eq.~\eqref{eq:reconstruction_power_biases}; $q_L\lesssim 0.002$]
\begin{equation}
  \label{eq:CphiphihatNoiseAvgDetails}
\la  \hat C^{\hat\phi\hat\phi}_{L,\mathrm{noise}} \ra = 
N^{(0)}_L + q_L (C^{\phi\phi}_L + N^{(1)}_L),\qquad
\la  \hat C^{\hat\phi\hat\phi}_{L,\mathrm{rest}} \ra = 
(1-q_L)(C_L^{\phi\phi} + N^{(1)}_L).
\end{equation}
},
\begin{equation}
  \label{eq:CphiphihatNoiseAvg}
\la  \hat C^{\hat\phi\hat\phi}_{L,\mathrm{noise}} \ra \approx
N^{(0)}_L,\qquad
\la  \hat C^{\hat\phi\hat\phi}_{L,\mathrm{rest}} \ra \approx
C_L^{\phi\phi} + N^{(1)}_L,
\end{equation}
To assess the covariance properties of the noise-corrected reconstruction power spectrum
$\hat C^{\hat\phi\hat\phi}-\hat C^{\hat\phi\hat\phi}_\text{noise}$, 
note that 
the $\mathcal{O}(\phi^0)$ noise contribution
\eqref{eq:noisecontri_cov_CTT_Cphiphi_rec} to the temperature-lensing
power cross-covariance is sourced by contractions 
\begin{equation}
  \label{eq:CouplingsCovCphirecCTTNoiseContri}
\contraction{}{\tilde T_{\lbar{1}}}{}{\tilde T_{\lbar{3}}}
\contraction{\tilde T_{\lbar{1}}\tilde T_{\lbar{3}}}{\tilde T_{\lbar{2}}}
{\tilde T_{\lbar{4}} \ \mid \ }{\tilde T_{L'M'}}
\contraction[2ex]{\tilde T_{\lbar{1}}\tilde T_{\lbar{3}}\tilde T_{\lbar{2}}}  
{\tilde T_{\lbar{4}}}{ \ \mid \  \tilde T_{L'M'}}{ \tilde T_{L',-M'}}
\tilde T_{\lbar{1}}\tilde T_{\lbar{3}}\tilde T_{\lbar{2}}  
\tilde T_{\lbar{4}} \ \mid \  \tilde T_{L'M'} \tilde T_{L',-M'},
\end{equation}
which imply $l_1=l_3$ and $l_2=l_4$. Since all terms with coinciding
multipoles of this form are contained in the noise term
\eqref{eq:CphiphihatNoise}, 
this covariance is  cancelled 
 if 
 $\hat C^{\hat\phi\hat\phi}-\hat C^{\hat\phi\hat\phi}_\text{noise}$
is used instead of $\hat C^{\hat\phi\hat\phi}$.
To see this explicitly note that
\begin{equation}
  \label{eq:covCphirecNoisecorrectionCTTexpt}
  \mathrm{cov}(\hat
  C^{\hat\phi\hat\phi}_{L,\mathrm{noise}}, \hat C^{\tilde T\tilde
    T}_{L',\expt})_\mathrm{disconn.}
=
  \mathrm{cov}(\hat
  C^{\hat\phi\hat\phi}_{L}, \hat C^{\tilde T\tilde
    T}_{L',\expt})_\mathrm{disconn.}.
\end{equation}
The dominant
non-diagonal contribution of Eq.~\eqref{eq:cov-phi-power} to the
auto-covariance of the uncorrected reconstruction power  $\hat C^{\hat\phi\hat\phi}$
is due to couplings [see Eq.~(46c) in Ref.~\cite{hanson1008}]
\begin{equation}
  \label{eq:couplingsLeadingToCovCphirecCphirecNonDiag}
  \contraction{}{\tilde T_{\lbar{1}}}{}{\tilde T_{\lbar{3}}}
\contraction[2ex]{\tilde T_{\lbar{1}}\tilde T_{\lbar{3}}}{\tilde T_{\lbar{2}}}{\tilde T_{\lbar{4}}
\ \mid \ \tilde T_{\lbar{7}}}{\tilde T_{\lbar{5}}}
\contraction{\tilde T_{\lbar{1}}\tilde T_{\lbar{3}}\tilde T_{\lbar{2}}}{\tilde T_{\lbar{4}}}
{\ \mid \ }{\tilde T_{\lbar{7}}}
\contraction{\tilde T_{\lbar{1}}\tilde T_{\lbar{3}}\tilde T_{\lbar{2}}\tilde T_{\lbar{4}}
\ \mid \ \tilde T_{\lbar{7}} \tilde T_{\lbar{5}}}{\tilde T_{\lbar{6}}}{}{\tilde T_{\lbar{8}}}
\tilde T_{\lbar{1}}\tilde T_{\lbar{3}}\tilde T_{\lbar{2}}\tilde T_{\lbar{4}}
\ \mid \ \tilde T_{\lbar{7}} \tilde T_{\lbar{5}}\tilde T_{\lbar{6}}\tilde T_{\lbar{8}},
\end{equation}
which implies $l_1=l_3$ and $l_6=l_8$.
Writing out the contractions following
\cite{hanson1008} and summing over $m_1$ and $M$ in Eq.~(42) of \cite{hanson1008} enforces
$l_2=l_4$, and therefore also $l_5=l_7$. Since all terms of the
reconstruction power \eqref{eq:Cphiphihat} with these
coinciding multipoles are in the noise
term \eqref{eq:CphiphihatNoise},
the auto-covariance of
 $\hat
C^{\hat\phi\hat\phi}-\hat C^{\hat\phi\hat\phi}_\text{noise}$
does not contain the non-diagonal contribution
\eqref{eq:cov-phi-power}. 
In contrast, the covariance contributions that involve a product of two trispectra and
 lead to the dominant diagonal variance \eqref{eq:var-phi-power}
\cite{hanson1008} are still present, so that
\begin{equation}
  \label{eq:CovCphiphiNoiseCorrected}
\mathrm{cov}(\hat C^{\hat\phi\hat\phi}_L-\hat C^{\hat\phi\hat\phi}_{L,\mathrm{noise}},\hat
C^{\hat\phi\hat\phi}_{L'}-\hat C^{\hat\phi\hat\phi}_{L',\mathrm{noise}})\approx
\delta_{LL'}\frac{2}{2L+1}\la C^{\hat\phi\hat\phi}_L\ra^2.
\end{equation}

Thus, the chance-aligned noise terms  \eqref{eq:CphiphihatNoise}, which lead to
the $N^{(0)}$ bias, are
responsible for both the dominant non-diagonal reconstruction power
auto-covariance \eqref{eq:cov-phi-power} and the noise
contribution \eqref{eq:noisecontri_cov_CTT_Cphiphi_rec} to the
temperature-lensing power cross-covariance. It is therefore desirable
to avoid these noise terms. 
 In practice, a brute-force way to achieve this
would be to subtract
 $\hat C^{\hat\phi\hat\phi}_\text{noise}$ from $\hat
 C^{\hat\phi\hat\phi}$ directly, or, equivalently, to restrict the
 summation over the $l_i$ in \eqref{eq:Cphiphihat}
 appropriately.  
A more efficient method is obtained
from a ``partial-averaging'' procedure, which will turn out to be
equivalent to the empirical $\hat N^{(0)}$ subtraction.\footnote{An alternative would be ``phase randomization'' following \cite{ACTLensingDetectionDas1103}. Alternatively, one could split
  $\tilde T_{lm}$  into in- and out-annuli \cite{sherwinDasPowerWOBias1011}, which however 
   reduces the signal-to-noise \cite{SPTLensingDetectionVanEngelen1202}.  Note that while the goal of Ref.~\cite{sherwinDasPowerWOBias1011} was
  to avoid the Gaussian $N^{(0)}$ noise bias for any
  CMB realisation, our goal is to simplify the
  reconstruction power auto- and cross-covariance.  } 
Guided by the fact that both covariances \eqref{eq:cov-phi-power} and
 \eqref{eq:noisecontri_cov_CTT_Cphiphi_rec} are due to disconnected
 terms containing the contraction 
 \begin{equation}
   \label{eq:couplings13}
\contraction{}{\tilde T_{\lbar{1}}}{}{\tilde T_{\lbar{3}}}
   \tilde T_{\lbar{1}}\tilde T_{\lbar{3}}
 \end{equation}
[see Eqs.~\eqref{eq:CouplingsCovCphirecCTTNoiseContri} and \eqref{eq:couplingsLeadingToCovCphirecCphirecNonDiag}],
we take the corresponding expectation value already at the level of evaluating
\eqref{eq:CphiphihatNoise}, before computing covariances, by defining
the partial averaging operation, $\mathcal{R}$, by
 \begin{equation}
   \label{eq:RoperatorDef}
\mathcal{R}[
\tilde T_{\underline l_1}\tilde T_{\underline l_2}
\tilde T_{\underline l_3}\tilde T_{\underline l_4}
  ] \equiv
\la \tilde T_{\underline l_1}\tilde T_{\underline l_2}\ra\tilde T_{\underline l_3}\tilde
T_{\underline l_4}
+
5 \mbox{ perms.},
 \end{equation}
i.e.~we average out two of the four modes while keeping the
 unaveraged realisation of the other two modes.
 This leaves
 contractions of the form 
\eqref{eq:CouplingsCovCphirecCTTNoiseContri} in 
 the disconnected (4+2)-point function and contractions of
 the form \eqref{eq:couplingsLeadingToCovCphirecCphirecNonDiag} in the
 disconnected (4+4)-point function
invariant in the
 sense that
\begin{equation}
  \label{eq:RoperationNpointfctInvariance42}
  \la \mathcal{R}[
\tilde T_{\underline l_1}\tilde T_{\underline l_2}
\tilde T_{\underline l_3}\tilde T_{\underline l_4}
  ] \tilde T_{L'M'}  \tilde
  T_{L',-M'}\ra_\mathrm{disconn.}^{\eqref{eq:CouplingsCovCphirecCTTNoiseContri}\;
    \mathrm{terms}} 
\;=\;
\la
\tilde T_{\underline l_1}\tilde T_{\underline l_2}
\tilde T_{\underline l_3}\tilde T_{\underline l_4}
   \tilde T_{L'M'}  \tilde
   T_{L',-M'}\ra_\mathrm{disconn.}^{\eqref{eq:CouplingsCovCphirecCTTNoiseContri}
   \;\mathrm{terms}}
\end{equation}
and
\begin{equation}
  \label{eq:RoperationNpointfctInvariance44}
  \la \mathcal{R}[
\tilde T_{\underline l_1}
\cdots \tilde T_{\underline l_4}
  ]\mathcal{R}[ \tilde T_{\underline l_5}\cdots  \tilde T_{\underline l_8}]
\ra_\mathrm{disconn.}^{\mathrm{\eqref{eq:couplingsLeadingToCovCphirecCphirecNonDiag}}\;\mathrm{terms}}
\;=\;
  \la \mathcal{R}[
\tilde T_{\underline l_1}
\cdots\tilde T_{\underline l_4}
  ]( \tilde T_{\underline l_5}\cdots  \tilde T_{\underline l_8})
\ra_\mathrm{disconn.}^{\mathrm{\eqref{eq:couplingsLeadingToCovCphirecCphirecNonDiag}}\;\mathrm{terms}}
\;=\;
\la
\tilde T_{\underline l_1}
\cdots
\tilde T_{\underline
     l_8}
\ra_\mathrm{disconn.}^{\mathrm{\eqref{eq:couplingsLeadingToCovCphirecCphirecNonDiag}}\;\mathrm{terms}}.
\end{equation}
Therefore both covariance contributions   \eqref{eq:cov-phi-power} and
 \eqref{eq:noisecontri_cov_CTT_Cphiphi_rec}  are still eliminated if 
instead of subtracting the  full noise term
\eqref{eq:CphiphihatNoise} from the reconstruction power we subtract the
partial average of this noise term, i.e.~if
we consider (up to realisation-independent bias mitigation terms)
$\hat C^{\hat \phi\hat\phi}_{L} -   \mathcal{R}[\hat C^{\hat\phi\hat\phi}_{L,\mathrm{noise}} ]$.
We find that this partially-averaged noise mitigation
 reduces to the empirical
$\hat N^{(0)}$ subtraction,
\begin{equation}
  \label{eq:RoperatorOnCphiphihatNoise}
  \mathcal{R}[\hat C^{\hat\phi\hat\phi}_{L,\mathrm{noise}} ] = 2\hat
  N^{(0)}_L.
\end{equation}

Thus, the empirical $\hat N^{(0)}$ subtraction
can be interpreted as an efficient method to mitigate disconnected
(auto- and cross-) covariance contributions generated by the chance-aligned 
noise terms \eqref{eq:CphiphihatNoise}
in the reconstruction power spectrum.  In contrast to the noise terms
\eqref{eq:CphiphihatNoise}, the empirical $\hat N^{(0)}$ defined in
Eq.~\eqref{eq:empirical_N0} can be evaluated very efficiently because the
empirical temperature power spectrum is isolated in the sum.

\section{Temperature-lensing power covariance from the 
    CMB trispectrum}
\label{se:Conn4ptAppendix}
We will show here that contributions from the lensed CMB trispectrum
to the temperature-lensing power
covariance have a sub-dominant effect on parameter estimation compared to the fully disconnected
$\mathcal{O}(\phi^0)$ contribution and the $\mathcal{O} (\phi^4)$
contribution from the connected $6$-point function. 
The connected $4$-point function contributes to
Eq.~\eqref{eq:cov_CTT_Cphiphi_rec} with couplings of the form
$\la\tilde T_1\tilde T_3 \ra\la \tilde T_2\tilde T_4\tilde
T_{L'M'}\tilde T_{L',-M'} \ra_c$ (`type A') and $\la \tilde T_1\tilde
T_{L'M'}\ra\la \tilde T_2\tilde T_3\tilde T_4\tilde T_{L',-M'}\ra_c$
(`type B').  Other couplings either cancel in
Eq.~\eqref{eq:cov_CTT_Cphiphi_rec} or vanish because $\la
\hat\phi_{LM}\ra=0$. The contribution from type A can be expressed
non-perturbatively in terms of the derivative
in Eq.~\eqref{eq:Deriv2EmpN0CTT} and the connected $4$-point contribution to
the temperature power auto-covariance as
\begin{equation}
  \label{eq:4ptConnAcontriToTempLensingPowerCovIntermsofFullCov}
    \mathrm{cov}(\hat C^{\hat\phi\hat\phi}_L, \hat C^{\tilde T\tilde
    T}_{L',\expt})^{\mathrm{conn.4pt.A}} =
\sum_{l_2}   \frac{\partial (2\hat N^{(0)}_L)}{\partial\hat C^{\tilde
    T\tilde T}_{l_2,\expt}} 
\mathrm{cov}(\hat C^{\tilde T\tilde T}_{l_2,\expt}, \hat C^{\tilde
  T\tilde T}_{L',\expt})^{\mathrm{conn. } 4\mathrm{pt.}} .
\end{equation}
Perturbatively, up to $\mathcal{O} (\phi^4)$, we have
\begin{equation}
  \label{eq:4ptConnAcontriToTempLensingPowerCov}
  \mathrm{cov}(\hat C^{\hat\phi\hat\phi}_L, \hat C^{\tilde T\tilde
    T}_{L',\expt})^{\mathrm{conn.4pt.A}} =
\sum_{l_2}   \frac{\partial (2\hat N^{(0)}_L)}{\partial\hat C^{\tilde
    T\tilde T}_{l_2,\expt}} 
\left[
\mathrm{cov}(\hat C^{\tilde T\tilde T}_{l_2,\expt}, \hat C^{\tilde
  T\tilde T}_{L',\expt})|_{\mathcal{O} (\phi^2)}
+
\mathrm{cov}(\hat C^{\tilde T\tilde T}_{l_2,\expt}, \hat C^{\tilde
  T\tilde T}_{L',\expt})|_{\mathcal{O} (\phi^4)}
\right],
  \end{equation}
where the perturbative temperature covariances are given by Eq.~\eqref{eq:CTTexpt-cov}.
The covariance \eqref{eq:4ptConnAcontriToTempLensingPowerCov} can  be
  interpreted as the correction to the noise contribution
  of Eq.~\eqref{eq:noise_correlation_heuristic} due to the non-diagonal
  $\mathcal{O} (\phi^2)$ and   $\mathcal{O} (\phi^4)$ parts of the temperature power
  auto-covariance \eqref{eq:CTTexpt-cov}.  The
  correlation corresponding to
  \eqref{eq:4ptConnAcontriToTempLensingPowerCov} is at most $5\times 10^{-5}$, which is two orders of magnitude smaller than the
  dominant noise contribution.
  The induced correlation of the lensing amplitude estimates $\hat A$ and
  $\hat A'$ is less than $0.15\%$ (see \fig{correl_AAprime_plot}),
  i.e.~the effect of \eqref{eq:4ptConnAcontriToTempLensingPowerCov} is
  negligible. 

The type-A contribution to the covariance is removed by the empirical $\hat N^{(0)}$
  correction introduced in Sec.~\ref{se:fullSkyLensing}.
 As for the disconnected
contribution to the covariance (see Appendix~\ref{app:edgeworth}), this is actually a more general result that applies for anisotropic surveys and for an arbitrary quadratic estimate of the temperature power spectrum. To see this, we generalise to the $\hat{C}_L^{\phi\phi}$ estimator of Eq.~(\ref{eq:app2}), and to quadratic temperature
power spectrum estimates of the form $T^{(1)}_p T^{(2)}_q$. The part of the $\hat{C}_L^{\phi\phi}$ estimator that is quartic in the data has a type-A, 4-point covariance with $T^{(1)}_p T^{(2)}_q$ given by
\begin{align}
\mathrm{cov}(\bar{T}_i \bar{T}_j \bar{T}_k \bar{T}_l, T^{(1)}_p T^{(2)}_q )^{\mathrm{conn.4pt.A}} &= \langle \bar{T}_i \bar{T}_j T^{(1)}_p T^{(2)}_q \rangle_c \langle \bar{T}_k \bar{T}_l \rangle + \text{5 terms} ,
\label{eq:appgen1}
\end{align}
where the other five terms are inequivalent permutations of $i$, $j$, $k$ and $l$.
The part of the $\hat{C}_L^{\phi\phi}$ estimator that is quadratic in the data is composed of six terms of the form $- \langle \bar{T}_i \bar{T}_j \rangle \bar{T}_k \bar{T}_l$. The contribution from the connected 4-point function to the covariance of these terms with $T^{(1)}_p T^{(2)}_q$ exactly cancels with the terms of the right-hand side of Eq.~(\ref{eq:appgen1}), removing the type-A covariance.

Terms of coupling type B have contributions from the primary
trispectrum term, where sums over $m_i$ simplify due to orthogonality
relations of $3j$-symbols, and from non-primary trispectrum terms,
where sums over $m_i$ lead to a non-trivial $6j$-symbol:
\begin{eqnarray}
  \label{eq:4ptConnBprimaryContriToTempLensingPowerCov}
&&  \mathrm{cov}(\hat C^{\hat\phi\hat\phi}_L,\hat C^{\tilde T\tilde
    T}_{L',\expt})^{\mathrm{conn.4pt.B }
    }_{\mathrm{primary}} 
=
2\frac{C_L^{\phi\phi}}{A_L}\frac{\partial (2\hat N^{(0)}_L)}{\partial\hat C^{\tilde
    T\tilde T}_{L',\expt}}\frac{2}{2L'+1} (C^{\tilde T\tilde
  T}_{L',\expt})^2
=2\frac{C_L^{\phi\phi}}{A_L}\mathrm{cov}(\hat
C^{\hat\phi\hat\phi}_L,\hat C^{\tilde T\tilde T}_{L',\expt})_\mathrm{disconn.}\,,\\
&&  \mathrm{cov}(\hat C^{\hat\phi\hat\phi}_L,\hat C^{\tilde T\tilde
    T}_{L',\expt})^{\mathrm{conn.4pt.B}}_{\text{non-primary}} 
=
\sum_{\stackrel{m'L''}{l_2l_3l_4}}
 \frac{A_L^2(-1)^{m'}C^{\phi\phi}_{L''}
\tilde f^\mathrm{no3j}_{L'Ll_2}\tilde f^\mathrm{(u)}_{l_3Ll_4}
\tilde f^\mathrm{(u)}_{L'L''l_3}\tilde f^\mathrm{(u)}_{l_2L''l_4}\;
\mathcal{G}^{l_3Ll_4}_{-m'0m'}
\mathcal{G}^{L'L''l_3}_{0-m'm'}
\mathcal{G}^{l_2L''l_4}_{0m'-m'}
} 
 {4 (2L+1) (2L'+1)C^{\tilde T\tilde T}_{l_2,\expt}C^{\tilde T\tilde
     T}_{l_3,\expt}C^{\tilde T\tilde T}_{l_4,\expt}}
  \label{eq:4ptConnBNonprimaryContriToTempLensingPowerCov}
 + (l_3\leftrightarrow l_4).\qquad\;
\end{eqnarray}
Here we used
Eq.~\eqref{eq:lensing_trispectrum} and expressed the product of a
$6j$-symbol with a $3j$-symbol as a sum over a product of three
$3j$-symbols \cite{NISTWigner}. 
The $\tilde{f}$ factors are defined in Appendix~\ref{se:N0BiasAppendix}.
Neither contributions in
Eqs.~\eqref{eq:4ptConnBprimaryContriToTempLensingPowerCov} and
\eqref{eq:4ptConnBNonprimaryContriToTempLensingPowerCov} are
cancelled by empirical $\hat N^{(0)}$ subtraction.  

\begin{figure}[t]
\includegraphics[width=0.45\textwidth]{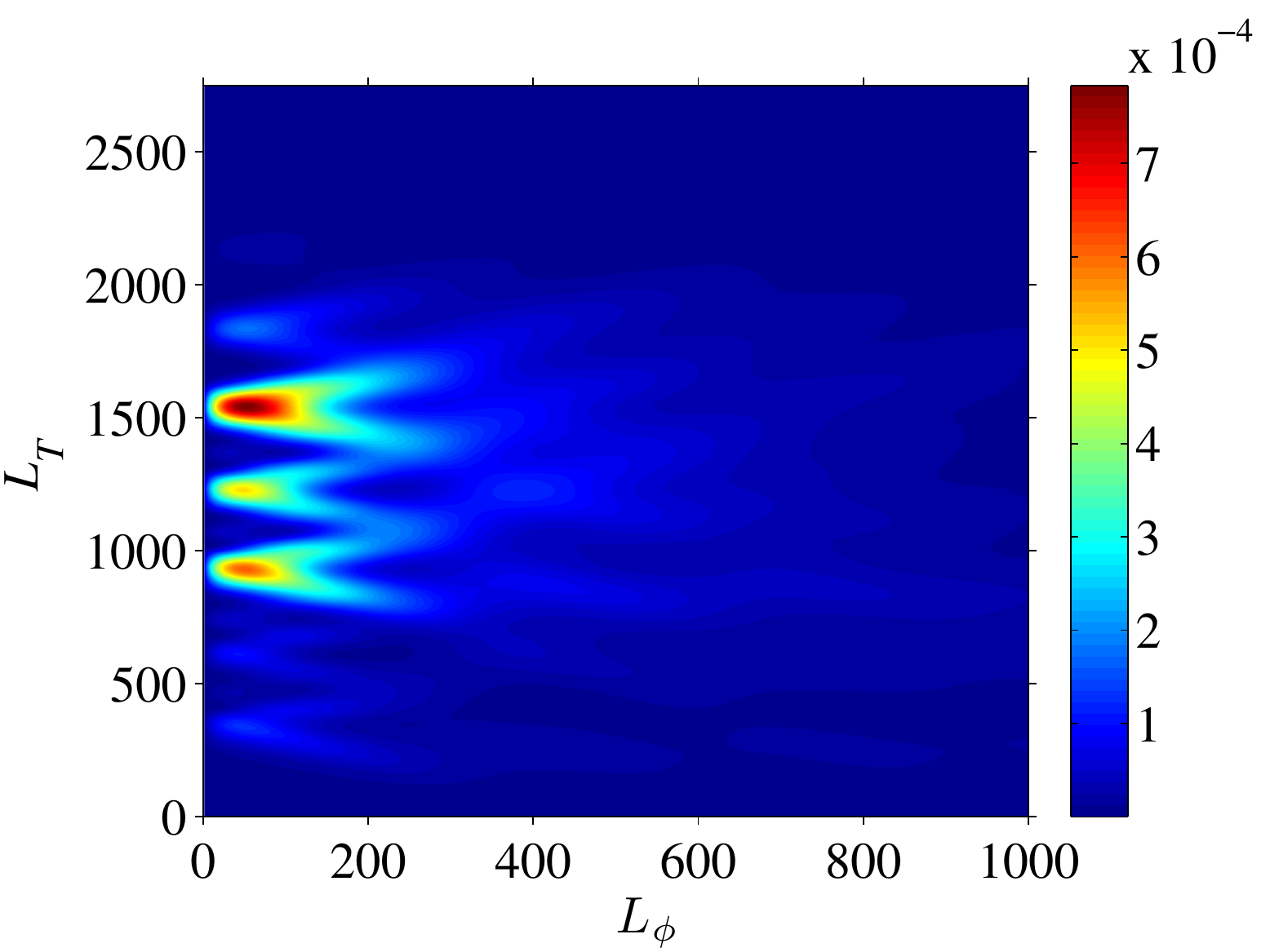}
\caption{Theoretical contribution from the primary type-B connected
  $4$-point function to the correlation of the unbinned power spectra
  of the lens reconstruction and the (lensed) temperature power spectrum.  The
  covariance \eqref{eq:4ptConnBprimaryContriToTempLensingPowerCov} is
  converted to a correlation using the same conversion factor as in
  Eq.~\eqref{eq:correl_noise_contri}. Correlations for $L_\phi>1000$ are
  too small to be visible in this plot. }
\label{fig:TempLensingCorrel4ptBPrimary} 
\end{figure}

The correlation of unbinned power spectra from the primary term
of Eq.~\eqref{eq:4ptConnBprimaryContriToTempLensingPowerCov} is shown in
\fig{TempLensingCorrel4ptBPrimary}. It is at most
$0.08\%$, which is almost a factor of two larger than the maximum of
the matter cosmic variance contribution [Eq.~\eqref{eq:matter_contri_theo}]
shown in \fig{cov-inphi-t-theory}.  The structure is very similar to that of the
disconnected (noise) contribution [Eq.~\eqref{eq:noisecontri_cov_CTT_Cphiphi_rec}] shown in \fig{cov-recphi-t-theory},
but with the additional signal-to-noise factor of $C_L^{\phi\phi}/A_L$ that falls
off rapidly for $L > 200$. (Recall, $A_L = N_L^{(0)}$ for our choice of optimal weights.)
Therefore the contribution of the primary type-B covariance to correlations between the lensing amplitudes $\hat A$ and $\hat A'$ is suppressed compared to that of the disconnected noise contribution, reaching at most $2\%$ (see \fig{correl_AAprime_plot}). 
It is also suppressed compared to the matter cosmic variance contribution because $\hat A'$ gives most weight to CMB modes at the acoustic peaks and troughs where the primary type-B power correlation is small.

While the measured correlation of $\hat A$ and $\hat A'$ is consistent with our
theoretical expectations, the measurements are too noisy to
test sub-dominant terms such as that in Eq.~\eqref{eq:4ptConnBprimaryContriToTempLensingPowerCov}.
In \fig{cov-recphi_SUBempN0-t-sims}, the lowest $L_\phi$ pixels show hints
of the structure plotted
in \fig{TempLensingCorrel4ptBPrimary}, but the measurements are, again, too
noisy to provide a conclusive test of Eq.~\eqref{eq:4ptConnBprimaryContriToTempLensingPowerCov}. The simulations do however imply that there cannot be
significantly larger covariance contributions than the ones we model
analytically. 

Although the summation in the expression
\eqref{eq:4ptConnBNonprimaryContriToTempLensingPowerCov} for the
non-primary type-B terms is restricted by triangle inequalities, its evaluation is still numerically
challenging. We could only evaluate
Eq.~\eqref{eq:4ptConnBNonprimaryContriToTempLensingPowerCov} for $30$
$(L,L')$ pairs at $L\lesssim 400$, which is relevant for the current Planck
lensing likelihood~\cite{2013arXiv1303.5077P}.  At these points the power correlation from Eq.~\eqref{eq:4ptConnBNonprimaryContriToTempLensingPowerCov} is at most
around $4\times 10^{-5}$, which is an order of magnitude smaller than the maximum correlation from the
primary term [Eq.~\eqref{eq:4ptConnBprimaryContriToTempLensingPowerCov}].  The structure of
the correlation matrix seems similar to that of the primary term in the $L_T$
direction, but seems to peak towards higher $L_\phi$ (but we cannot assess the
structure reliably from the small number of evaluation points).
Moreover, given that our simulations do not show any
significant excess covariance we expect the non-primary term
of Eq.~\eqref{eq:4ptConnBNonprimaryContriToTempLensingPowerCov} to be
negligible.

\section{Temperature-lensing power covariance from the connected CMB 6-point function}
\label{se:connected_6point_orderphi4}
The $\mathcal{O} (\phi^2)$ contribution to the connected part of the
lensed temperature $6$-point function vanishes
\cite{KCK0208cumulants}. Here we will compute the $\mathcal{O}
(\phi^4)$
terms. 
There are five types of terms, which involve $\delta^{4}T$,
$\delta^3 T\delta T$, $\delta^2T\delta^2T$, $\delta^2 T\delta T\delta
T$ or $\delta T\delta T\delta T\delta T$ with the appropriate number of factors of the
unlensed CMB. Only the last two types of
terms contribute to the connected 6-point function; the remaining terms contribute to the full six-point function but are cancelled when the connected 4-point and fully disconnected parts are removed.

Writing out only the two
relevant types of terms, the full $6$-point function at
$\mathcal{O}(\phi^4)$ is ($\tilde T_i\equiv\tilde T_{l_im_i}$)
\begin{equation}
  \label{eq:6pt_orderphi4_full}
  \la \tilde T_1\tilde T_2\tilde T_3\tilde T_4\tilde T_5\tilde T_6
  \ra^{(4)}=
\frac{1}{12}\left[
\la \delta^2 T_1\delta T_2 \delta T_3 T_4 T_5 T_6\ra+\mbox{all perms}
\right]+
\frac{1}{48}\left[
\la \delta T_1\delta T_2 \delta T_3 \delta T_4 T_5 T_6\ra+\mbox{all perms}
\right].
\end{equation}
From this we must subtract the following terms involving the connected 4-point function,
\begin{eqnarray}
\nonumber
  \tfrac{1}{48}\big[
\la\tilde T_1\tilde T_2\tilde T_3\tilde T_4\ra_c\la \tilde T_5\tilde
T_6\ra+\mbox{all perms}
\big] &=
  \frac{1}{48}\big[  &
6\la \delta T_1\delta T_2 T_3T_4\ra \la \delta T_5\delta T_6\ra
+12 \la\delta T_1\delta T_2 T_3 T_4\ra\la \delta^2T_5T_6\ra \\
\nonumber
& &+ 12 \la \delta^2T_1\delta T_2\delta T_3 T_4\ra \la T_5T_6 \ra
+ \la \delta T_1 \delta T_2\delta T_3\delta T_4\ra\la T_5T_6\ra \\
  \label{eq:6pt_orderphi4_from_conn4pt}
& & -\left(9\la\delta T_1\delta T_2\ra+24\la\delta^2 T_1 T_2\ra\right)\la\delta T_3\delta T_4\ra\la T_5 T_6\ra
+\mbox{all perms}
\big]+\cdots,
\end{eqnarray}
and the fully disconnected part,
\begin{eqnarray}
\nonumber
  \la \tilde T_1\tilde T_2\tilde T_3\tilde T_4\tilde T_5\tilde T_6
  \ra^{(4)}_\mathrm{disc}&=&
  \tfrac{1}{2}\left[
\la\delta^2T_1\delta T_2\ra \la \delta T_3 T_4\ra \la
T_5T_6\ra+\mbox{all perms}\right]+
  \tfrac{1}{4}\left[
\la\delta^2T_1T_2\ra \la \delta T_3 \delta T_4\ra \la
T_5T_6\ra+\mbox{all perms}\right]\\
\nonumber
&&+
  \tfrac{1}{2}\left[
\la\delta^2T_1T_2\ra \la \delta T_3  T_4\ra \la
\delta T_5T_6\ra+\mbox{all perms}\right]+
  \tfrac{1}{16}\left[
\la\delta T_1\delta T_2\ra \la \delta T_3 \delta T_4\ra \la
 T_5T_6\ra+\mbox{all perms}\right]\\
  \label{eq:6pt_orderphi4_totally_disconnected}
&&+
  \tfrac{1}{4}\left[
\la\delta T_1\delta T_2\ra \la \delta T_3 T_4\ra \la
\delta T_5T_6\ra+\mbox{all perms}\right]+\cdots.
\end{eqnarray}
To evaluate these expressions, we use, for example,
\begin{eqnarray}
  \label{eq:6pt_gaussian_example_lss_cmb_split}
  \la\delta T_1\delta T_2T_3T_4\ra=
\la\delta T_1\delta T_2\ra\la T_3T_4\ra
+ \la \la \delta T_1T_3\ra_\mathrm{CMB}\la \delta
T_2T_4\ra_\mathrm{CMB}\ra_\mathrm{LSS}
+ \la \la \delta T_1T_4\ra_\mathrm{CMB}\la \delta
T_2T_3\ra_\mathrm{CMB}\ra_\mathrm{LSS},
\end{eqnarray}
where $\la\ra_\mathrm{CMB}$ denotes averaging over unlensed CMB
realisations and $\la\ra_{\mathrm{LSS}}$ denotes averaging over
realisations of the lensing potential $\phi$. Rewriting the other
$4$- and $6$-point functions in a similar way gives  the following final
expression for the connected 6-point function:
\begin{eqnarray}
\nonumber
    \la \tilde T_1\tilde T_2\tilde T_3\tilde T_4\tilde T_5\tilde T_6
  \ra^{(4)}_c&=&
\tfrac{1}{2}\mathrm{cov}_\mathrm{LSS}\left(
\la \delta^2T_1T_2\ra_\mathrm{CMB}, 
\la \delta T_3 T_4\ra_\mathrm{CMB}\la \delta T_5 T_6\ra_\mathrm{CMB}
\right) + \mbox{all perms} \\
\nonumber
&& +\tfrac{1}{4}\mathrm{cov}_\mathrm{LSS}\left(
\la \delta T_1\delta T_2\ra_\mathrm{CMB}, 
\la \delta T_3 T_4\ra_\mathrm{CMB}\la \delta T_5 T_6\ra_\mathrm{CMB}
\right) + \mbox{all perms}\\
  \label{eq:6pt_orderphi4_connected_final1}
&=&\tfrac{1}{16}\mathrm{cov}_\mathrm{LSS}\left(
\la \tilde T_1 \tilde T_2\ra^{(2)}_\mathrm{CMB}, 
\la \tilde T_3 \tilde T_4\ra^{(1)}_\mathrm{CMB}\la \tilde T_5 \tilde T_6\ra^{(1)}_\mathrm{CMB}
\right) + \mbox{all perms}.
\end{eqnarray}
The result is written in terms of the covariance over
different realisations of the lensing potential,
$\mathrm{cov}_\mathrm{LSS}(X,Y)\equiv \la XY\ra_\mathrm{LSS}-\la
X\ra_\mathrm{LSS}\la Y\ra_\mathrm{LSS}$.

We expect only a few of the $720$ permutations in
Eq.~\eqref{eq:6pt_orderphi4_connected_final1} to be relevant for the
lensing-temperature power covariance
[Eq.~\eqref{eq:cov_CTT_Cphiphi_rec}]. The weights $\tilde g$ in
Eq.~\eqref{eq:cov_CTT_Cphiphi_rec} impose triangle conditions on
$l_1,l_2,L$ and $l_3,l_4,L$, which constrain the summation volume
(especially at low $L$ where the lensing power dominates). Terms that
couple e.g.~$\la\tilde T_1\tilde T_3\ra_{\text{CMB}}$ impose an additional triangle
constraint on $l_1, l_3$ and a lensing multipole $\tilde L$, which
reduces the summation volume further. This is not the case for
couplings of the type $\la \tilde T_1 \tilde T_2\ra_{\text{CMB}}$ and $\la \tilde T_3
\tilde T_4\ra_{\text{CMB}}$. For this reason we expect the dominant terms to come from the
couplings $12$, $34$ and $(L',M';L',-M')$ in
\eqref{eq:cov_CTT_Cphiphi_rec}, i.e.~terms which factor most under the
weights \cite{hanson1008}. The dominant contribution of the connected
$6$-point function at $\mathcal{O} (\phi^4)$ to the
lensing-temperature power covariance is
therefore expected to come from
\begin{eqnarray}
  \nonumber
  (-1)^{M'}\la\tilde T_1\tilde T_2\tilde T_3\tilde T_4 \tilde T_{L'M'}\tilde
T_{L',-M'}\ra_{c,\mathrm{dom.}}^{(4)} &=&
\mathrm{cov}_\mathrm{LSS}\left(
\la \tilde T_{L'M'}\tilde T^*_{L'M'}\ra^{(2)}_\mathrm{CMB},
\la \tilde T_1 \tilde T_2\ra^{(1)}_\mathrm{CMB}\la \tilde T_3 \tilde T_4\ra^{(1)}_\mathrm{CMB}
\right)
\\
\label{eq:6pt_orderphi4_connected_dominant}
&&
+ 2\,
\mathrm{cov}_\mathrm{LSS}\left(
\la \tilde T_1\tilde T_2\ra_\mathrm{CMB}^{(2)}, 
\la \tilde T_{L'M'} \tilde T^*_{L'M'}\ra^{(1)}_\mathrm{CMB}
\la \tilde T_3 \tilde T_4\ra^{(1)}_\mathrm{CMB}
\right) .
\end{eqnarray}
Here, we have used the result that permutations of the form
$1\leftrightarrow 2$,  $3\leftrightarrow 4$,
$(12)\leftrightarrow (34)$ and $(L',M')\leftrightarrow (L',-M')$
 all lead to the same contribution to
 Eq.~\eqref{eq:cov_CTT_Cphiphi_rec}. 

The $\la \tilde T_1 \tilde T_2 \ra^{(1)}_{\text{CMB}}$ term on the right of Eq.~\eqref{eq:6pt_orderphi4_connected_dominant} evaluates to
 \begin{eqnarray}
    \label{eq:deltaT_T_avg_over_CMB}
    \la \tilde T_1 \tilde T_2\ra^{(1)}_\mathrm{CMB} &=&
\sum_{L_1 M_1} (-1)^{M_1}  
\left( \begin{matrix} 
l_1 & l_2 & L_1  \\ 
m_1 & m_2 & -M_1
\end{matrix} \right)
f_{l_1L_1l_2} \phi_{L_1 M_1} ,
  \end{eqnarray}
which combines with the weights, normalisation and one of the $3j$ symbols in Eq.~\eqref{eq:cov_CTT_Cphiphi_rec} to give $\phi_{LM} + \mathcal{O}(\phi^2)$.
Similarly, the $\la \tilde T_3 \tilde T_4 \ra^{(1)}_{\text{CMB}}$ term returns $\phi_{L,-M}$.
Equation~\eqref{eq:deltaT_T_avg_over_CMB} also shows that the second term on the
right of Eq.~\eqref{eq:6pt_orderphi4_connected_dominant} does not contribute to
the power covariance since $\la \tilde T_{L'M'} \tilde T^*_{L' M'} \ra^{(1)}_{\text{CMB}}$ returns the monopole of $\phi$ on summing over $M'$.

Putting these pieces together, and using Eq.~\eqref{eq:TTpower_avg_only_over_CMB}, finally gives
\begin{eqnarray}
\nonumber
  \mathrm{cov}(\hat C^{\hat\phi\hat\phi}_L,\hat C^{\tilde T\tilde
    T}_{L'})_{\la\tilde T_1\cdots\tilde T_6\ra_{c,\mathrm{dom.}}^{(4)}} &=&
\sum_{L''} \frac{\partial C^{\tilde T\tilde T}_{L'}}{\partial C^{\phi\phi}_{L''}}
\text{cov}_{\text{LSS}}(\hat{C}^{\phi\phi}_{L''},\hat{C}_L^{\phi\phi}) +\mathcal{O}(\phi^5) \\
&=& 
\frac{2}{2L+1}(C^{\phi\phi}_L)^2
\frac{\partial C^{\tilde T\tilde T}_{L'}}{\partial C^{\phi\phi}_{L}}
+\mathcal{O}(\phi^5).
\label{eq:matter_contri_theo}
\end{eqnarray}

\section{\texorpdfstring{Effect of $C^{T\phi}$}{Effect of CTphi}}
\label{se:CTPhiEeffect}
In this appendix we discuss the contribution of the ISW-induced large-scale $C^{T\phi}$
correlation on the temperature-lensing power
covariance. We first compute corrections due to the ISW-lensing
bispectrum and then the ones due to corrections of the lensed temperature
power spectrum.

\begin{figure}[t]
\includegraphics[width=0.4\textwidth]{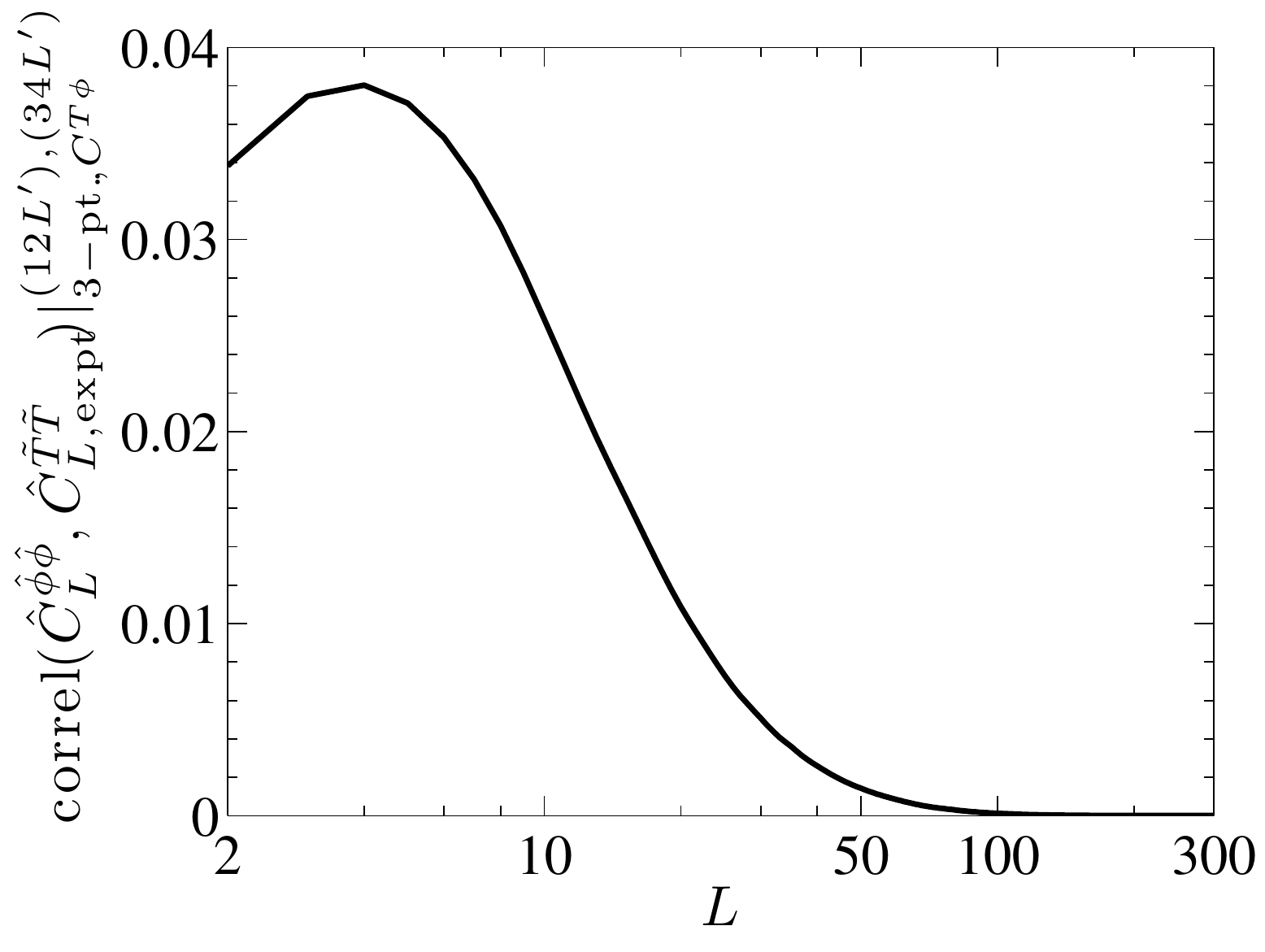}
\caption{ISW-lensing bispectrum
  contribution of Eq.~\eqref{eq:cov_CTT_CPhiphi_rec_CTphi_bispbetaterms} to
  the diagonal unbinned temperature-lensing power correlation. The
  difference between the full expression and the approximation in
  Eq.~\eqref{eq:cov_CTT_CPhiphi_rec_CTphi_bispbetaterms} is too small to
  be visible in this plot. }
\label{fig:correl_CPhirec_CTTlensed_CPhiT_bispbetaterms_theo_contribution}
\end{figure}

The lensed temperature $6$-point function in the temperature-lensing
power covariance of Eq.~\eqref{eq:cov_CTT_Cphiphi_rec} involves the following 3-point terms:
\begin{equation}
  \label{eq:lensed6point_BTphi_pieces}
   \langle \tilde T_{\underline l_1} 
\tilde T_{\underline l_2} \tilde T_{\underline l_3} \tilde
T_{\underline l_4} \tilde T_{L'M'} \tilde
T_{L'\,-M'}\rangle|_{\text{3-pt}, C^{T\phi}}
=
   \langle \tilde T_{\underline l_1} 
\tilde T_{\underline l_2} \tilde T_{\underline l_3} \rangle^\mathrm{ISW}
\langle
\tilde T_{\underline l_4} \tilde T_{L'M'} \tilde
T_{L'\,-M'}\rangle^\mathrm{ISW} + 9\,\mathrm{perms},
\end{equation}
where the non-perturbative ISW-lensing bispectrum
is approximately given by \cite{lewis1101,Hanson0905ISWLensing}
\begin{equation}
  \label{eq:TTTlensed_CTphi}
   \langle \tilde T_{\underline l_1} 
\tilde T_{\underline l_2} \tilde T_{\underline l_3} \rangle^\mathrm{ISW}
=
\left( \begin{matrix} 
l_1 & l_2 & l_3 \\ 
m_1 & m_2 & m_3
\end{matrix} \right)
B^\mathrm{ISW}_{l_1l_2l_3}
= 
\left( \begin{matrix} 
l_1 & l_2 & l_3 \\ 
m_1 & m_2 & m_3
\end{matrix} \right)
C^{T\phi}_{l_1}\tilde f_{l_2l_1l_3} 
+2\,\mbox{perms}.
\end{equation}
While the contribution to Eq.~\eqref{eq:cov_CTT_Cphiphi_rec} 
from the coupling $(123)(4L'L')$, which is written out explicitly in
Eq.~\eqref{eq:lensed6point_BTphi_pieces}, vanishes on summing over $M'$,
the couplings of type 
$(12L')(34L')$ yield the diagonal covariance contribution\footnote{To obtain the approximate result in the last step we use
Eq.~\eqref{eq:A_L_def} and neglect all terms depending on
$C^{T\phi}_{l_1}$ or $C^{T\phi}_{l_2}$, because they enforce $l_1$ or
$l_2$ to be small, which reduces the remaining summation volume
due to the triangle condition on $l_1$, $l_2$ and $L$. We find that using
the approximation instead of the full expression leads a
temperature-lensing correlation which is wrong by at most $3\times
10^{-6}$ for $L\le 300$.}
\begin{equation}
  \label{eq:cov_CTT_CPhiphi_rec_CTphi_bispbetaterms}
  \mathrm{cov}(\hat C^{\hat\phi\hat\phi}_L, \hat C^{ \tilde T
   \tilde T}_{L',\expt})|_{\text{3-pt.},C^{T\phi}}^{(12L'),(34L')} = 
\delta_{LL'}\frac{2A_L^2}{(2L+1)^3}\left[
\sum_{l_1l_2}\tilde g_{l_1l_2}(L)
B^\mathrm{ISW}_{l_1l_2L}
\right]^2\approx
\delta_{LL'}\frac{2}{2L+1}(C^{T\phi}_L)^2.
\end{equation}
The approximation on the right is simply the covariance between the power spectra
of the input lensing potential and the lensed temperature, an intuitive result that we might have anticipated.
For our Planck-like parameters, the corresponding power correlation
 is always less than $5\%$ and rapidly decreases
with increasing multipoles, being less than $0.1\%$ for $L\geq
60$ (see \fig{correl_CPhirec_CTTlensed_CPhiT_bispbetaterms_theo_contribution}).  
The induced correlation of the lensing amplitudes $A$ and $A'$ is
less than $10^{-5}$ and therefore negligible.  This small correlation arises because most of the information on the lensing amplitude from the temperature power spectrum comes from small scales. The power correlation due to the ISW-lensing effect is small on these scales, and, additionally, there is limited information in the lensing reconstruction on such scales as the signal-to-noise is very low there.
We expect that couplings of the type $(13L')(24L')$ in
Eq.~\eqref{eq:lensed6point_BTphi_pieces}, that do not factor under the weights, are further suppressed as they limit the summation volume.

The ISW-induced change in the lensed temperature power spectrum,
\begin{equation}
  \label{eq:CTTlensed_CTphi_Delta}
  \Delta C^{\tilde T\tilde T}_{L} = 
C^{\tilde T\tilde T}_{L}-C^{\tilde T\tilde T}_{L}|_{C^{T\phi}=0}
= \frac{1}{2L+1} 
\sum\limits_{L_1,L_2} F^2_{LL_1L_2} C_{L_1}^{T\phi}C_{L_2}^{T\phi},
\end{equation}
leads to a correction of the disconnected
noise contribution [Eq.~\eqref{eq:noisecontri_cov_CTT_Cphiphi_rec}] to the
temperature-lensing power covariance which is also second order in
$C^{T\phi}$.  The correction has a similar
structure to the noise contribution itself (shown in
\fig{cov-recphi-t-theory}), but it is around $10^{6}$ times smaller and
induces a lensing amplitude correlation of $\mathcal{O}(10^{-9})$ that is totally negligible.

\bibliography{marcel_lensing}

\label{lastpage}

\end{document}